\documentclass[traditabstract]{aa} % for the abstract without structuration
\usepackage{graphicx}
\usepackage{txfonts}
\usepackage{mathtools}
\usepackage{numprint}
\usepackage{color}
\usepackage{placeins}
\npdecimalsign{.}
\nprounddigits{0}
\newcolumntype{M}[2]{>{\hfill}n{#1}{#2}<{\hfill}}
\newcolumntype{N}[2]{>{\hfill}n{#1}{#2}<{\hfill}}
\usepackage[colorlinks,linkcolor=blue,anchorcolor=blue,citecolor=blue]{hyperref}

\def\la{\lower.5ex\hbox{$\; \buildrel < \over \sim \;$}}
\def\ga{\lower.5ex\hbox{$\; \buildrel > \over \sim \;$}} 
\begin{document}
   \title{Deciphering the radio-star formation correlation on kpc scales}

   \subtitle{IV. Radio halos of highly-inclined Virgo cluster spiral galaxies}

   \author{B.~Vollmer\inst{1}, M.~Soida\inst{2}, and V.~Heesen\inst{3}}

   \institute{Universit\'e de Strasbourg, CNRS, Observatoire Astronomique de Strasbourg, UMR 7550, 67000 Strasbourg, France \and
     Astronomical Observatory, Jagiellonian University, ul. Orla 171, 30-244 Krak\'ow, Poland \and
     Hamburg University, Hamburger Sternwarte, Gojenbergsweg 112, 21029 Hamburg, Germany}

   \date{Received ; accepted }

% \abstract{}{}{}{}{}
% 5 {} token are mandatory

  \abstract
      {In addition to the radio continuum emission of the thin galactic disk, vertically extended emission is ubiquitous in
        starforming disk galaxies. This halo emission can represent an important fraction of the total emission of the galaxy.
        The cosmic ray electrons (CRe) responsible for the radio continuum emission are produced within the thin disk and
        transported into the halo. They might interact with the warm neutral and ionized medium, which are also present in the
        halo region. We made an attempt to reconstruct the radial properties of radio continuum halos in nearly edge-on
        galaxies where the star formation rate (SFR) distribution can be deprojected and the vertical radio continuum emission is
        well distinct from the disk emission. The deprojected SFR distribution is convolved with a Gaussian kernel to take CRe
        diffusion within the galactic disk into account and a vertical profile of the radio continuum emissivity is added to the
        disk emission. The three-dimensional emission distribution is then projected on the sky and compared to VLA radio
        continuum observations at 20 and 6 cm. We found that overall the halo emission contains information on the underlying
        distribution of the star formation rate. The majority of our galaxies show flaring radio continuum halos. Except for
        one galaxy, our Virgo galaxies follow the trend of increasing effective height with increasing radio continuum size
        found by the CHANG-ES collaboration. We confirm that radio continuum halos can represent a significant fraction
        of the total radio continuum emission of a starforming spiral galaxy. At 20 cm and 6 cm between 30 and 70\,\% of the
        total radio continuum emission originate in the halo. We propose a halo classification based on the height ratio and SI
        between $20$ and $6$~cm. If we interpret the vertical structures of the large-scale magnetic
        field within the disk-halo and halo types as a sign of a galactic outflow or wind, all galaxies except one most probably
        harbor an advection-dominated halo.
      }

   \keywords{galaxies: ISM -- galaxies: magnetic fields -- radio continuum: galaxies}

   \authorrunning{Vollmer et al.}
   %\titlerunning{Large-scale radio continuum properties of 19 Virgo galaxies}

   \maketitle
%
%________________________________________________________________
\nolinenumbers

\section{Introduction\label{sec:introduction}}

The vertical scale heights of the different phases of the Galaxy's interstellar medium (ISM) vary between $\sim 70$~pc for
the molecular gas, $\sim 140$~pc for the cold neutral medium, and $\sim 400$~pc for the warm neutral medium (Boulares \& Cox 1990).
In addition, the warm ionized medium (WIM) or diffuse ionized gas (DIG) extends to a height of $\sim 1$~kpc (e.g., Haffner 2009).
Beyond the Galaxy, similar distributions of the DIG have been observed in many nearby disk galaxies (e.g., Dettmar 1990; Rand et al.
1990; Zurita et al. 2000; Levy et al. 2019). Diffuse radio continuum emission, which is generally associated to the high-latitude WIM
(Krause et al. 2018) and is called a radio continuum halo. The mechanism of the radio continuum emission is synchrotron emission stemming from
cosmic ray (CR) electrons interacting with the magnetic field of a galaxy. 
Cosmic ray particles are mainly produced in supernova shocks via Fermi acceleration. However, the relativistic electrons
do not stay at the location of their creation. They propagate either via diffusion, or by streaming with the Alfv\'en velocity.
In addition, cosmic ray electrons can be transported into the halo by advection, meaning a galactic wind or outflow.
The CHANG-ES project (Irwin et al. 2012) showed that for starforming disk galaxies the vertical transport of CR electrons
is mostly caused by advection (Heesen 2021; Irwin et al. 2024):
only two out of eleven analyzed CHANG-ES galaxies show a diffusion-dominated
radio halo. The best-known discriminator thus far is the star-formation rate surface density.
Galaxies at low star formation rate (SFR) surface densities have diffusive halos. The heights of the radio halos are set by either diffusion and advection
together with CR electron escape or synchrotron losses. The typical scale height of radio continuum halos is about $1$~kpc
(Krause et al. 2018), very close to that of the WIM.

The CHANG-ES collaboration studied the radio continuum halos of edge-on galaxies ($i=76$-$90^{\circ}$) by fitting Gaussian and exponential functions to
the vertical profiles of the radio continuum emission at $6$ (C-band) and $20$cm (L-band).
The CHANG-ES project is described in Irwin et al. (2012) and details about the first data release are provided by Wiegert et al. (2015).
In this work, we go a step further by looking at highly-inclined galaxies ($70^\circ \le i \le 78^\circ$) for which maps
of the SFR surface density are available.
Vollmer et al. (2020) predicted radio continuum maps by convolving the source map, represented by that of the SFR surface density,
with Gaussian (diffusion) and exponential (streaming) kernels. Here, we add a vertical component to the predicted radio continuum maps,
which is anchored in the starforming disk.
We would like to know if the halo radio continuum emission carries information on the underlying disk and if the radio continuum halo
has a constant height or if it is flaring.
Extending the CHANG-ES methodology, we use the spectral index (SI) of the halo emission to investigate if a halo is (i) advection or
diffusion and (ii) escape or synchrotron loss-dominated.

This article is structured in the following way: the radio continuum data and the those used to calculate the star formation maps
are presented in Sect.~\ref{sec:observations}. Our analytical model for the 3D distribution of the radio continuum emissivity
and its projection of the sky are explained in Sect.~\ref{sec:method}. The presentation of the results (Sect.~\ref{sec:results})
is followed by a classification of the radio halos (Sect.~\ref{sec:physics}). We discuss our results in Sect.~\ref{sec:discussion}
and give our conclusions in Sect.~\ref{sec:conclusions}.

\section{Data\label{sec:observations}}

The integrated SFR, the radio continuum flux at $4.85$~GHz from Vollmer et al. (2013), and the stellar mass
from Boselli et al. (2015) of each galaxy are presented in Table~\ref{tab:sample}.
\begin{table*}
      \caption{Galaxy sample}
         \label{tab:sample}
      \[
       \begin{tabular}{lcccccccccc}
        \hline
         & D & i & $S_{4.85~{\rm GHz}}$ & $f^{\rm therm}_{4.85~{\rm GHz}}$ &$S_{1.4~{\rm GHz}}$ & $f^{\rm therm}_{1.4~{\rm GHz}}$ & SI & $D_{4.85~{\rm GHz}}$ & SFR & log($M_*$) \\
         & (Mpc) & (deg) & (mJy) & & (mJy) & & & (kpc) & (M$_{\odot}$yr$^{-1}$) & (M$_{\odot}$) \\
        \hline
        NGC4178 & 17 & 70 & 13.0 & 0.07 & 30.7 & 0.04 & -0.69 & 23.5 & 0.8 & 9.6 \\
        NGC4192 & 17 & 78 & 35.0 & 0.21 & 120.9 & 0.06 & -1.00 & 30.5 & 1.8 & 10.7 \\       
        NGC4294 & 17 & 70 & 11.5 & 0.17 & 32.3 & 0.09 & -0.83 & 12.8 & 0.5 & 9.2 \\        
        NGC4419 & 17 & 74 & 23.6 & 0.32 & 59.8 & 0.14 & -0.74 & 10.7 & 1.4 & 10.2 \\       
        NGC4532 & 17 & 70 & 51.2 & 0.11 & 119.9 & 0.06 & -0.68 & 15.2 & 1.3 & 9.2 \\      
        NGC4808 & 17 & 68 & 19.4 & 0.22 & 62.6 & 0.08 & -0.94 & 13.2 & 0.9 & 9.5 \\
        \hline
        \end{tabular}
      \]
\end{table*}

\subsection{Radio continuum}

For the radio continuum maps we used published VLA data at $4.85$ and $1.4$~GHz at $15$-$22''$ resolution for all galaxies.
The Virgo spiral galaxies were observed at 4.85 GHz between October 12, 2009 and December 23, 2009 with the Very Large
Array (VLA) of the National Radio Astronomy Observatory (NRAO) in the D array configuration. The bandwidths were
$2 \times 50$~MHz. The final cleaned maps were convolved to a beam size of $22'' \times 22''$ (Vollmer et al. 2013). 
In addition, we observed the galaxies at $1.4$~GHz on March 21, 2008
in the C array configuration. The bandwidths were $2 \times 50$~MHz. The final cleaned maps were convolved to a beam size
of $22'' \times 22''$ (Vollmer et al. 2013). At a distance of $17$~Mpc, $1''$ corresponds to $82$~pc.
Furthermore, we used the C-band D-array and L-band C-array data from CHANG-ES (Irwin et al. 2012).
The C-band and L-band images of our galaxy sample are presented in Figs.~\ref{fig:galaxies_rad} and \ref{fig:galaxies_rad20}.
We used the $1.4$~GHz data of NGC~4294 from VIVA (Chung et al. 2009) because these data significantly improved the model results. 

For our analysis, we subtracted the thermal free-free radio emission according to the recipe of Murphy et al. (2008),
\begin{equation}
\big(\frac{S_{\rm therm}}{\rm Jy}\big) = 7.9 \times 10^{-3} \big(\frac{T}{10^4~{\rm K}}\big)^{0.45} \big(\frac{\nu}{\rm GHz}\big)^{-0.1}
\big(\frac{I_\nu(24~\mu{\rm m})}{\rm Jy}\big)\ ,
\end{equation}
where $T$ is the electron temperature, and $I_\nu(24~\mu{\rm m})$ is the flux density at a wavelength of $24~\mu$m.
We note that there are other alternative methods to
account for the thermal free-free emission using, for example, the extinction-corrected H$\alpha$ emission
(Tabatabaei et al. 2007, Heesen et al. 2014). Since information on the spatially resolved H$\beta$ lines was not available to us,
we could not calculate the extinction correction of the H$\alpha$ emission based on the Balmer decrement.

\subsection{Star formation rate}

The SFR surface density was calculated from the FUV luminosities corrected by the total infrared to FUV luminosity ratio
(Hao et al. 2011). This method takes into account the UV photons from young massive stars that escape the galaxy, and those which
are absorbed by dust and re-radiated in the far infrared:
\begin{equation}
\dot{\Sigma}_{*} = 8.1 \times 10^{-2}\ (I({\rm FUV}) + 0.46\ I({\rm TIR}))\ ,
\end{equation}
where $I({\rm FUV})$ is the GALEX far ultraviolet and $I({\rm TIR})$ the total infrared intensity based on Spitzer IRAC and MIPS data
in units of MJy\,sr$^{-1}$. $\dot{\Sigma}_{*}$ has the units of M$_{\odot}$kpc$^{-2}$yr$^{-1}$.
This prescription only holds for a constant SFR surface density over the last few $100$~Myr.

The deprojected star formation maps of our galaxy sample are presented in Fig.~\ref{fig:galaxies_sfr}.
Since the deprojection is not unique, our deprojected maps do not look like typical face-on galaxies: spiral structures are patchy
and the overall disk structure is more oval than round (especially for NGC~4419). Nevertheless, given the simplicity
of our halo model, we are confident that the quality of our simple deprojection is good enough to constrain the major halo properties:
its dependence on the morphology of the underlying starforming disk, the radial profile of the halo height, and the halo flux fraction.

\section{Method\label{sec:method}}

As in Vollmer et al. (2020), we convolved the star formation maps of the six highly-inclined galaxies with adaptive
Gaussian smoothing kernels in two dimensions to obtain model radio continuum emission maps. These authors showed that the smoothing length scales depend
on the observation frequency. For simplicity and in line with the results of Vollmer et al. (2020), we used
a constant smoothing length scale of $2.0$~kpc at $20$~cm and $1.3$~kpc at $6$~cm. We did not vary these scale lengths because
the models with different halo types were quite time-consuming.

If the radio halo is diffusion-dominated a Gaussian vertical profile is expected, where if the halo is advection dominated
an exponential halo profile is expected. A thin and a thick disk give rise to two Gaussians or exponentials.
We also used the empirical function of Oosterloo et al. (2007), which they fitted to the H{\sc i} halo of NGC~891.
For most of the models the height depends of the galactic radius $R$ but it might also depend
on the local SFR surface density $\dot{\Sigma_*}$. The specific intensity $I_{\nu}$ depends on $\dot{\Sigma_*}$
in most of the models but it may also depend on $R$. We also allow for a small offset $o$ of the vertical profiles.
In this way, we can correct for possible small errors of the position angle and warps of the galactic disk.
In general, the offset was found to be $o \la 300$~pc.

In total, we used $10$ different halo prescriptions (Table~\ref{tab:halopresc}).
It is expected that the CRe follow the vertical halo magnetic field when they leave the thin galactic disk.
By applying our vertical halo profiles, we implicitly assume that the halo magnetic fields are vertical without any bending
(see Heald et al. 2022 for a more sophisticated model).
For the production of the model radio continuum map the following steps were performed: (i) insertion of the SFR surface density map
as the thin disk into a model cube, (ii) convolution with a Gaussian to generate the predicted radio continuum map
of the thin disk, (iii) addition of the halo, (iv) projection according to the inclination and position angle of the galaxy,
and (v) convolution with a Gaussian to obtain the spatial resolution of the observations.
Our models have up to $12$ variables, which have to be determined by fitting the model
to the radio continuum map. For each model the reduced $\chi^2=\sqrt{1/N \sum_1^i(f_{\rm obs}-f_{\rm model})^2}$ was calculated using the rms of the radio continuum map, where $f_{\rm obs}$ and $f_{\rm model}$ are the observed and model flux densities, respectively.
\begin{table*}
      \caption{Radio continuum halo prescriptions.}
         \label{tab:halopresc}
         \begin{tabular}{llll}
%           \hline
        gauss & 
	$\begin{array}{rcl} h_1&=&x_2+x_3\,R^{x_4}\\ h_2&=&x_5+x_6\,\dot{\Sigma_*}^{x_7} \end{array}$ &
	$\begin{array}{rcl} o&=&x_8+x_9\,R^{x_{10}} \end{array}$ &
	$\begin{array}{rcl} I_{\nu}&=&x_{11}\,\dot{\Sigma_*}^{x_{12}}\,\exp(-(|z-o|/h_1)^2)+\\
	&+&x_0\,\dot{\Sigma_*}^{x_1}\,\exp(-(|z-o|/h_2)^2) \end{array}$ \\
%        \hline
        gausssinh &
	$\begin{array}{rcl} h_1&=&x_2+x_3\,R^{x_4} \\ h_2&=&x_5+x_6\,\dot{\Sigma_*}^{x_7} \end{array}$ &
	$\begin{array}{rcl} o&=&x_8+x_9\,R^{x_{10}} \end{array}$ &
	$\begin{array}{rcl} I_{\nu}&=&x_{11}\,\dot{\Sigma_*}^{x_{12}}\,\exp(-(|z-o|/h_1)^2)+\\
	&+&x_0\,\dot{\Sigma_*}^{x_1}\,\sinh(-|z-o|/h_2)/(\cosh(-|z-o|/h_2))^2 \end{array}$ \\
%        \hline
        exp &
	$\begin{array}{rcl} h_1&=&x_2+x_3\,R^{x_4} \\ h_2&=&x_5+x_6\,\dot{\Sigma_*}^{x_7} \end{array}$ &
	$\begin{array}{rcl} o&=&x_8+x_9\,R^{x_{10}} \end{array}$ &
	$\begin{array}{rcl} I_{\nu}&=&x_{11}\,\dot{\Sigma_*}^{x_{12}}\,\exp(-|z-o|/h_1)+\\
	&+&x_0\,\dot{\Sigma_*}^{x_1}\,\exp(-(|z-o|/h_2)) \end{array}$ \\
%        \hline
        expsinh &
	$\begin{array}{rcl} h_1&=&x_2+x_3\,R^{x_4} \\ h_2&=&x_5+x_6\,\dot{\Sigma_*}^{x_7} \end{array}$ &
	$\begin{array}{rcl} o&=&x_8+x_9\,R^{x_{10}} \end{array}$ &
	$\begin{array}{rcl} I_{\nu}&=&x_{11}\,\dot{\Sigma_*}^{x_{12}}\,\exp(-|z-o|/h_1)+\\
	&+&x_0\,\dot{\Sigma_*}^{x_1}\,\sinh(-(|z-o|/h_2))/(\cosh(-|z-o|/h_2))^2 \end{array}$ \\
%        \hline
        onegauss &
	$\begin{array}{rcl} h&=&x_2+x_3\,R^{x_4} \end{array}$ &
	$\begin{array}{rcl} o&=&x_5+x_6\,R^{x_{7}} \end{array}$ &
	$\begin{array}{rcl} I_{\nu}&=&x_0\,\dot{\Sigma_*}^{x_1}\,\exp(-(|z-o|/h)^2) \end{array}$ \\
%        \hline
        oneexp &
	$\begin{array}{rcl} h&=&x_2+x_3\,R^{x_4} \end{array}$ &
	$\begin{array}{rcl} o&=&x_5+x_6\,R^{x_{7}} \end{array}$ &
	$\begin{array}{rcl} I_{\nu}&=&x_0\,\dot{\Sigma_*}^{x_1}\,\exp(-|z-o|/h) \end{array}$ \\
%        \hline
        gausssfr &
	$\begin{array}{rcl} h&=&x_2+x_3\,\dot{\Sigma_*}^{x_4} \end{array}$ &
	$\begin{array}{rcl} o&=&x_5+x_6\,R^{x_{7}} \end{array}$ &
	$\begin{array}{rcl} I_{\nu}&=&x_0\,\dot{\Sigma_*}^{x_1}\,\exp(-(|z-o|/h)^2) \end{array}$ \\
%        \hline
        expsfr &
	$\begin{array}{rcl} h&=&x_2+x_3\,\dot{\Sigma_*}^{x_4} \end{array}$ &
	$\begin{array}{rcl} o&=&x_5+x_6\,R^{x_{7}} \end{array}$ &
	$\begin{array}{rcl} I_{\nu}&=&x_0\,\dot{\Sigma_*}^{x_1}\,\exp(-|z-o|/h) \end{array}$ \\
%        \hline
        onegausssmooth &
	$\begin{array}{rcl} h&=&x_2+x_3\,R^{x_4} \end{array}$ &
	$\begin{array}{rcl} o&=&x_5+x_6\,R^{x_{7}} \end{array}$ &
	$\begin{array}{rcl} I_{\nu}&=&x_0\,R^{x_1}\,\exp(-(|z-o|/h)^2) \end{array}$ \\
%        \hline
        oneexpsmooth &
	$\begin{array}{rcl} h&=&x_2+x_3\,R^{x_4} \end{array}$ &
	$\begin{array}{rcl} o&=&x_5+x_6\,R^{x_{7}} \end{array}$ &
	$\begin{array}{rcl} I_{\nu}&=&x_0\,R^{x_1}\,\exp(-|z-o|/h) \end{array}$ \\
%        \hline
        \end{tabular}
\end{table*}

For the fitting of the models to the data, we performed a multidimensional minimization using the downhill simplex method
(Nelder \& Mead 1965). The {\tt IDL} routine {\tt amoeba} (Press et al. 1992) was used for this purpose. We checked for a very
limited number of cases that different initial conditions lead to the same results. However, we cannot exclude that a local
minimum was found by {\tt amoeba}. All models converged to the presented solutions.
The emission of the four quadrants of the radio continuum maps were fitted separately. Since NGC~4192 harbors a strong central point source
associate to an active galactic nucleus, the model was not fitted to the data in a band around the center (Fig.~\ref{fig:ngc4192_profres}).
Each model needed about two days on a single CPU to converge.

We realized that for models, where we tried to fit the thin disk with a separate profile (gauss, gaussinh, exp, expsinh in Table~\ref{tab:halopresc}),
the fluxes were sometimes distributed between the two profiles in an arbitrary way. Therefore, we did not use the scaleheights $h$, $h_1$, and $h_2$
but determined an effective halo height $H$. For each galactic radius $R$ we calculated $H$ with
\begin{equation}
  \int_0^H I_{\nu} {\rm d}z=0.5\,\int_0^{z_{\rm max}}I_{\nu} {\rm d}z\ ,
\end{equation}
where $z_{\rm max}=8.2$~kpc. For NGC~4178, we realized that the fitting procedure lead to much larger or smaller
effective heights than the mean height. We thus decided to discard these solutions by applying {\tt resistant\_mean} in {\tt IDL}
to identify the outlying solutions. These are marked with parentheses in Table~C.1.

The final radial profile of $H$ were calculated (i) with all $10$ models and (ii) with 
the four best-fitting models. Furthermore, we calculated the thin disk and halo flux fraction for each model to determine the
spectral index (SI) of the halo emission.

Our rather time-consuming model calculations did not permit a proper estimation of the uncertainties of the different parameters
via Markov Chain Monte Carlo (MCMC) techniques. Instead, we investigated which parameter has the strongest influence on the reduced $\chi^2$.
We did this by decreasing and increasing each parameter separately by $50$\,\% for the NGC~4192 models at $6$~cm. We then identified the parameters,
which increased the reduced $\chi^2$ by at least a factor of two. As expected, the reduced $\chi^2$ is sensitive to the exponents of the radius in the descriptions of
the height and the vertical offset. In addition, the constant of the vertical offset and the exponent of the SFR surface density
also have a significant influence on the reduced $\chi^2$. We conclude that all exponents in the prescriptions of Table~\ref{tab:halopresc} are well constrained.

\section{Results\label{sec:results}}

We found $10 \la \chi^2 \la 50$ for most of the models of NGC~4178, NGC~4192, NGC~4294, and NGC~4808 (Tables~C.1,
C.2, C.4, and C.7).
For NGC~4192CHANG-ES, the range is $10 \la \chi^2 \la 100$ (Table~C.3), for NGC~4419 it is $100 \la \chi^2 \la 800$
(Table~C.5), and for NGC~4532 it is $30 \la \chi^2 \la 400$ (Table~C.6). Thus,
the models reproduce the radio emission distribution of NGC~4419 and NGC~4532 significantly less well than that of the other galaxies.
We note that the reduced $\chi^2$ of the smoothing experiments of Vollmer et al. (2020) were of the same order.

When we compared the three models with the lowest $\chi^2$ at the two wavelengths for a given quadrant in a given galaxy, we
did not find preferences for a Gaussian (diffusion) or exponential (advection) vertical profile (Tables~C.1 to C.7).
As for the transport of CRe within the galactic disk studied in Vollmer et al. (2020), it is not possible to discriminate between
diffusion- and advection-dominated radio continuum halos solely based on the type of the vertical intensity profile.

The vertical profiles, which exclusively depend on the galactic radius $R$ (onegausssmooth and oneexpsmooth; Table~\ref{tab:halopresc}),
are very rarely found amongst the three models with the lowest $\chi^2$ (Tables~C.1 to C.7).
When this is the case, it is found at one wavelength but not in the other.
We thus conclude that overall the halo emission contains information on the underlying distribution of the SFR surface density ($\dot{\Sigma}_*$).

\begin{figure*}[ht!]
  \centering
  \resizebox{15cm}{!}{\includegraphics{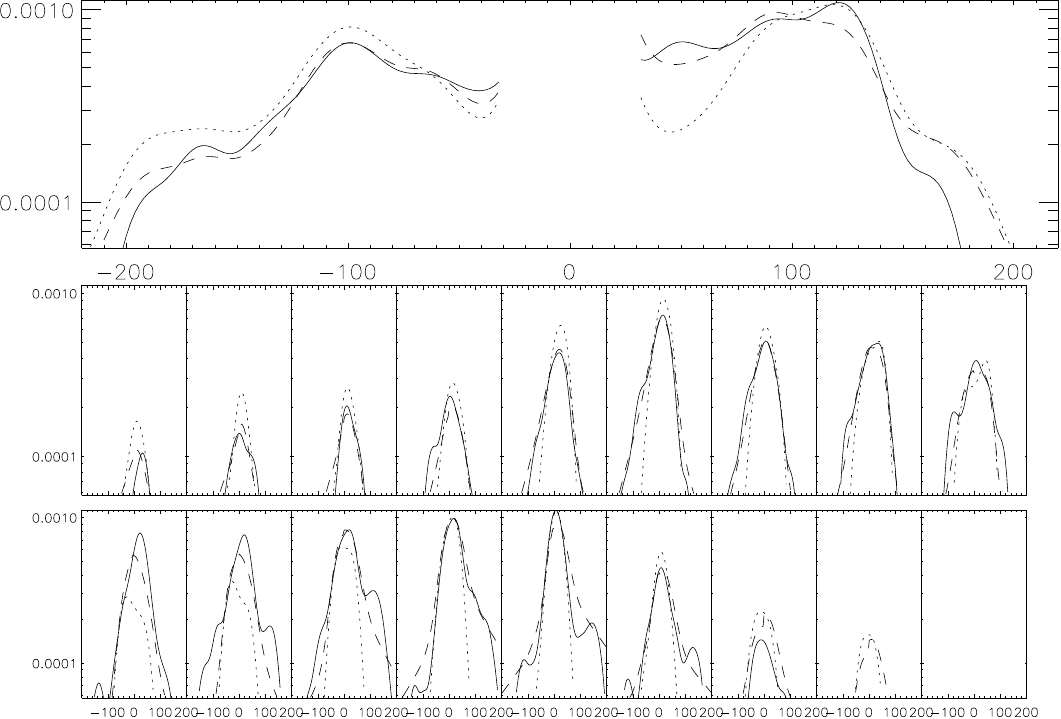}\put(-400,235){\bf \Large NGC 4192 6cm}}
  \resizebox{15cm}{!}{\includegraphics{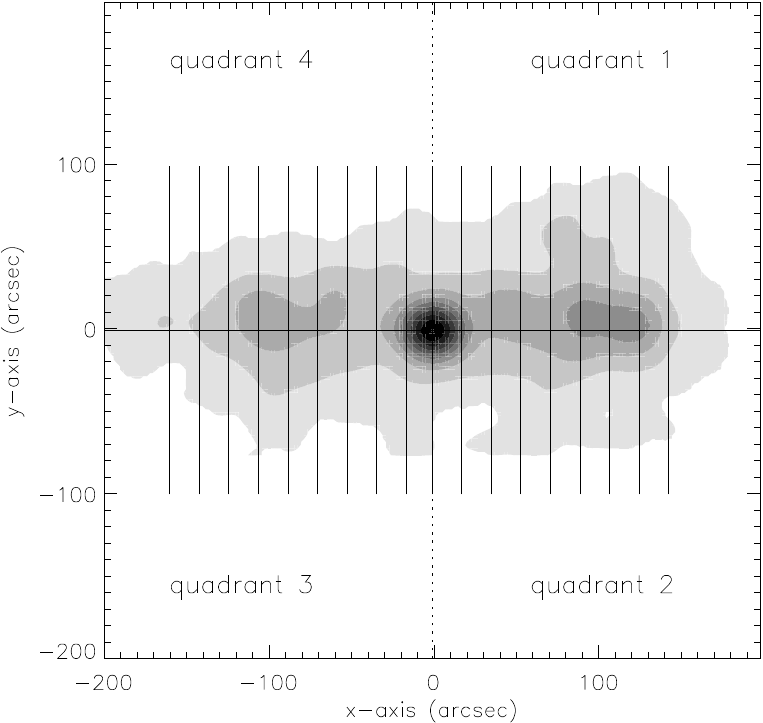}\includegraphics{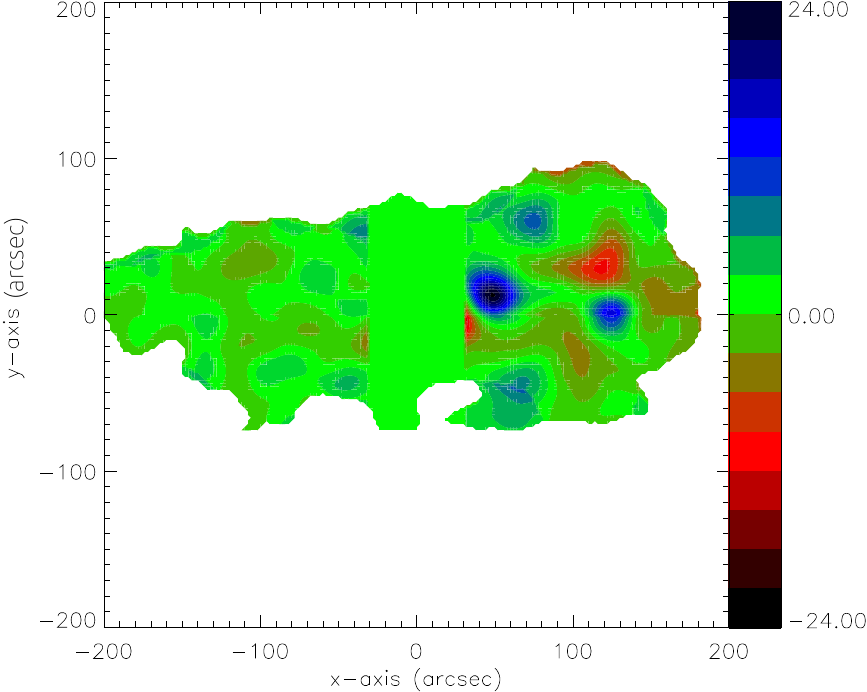}\put(-350,300){\bf \Large NGC4192 6cm}}
  \caption{Radio continuum halo model of NGC~4192. Upper panel: profiles along the minor and major axes. Solid lines: observations, dashed lines: model
    including a a thin disk and a halo, dotted lines: model including only a thin disk. All distance are in arcseconds. Lower left panel: locations of the slices along the major and minor axis. Lower right panel: maps of the model residuals in units of the rms.   
  \label{fig:ngc4192_profres}}
\end{figure*}

The surface brightness profiles along the major axis and along lines parallel to the minor axis for NGC~4192 based
on our and the CHANG-ES $20$~cm and $6$~cm data together with the residual maps are shown in Figs.~\ref{fig:ngc4192_profres} and \ref{fig:ngc4192_profres1}.
The best-fit models were selected in all four quadrants separately.
 
As expected, we found consistent overall results for the CHANG-ES and our data. However, the details of the residual maps are different.
For example, the negative residuals in the eastern half of the galactic disk based on the CHANG-ES data are not present in the
residual maps based on our data.
Halo radio continuum emission is clearly detected in all quadrants (dashed lines, which correspond to models with halo emission, compared to the dotted lines,
which correspond to models without halo emission). The halo emission also improves the profile along the major axis.
The halo emission is less prominent at $6$~cm than at $20$~cm. The residuals are larger on the western side of the galactic disk than on the eastern side.
%\FloatBarrier

\begin{figure*}[ht!]
  \centering
  \resizebox{\hsize}{!}{\includegraphics{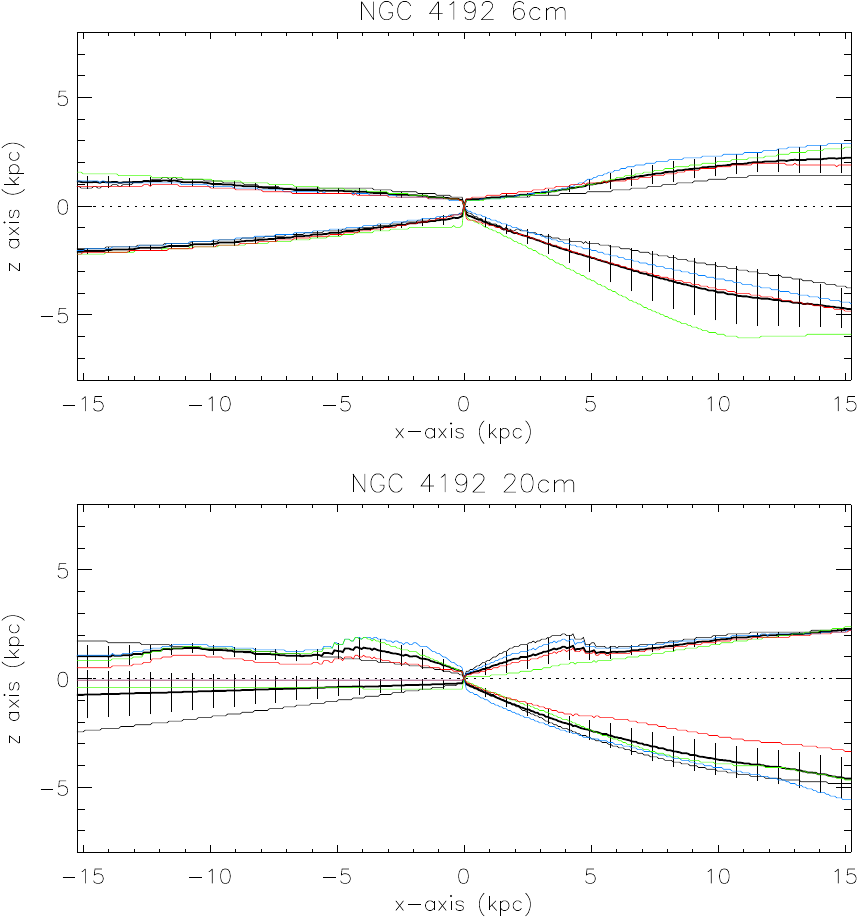}\includegraphics{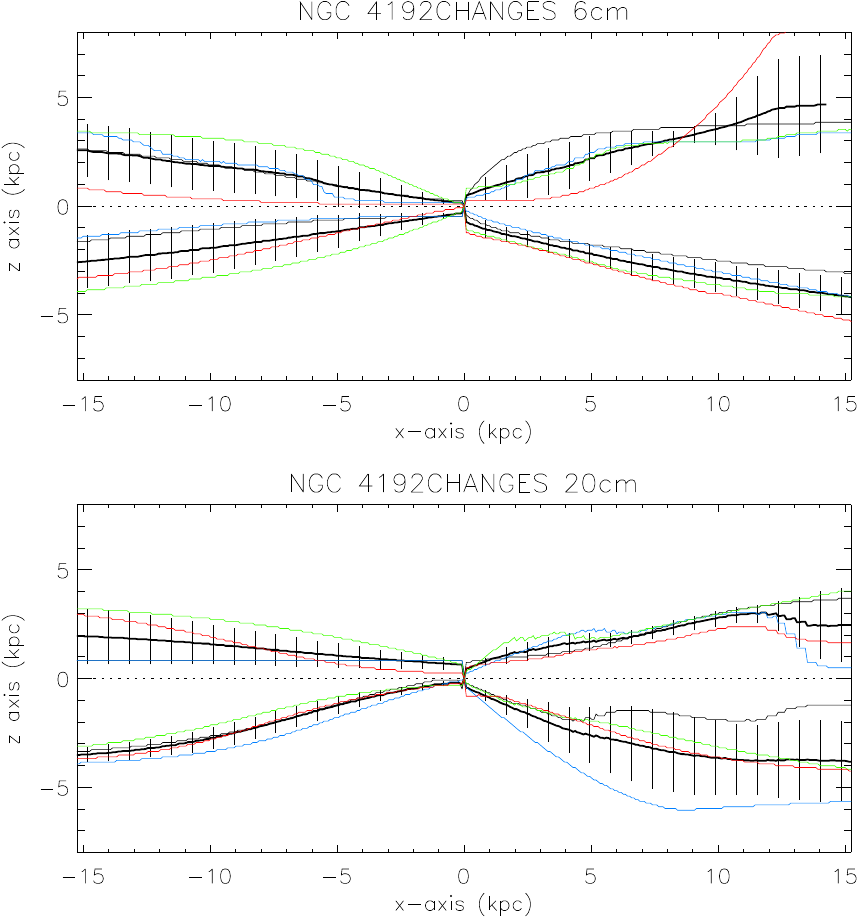}}
  \caption{Radial profiles of the halo scale height of NGC~4192. Upper panel: all ten models. Lower panel: best four models.
    Models with a low $\chi^2$ are red, those with a high $\chi^2$ are blue.
  \label{fig:ngc4192H20}}
\end{figure*}

The radial profiles of the halo effective height $H$ of NGC~4192 are presented in Fig.~\ref{fig:ngc4192H20} for the four best models and in
Fig.~\ref{fig:ngc4192H20a} for all ten models. The height profiles derived from the CHANG-ES
and our data are in good agreement with each other. The radio halo emission of all quadrants is flaring. The western half of the disk is somewhat
thicker than the eastern half at both wavelengths. Overall, the radially averaged halo heights and halo flux fractions based on the CHANG-ES and
our data are consistent (Table~\ref{tab:heights1}). Our model procedure is thus robust.
%\FloatBarrier

\begin{figure*}[ht!]
  \centering
  \resizebox{\hsize}{!}{\includegraphics{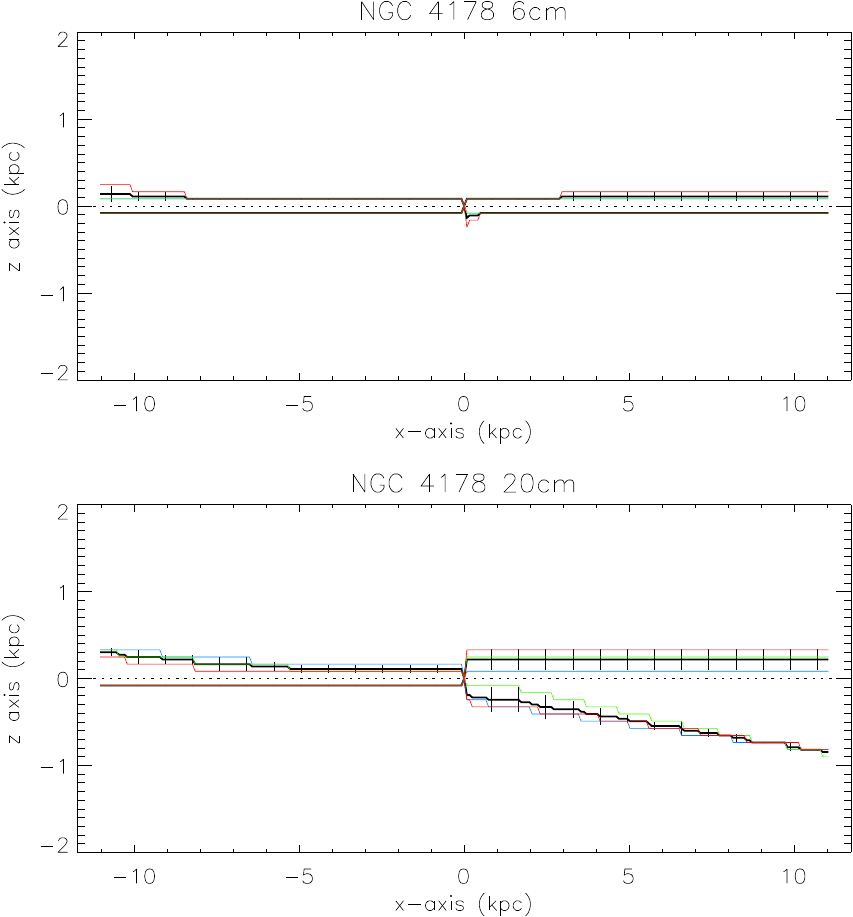}\includegraphics{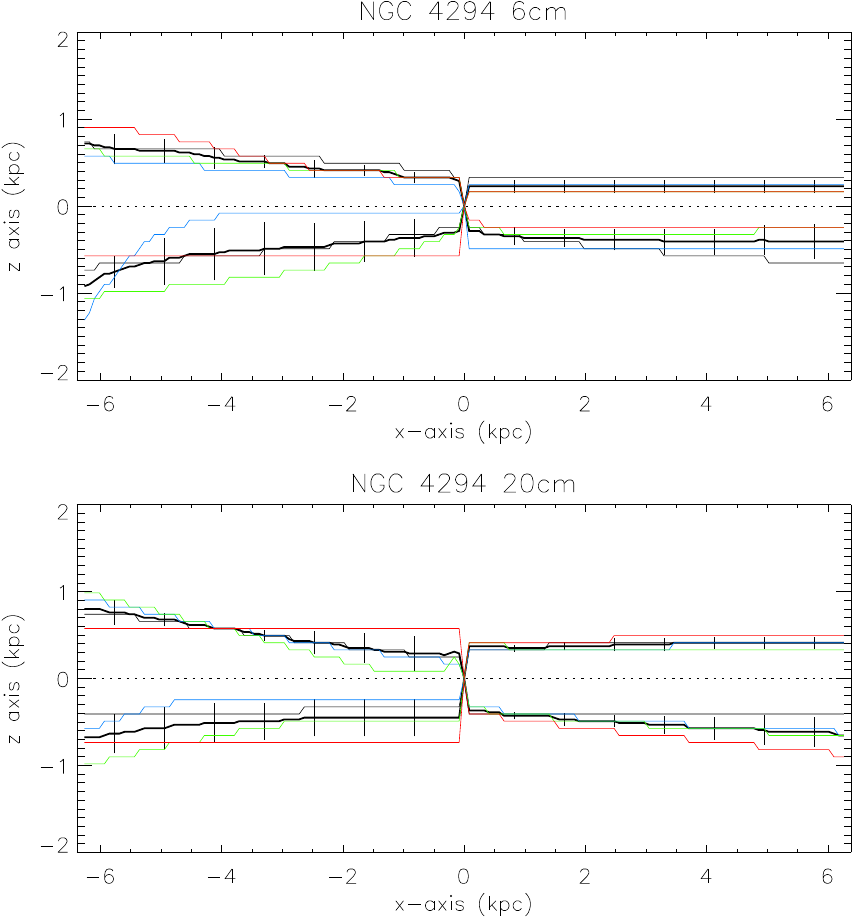}}
  \resizebox{\hsize}{!}{\includegraphics{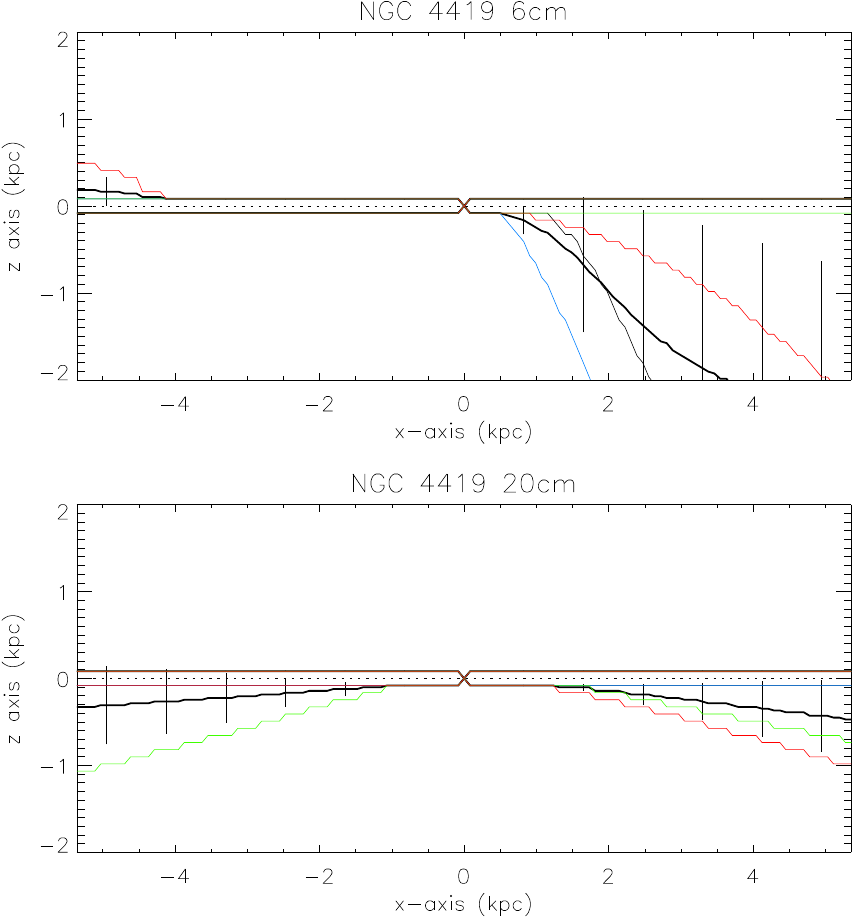}\includegraphics{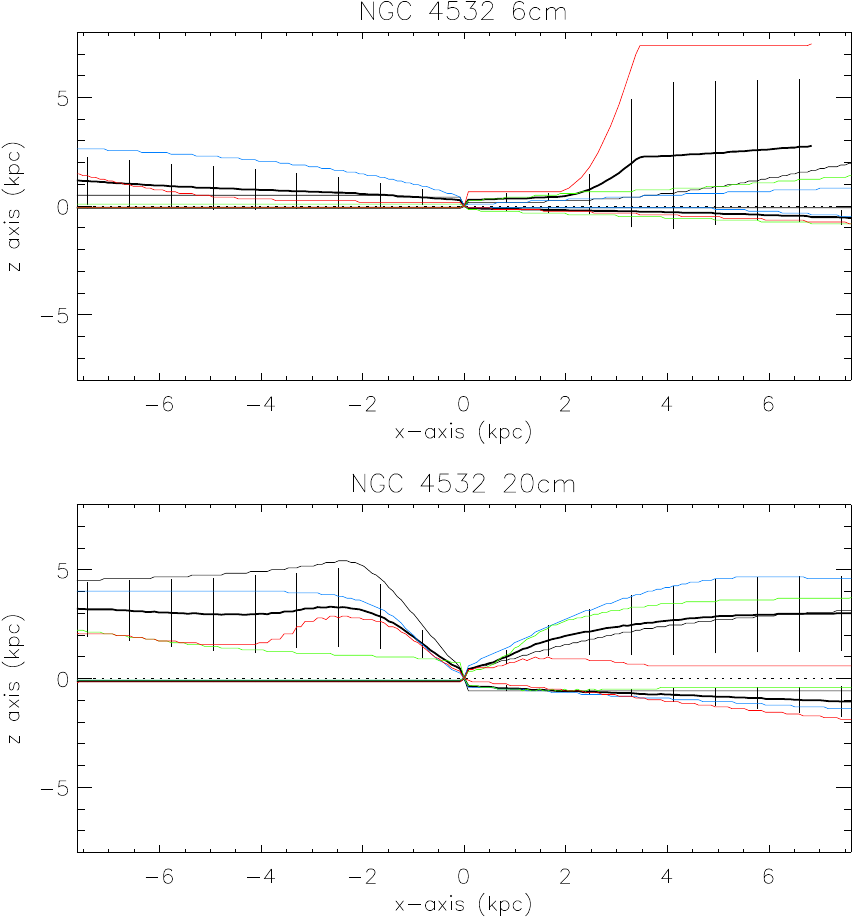}}
  \caption{Radial profiles of the halo scale height of NGC~4178, NGC~4294, NGC~4419, NGC~4532. Best four models.
    Models with a low $\chi^2$ are red, those with a high $\chi^2$ are blue.
  \label{fig:ngc4178H20}}
\end{figure*}
\begin{figure}[ht!]
  \centering
  \resizebox{\hsize}{!}{\includegraphics{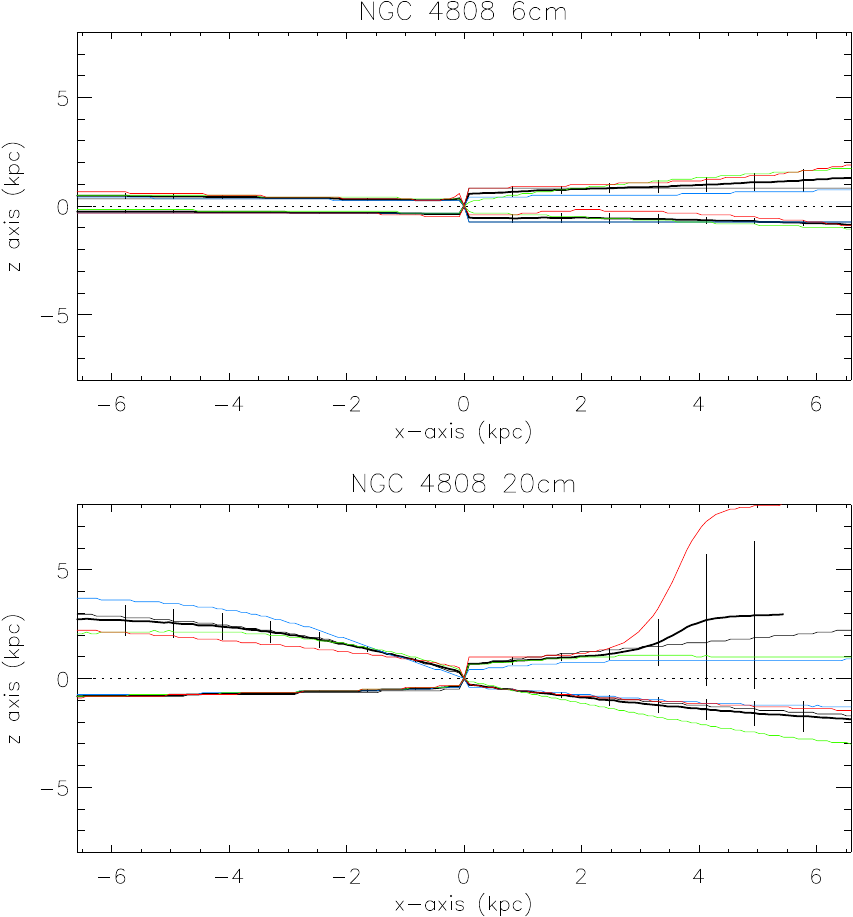}}
  \caption{Radial profiles of the halo scale height of NGC~4808. Best four models.
    Models with a low $\chi^2$ are red, those with a high $\chi^2$ are blue.
  \label{fig:ngc4178H20a}}
\end{figure}
 
\begin{table*}
      \caption{Mean heights and flux fractions.}
         \label{tab:heights1}
      \[
      \begin{tabular}{lcccccccc}
        \hline
        & $h_{\rm all}^{\rm 6cm}$ & $f_{\rm all}^{\rm 6cm}$ & $h_{\rm 4\ best}^{\rm 6cm}$ & $f_{\rm 4\ best}^{\rm 6cm}$ & $h_{\rm all}^{\rm 20cm}$ & $f_{\rm all}^{\rm 20cm}$ & $h_{\rm 4\ best}^{\rm 20cm}$ & $f_{\rm 4\ best}^{\rm 20cm}$ \\
         & (kpc) & & (kpc) & & (kpc) & & (kpc) &  \\
        \hline
        N4178 disk & $0.10  \pm  0.03$ & $0.58  \pm   0.13$ & $0.09  \pm  0.02$ & $0.61  \pm   0.17$ & $0.30   \pm  0.24$ & $0.62   \pm  0.17$ & $0.25   \pm  0.22$ & $0.65  \pm   0.19$ \\        
        N4192 halo & $1.59  \pm   1.03$ & $0.60  \pm   0.25$ & $1.73  \pm   0.98$ & $0.61  \pm  0.18$ & $1.40  \pm   1.19$ & $0.53  \pm   0.25$ & $1.55  \pm   1.05$ & $0.55   \pm  0.14$ \\
        N4192C halo & $1.78   \pm   1.02$ & $0.62  \pm   0.23$ & $2.19  \pm   0.92$ & $0.52    \pm 0.14$ & $1.41    \pm  1.19$ & $0.50   \pm  0.17$ & $2.11   \pm  0.93$ & $0.62   \pm  0.14$ \\
        N4294 halo & $0.37  \pm   0.15$ & $0.55   \pm  0.33$ & $0.40   \pm  0.17$ & $0.47  \pm   0.32$ & $0.41   \pm 0.14$ & $0.55  \pm   0.35$ & $0.41  \pm  0.17$ & $0.56  \pm   0.31$ \\
        N4419 halo & $0.33   \pm    0.79$ & $0.86   \pm   0.40$ & $0.67   \pm    1.19$ & $0.92   \pm   0.35$ & $0.30  \pm    0.90$ & $0.75  \pm   0.16$ & $0.29  \pm    0.45$ & $0.76  \pm   0.12$ \\
        N4532 halo & $0.65   \pm   0.91$ & $0.20 \pm    0.20$ & $1.13   \pm   1.29$ & $0.30   \pm  0.22$ & $1.89    \pm  1.85$ & $0.30  \pm   0.18$ & $1.81   \pm   1.56$ & $0.38  \pm   0.16$ \\
        N4808 halo & $0.90  \pm   1.00$ & $0.66  \pm   0.21$ & $0.79  \pm   0.57$ & $0.66  \pm   0.20$ & $2.29  \pm    1.56$ & $0.62  \pm   0.28$ & $1.82   \pm   0.95$ & $0.71  \pm   0.19$ \\
        \hline
        \end{tabular}
      \]
\end{table*}
%\FloatBarrier

NGC~4178 is the only galaxies in our sample, which does not show detectable radio halo emission at $6$~cm (Table~\ref{tab:heights1}
and Figs.~\ref{fig:ngc4178H20} and \ref{fig:ngc4178H20b}). The mean effective height is small ($\sim 100$~pc) at $6$~cm.
At $20$~cm the halo emission distribution has a effective height of $\sim 300$~pc.
NGC~4294 and NGC~4419 have relatively thin radio continuum halos ($300$-$400$~pc at both wavelengths). The halo flux fraction of NGC~4419 ($f > 0.7$)
is significantly higher than that of NGC~4294 ($f \sim 0.5$).
In NGC~4419 we detect somewhat detached extended $6$~cm emission at very low surface brightnesses, which cannot be properly reproduced by our models
(Figs.~\ref{fig:ngc4178H20} and \ref{fig:ngc4178H20b}). 
NGC~4532 has a thick ($H \sim 0.7$~kpc at $6$~cm and $H \sim 1.8$~kpc at $20$~cm) but not prominent radio continuum halo ($f \sim 0.3$) at
both wavelengths (Figs.~\ref{fig:ngc4178H20} and \ref{fig:ngc4178H20b}).
NGC~4808 has the thickest and prominent radio continuum halo ($H \sim 0.8$~kpc at $6$~cm and $H \sim 2.0$~kpc at $20$~cm; Figs.~\ref{fig:ngc4178H20a} and
\ref{fig:ngc4808H20}).

The majority of our galaxies show flaring radio continuum halos meaning that the effective height increases with galactic radius
(Figs.~\ref{fig:ngc4192H20} to \ref{fig:ngc4178H20a} and Fig.~\ref{fig:ngc4192H20a} to \ref{fig:ngc4808H20}).
In most of the cases the radial profiles of the effective height have different
behaviors in the four quadrants. There can be an east-west asymmetry (NGC~4178, NGC~4192) or a north-south asymmetry (NGC~4532).

\begin{figure}[ht!]
  \centering
  \resizebox{\hsize}{!}{\includegraphics{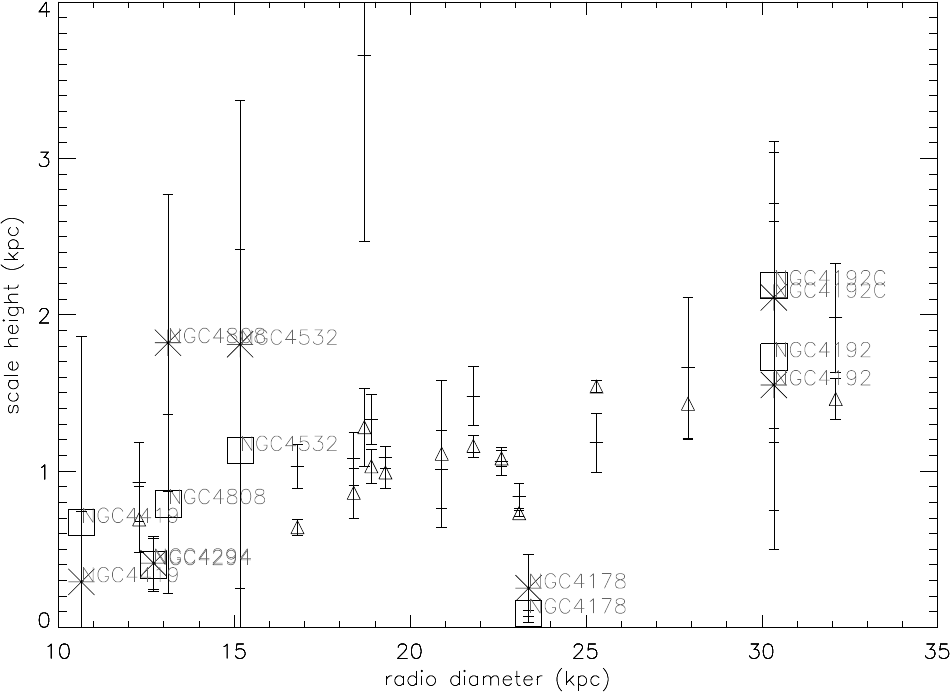}}
  \caption{Halo scale heights as function of the radio diameter. Triangles: CHANG-ES galaxies at $6$~cm, pluses: CHANG-ES galaxies at $20$~cm,
    boxes: Virgo galaxies at $6$~cm, asterisks: Virgo galaxies at $20$~cm.
  \label{fig:resultats_finaux_mesquita2}}
\end{figure}

The radially averaged effective heights as a function of the size of the radio continuum disk for our highly-inclined Virgo spiral galaxies
are compared to those of the CHANG-ES galaxies in Fig.~\ref{fig:resultats_finaux_mesquita2}.
Except for NGC~4178, our Virgo galaxies follow the trend of increasing effective height with increasing radio continuum size found
by the CHANG-ES collaboration (Krause et al. 2018). Moreover, the effective heights of NGC~4532 and NGC~4808 at $20$~cm are significantly
larger than expected from the correlation. As already stated before, a radio continuum halo is absent at $6$~cm in NGC~4178 and it is very thin
at $20$~cm given its large radio continuum size. 
%\FloatBarrier

\section{The classification of radio continuum halos \label{sec:physics}}

The vertical intensity profile of a radio continuum halo is determined by the transport model, which can be diffusion, streaming, or advection,
and the decrease of synchrotron emission, which can be due to synchrotron losses or CRe escape.
The ratio $r_H$ between the effective halo heights at $6$ and $20$~cm and the SI of the halo are different for different types of halos.
As stated by Krause et al. (2018), in a synchrotron energy loss-dominated halo the outer halo boundary is determined where the CRe have lost their energy by a
sufﬁcient amount that its radiation is below the observational detection limit. In an escape-dominated halo the halo boundary is observed where the number of CRe and the magnetic
ﬁeld strength are too small to emit detectable radio emission, which is also dependent on the sensitivity of the observations.
In practice both processes can occur simultaneously along the line of sight. 

\begin{figure}[ht!]
  \centering
  \resizebox{\hsize}{!}{\includegraphics{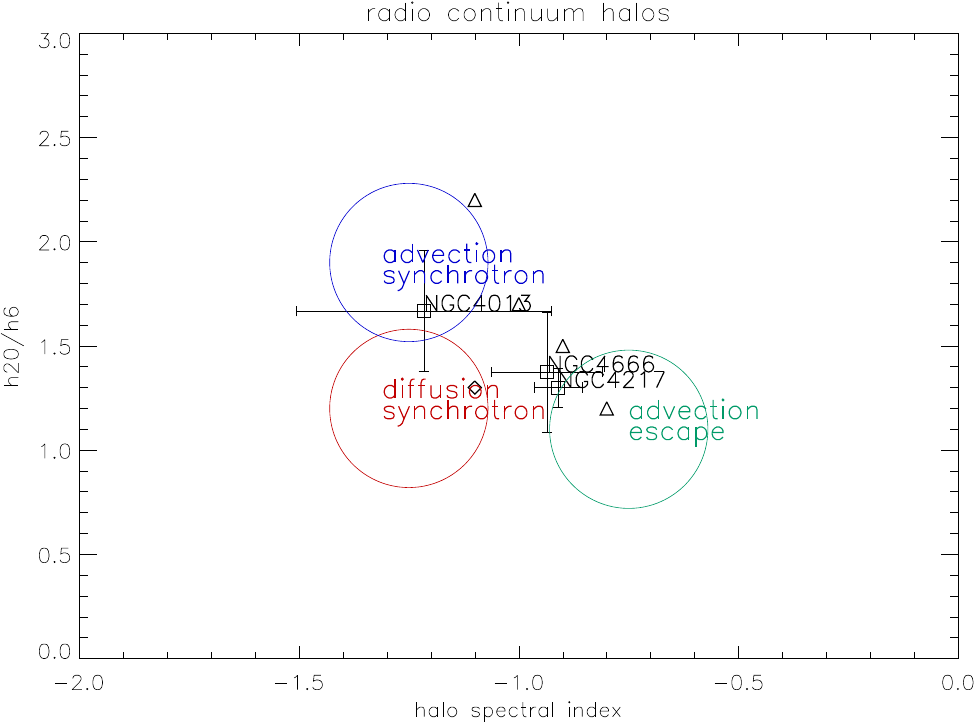}}
  \caption{CHANG-ES galaxies: height ratio at 20cm and 6cm as a function of the spectral index between 20cm and 6cm. Triangles: SPINNAKER models of
    halos with advection. Diamond: SPINNAKER model of a synchrotron-loss dominated halo with diffusion.
  \label{fig:finalplot_SIHR}}
\end{figure}

We followed Krause et al. (2018) and Heesen (2021) for the height ratios. Based on the height ratio a diffusion synchrotron loss-dominated halo
cannot be distinguished from an advection escape-dominated halo ($r_H \sim 1$). We suggest to break this degeneracy by including the
spectral index (SI) into the analysis: whereas the SI for synchrotron losses in about $-1.0$, it is $\sim -0.5$ if escape-losses
dominate the halo. Additional CRe aging is taken into account by a slight steepening of the radio continuum spectrum (SI decreased by $-0.25$).
We found the following halo characteristics: (i) synchrotron loss-dominated with advection: $r_H \sim 1.9$, SI$\sim -1.25$; (ii)
synchrotron loss-dominated with diffusion: $r_H \sim 1.2$, SI$\sim -1.25$; (iii) escape-dominated with diffusion: $r_H \sim 1.1$, SI$\sim -0.75$,
and (iv) escape-dominated with advection: $r_H \sim 1.1$, SI$\sim -0.75$. The two escape-dominated halo types (iii) and (iv) are indistinguishable.
It is expected that the preferred CRe transport mechanism for escape-dominated halos is advection by a galactic wind or outflow provided that the SFR surface density is
equal to or higher than the average of starforming galaxies. Since galaxies with lower SFR surface density (green valley galaxies) are rare, most of the
escape-dominated halos should be dominated by advection rather than by diffusion.
We arbitrarily draw circles around these points to delimit the different regions if the $r_H$--$SI$ diagram of Fig.~\ref{fig:finalplot_SIHR}.
Cosmic ray transport models for a synchrotron-loss dominated halo with diffusion and halos with advection calculated with SPINNAKER (Heesen et al. 2018)
are consistent with our choice of the three regions in SI--h20/h6 space.

The CHANG-ES galaxies for which the height ratios and halo SI are measured are NGC~4013 (Stein et al. 2019a), NGC~4217 (Stein et al. 2020), and
NGC~4666 (Stein et al. 2019b). Whereas diffusion is the CRe transport mechanism in NGC~4013 (Gaussian vertical profile),
it is advection in the other two galaxies (exponential vertical profile). Indeed, NGC~4013 lies within the region of
diffusion synchrotron loss-dominated halos whereas NGC~4217 and NGC~4666 lies at the edge but still within the region of advection escape-dominated halos.
This demonstrates that our halo classification based on $r_H$ and SI is meaningful.
%\FloatBarrier

The halo SI roughly follow the SI of the integrated emission. The mean deviation of the halo SI from the integrated SI is $0.12$.
As expected, most halo SI (5 out of 7) are somewhat lower (steeper) than the integrated SI.

\begin{figure*}[ht!]
  \centering
  \resizebox{\hsize}{!}{\includegraphics{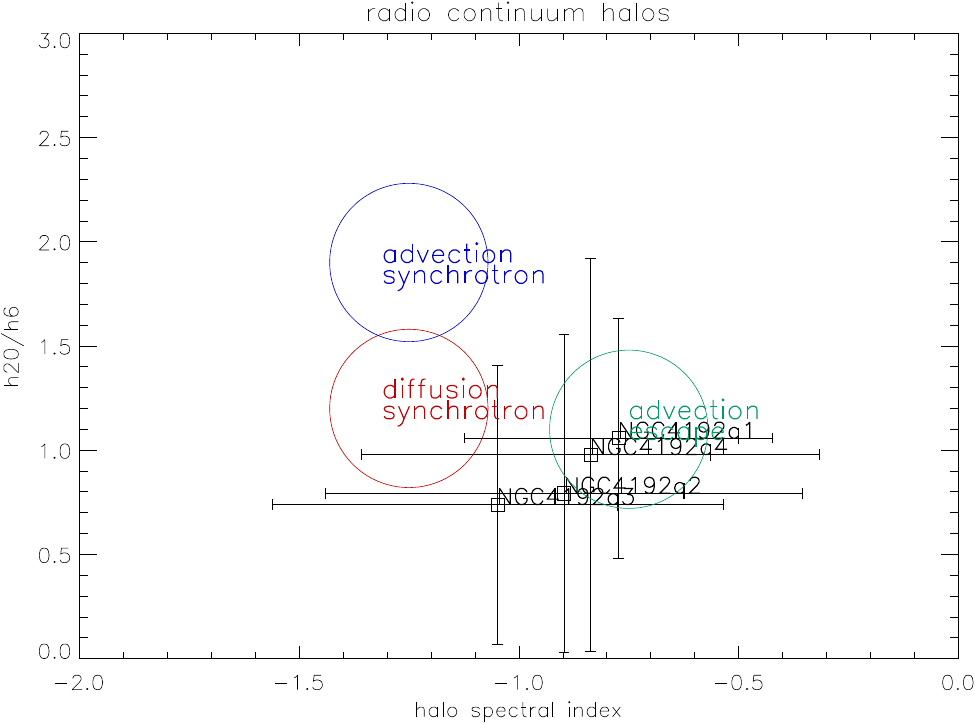}\put(-50,300){\bf \Huge a}\includegraphics{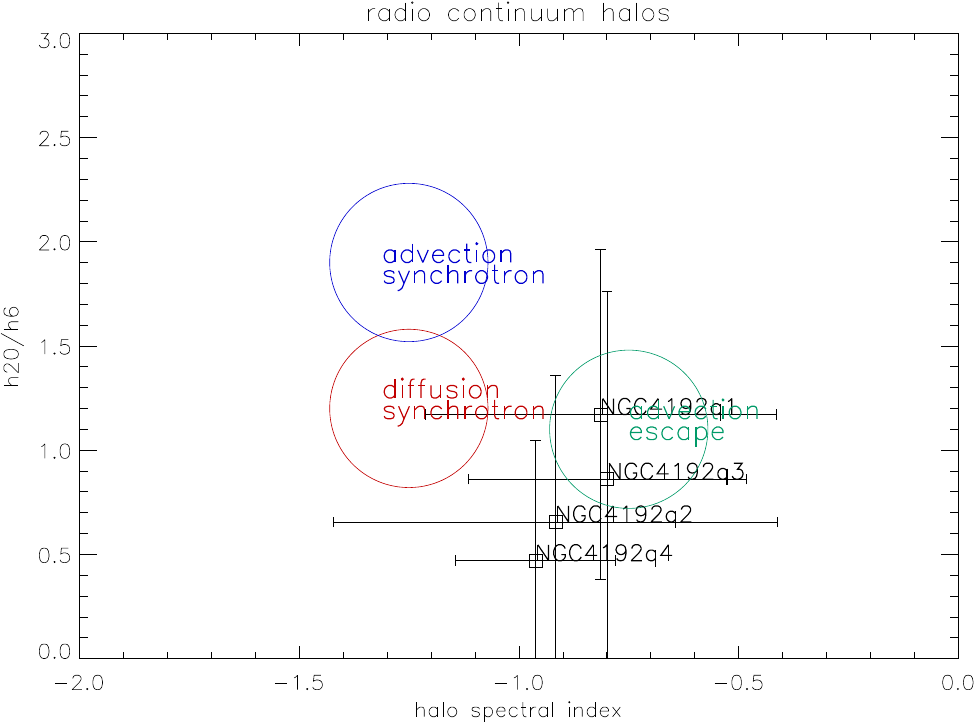}\put(-50,300){\bf \Huge b}\put(-400,300){\huge CHANG-ES}}
  \resizebox{\hsize}{!}{\includegraphics{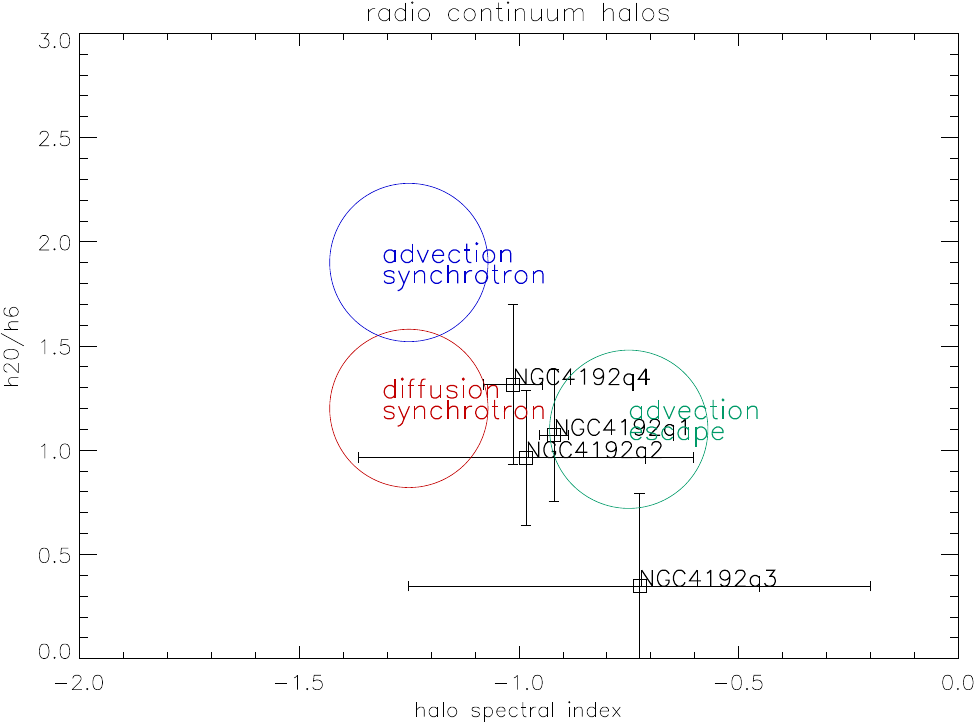}\put(-50,300){\bf \Huge c}\includegraphics{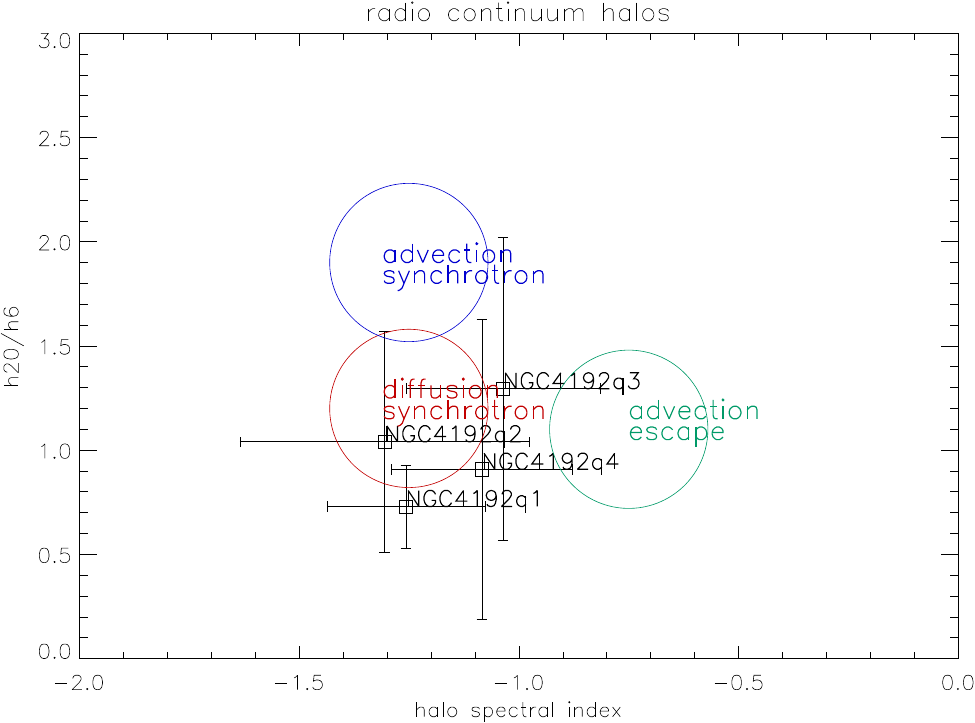}\put(-50,300){\bf \Huge d}\put(-400,300){\huge CHANG-ES}}
  \caption{NGC~4192. Height ratio as a function of the spectral index between 20cm and 6cm; (a,b) mean of all ten models, (c,d) mean of the four best-fit models.
  \label{fig:ngc4192diag}}
\end{figure*}

The corresponding halo classification for NGC~4192 is presented in Fig.~\ref{fig:ngc4192diag} based on the CHANG-ES and our data for all
ten models and the four best models. All results are in good agreement with each other.
The uncertainties are large but not too large to permit the classification of the halo, which is advection escape-dominated in NGC~4192.
%\FloatBarrier

The radio halo of NGC~4294 is also most probably advection escape-dominated (Fig.~\ref{fig:haloclass_n4294}). Only the third quadrant is
consistent with being diffusion synchrotron-dominated. Given that a vertical halo magnetic field is observed in this quadrant (Vollmer et al. 2013), 
we think that the halo of this quadrant is also advection escape-dominated (see also Sect.~\ref{sec:discussion}).

\begin{figure}[ht!]
  \centering
  \resizebox{\hsize}{!}{\includegraphics{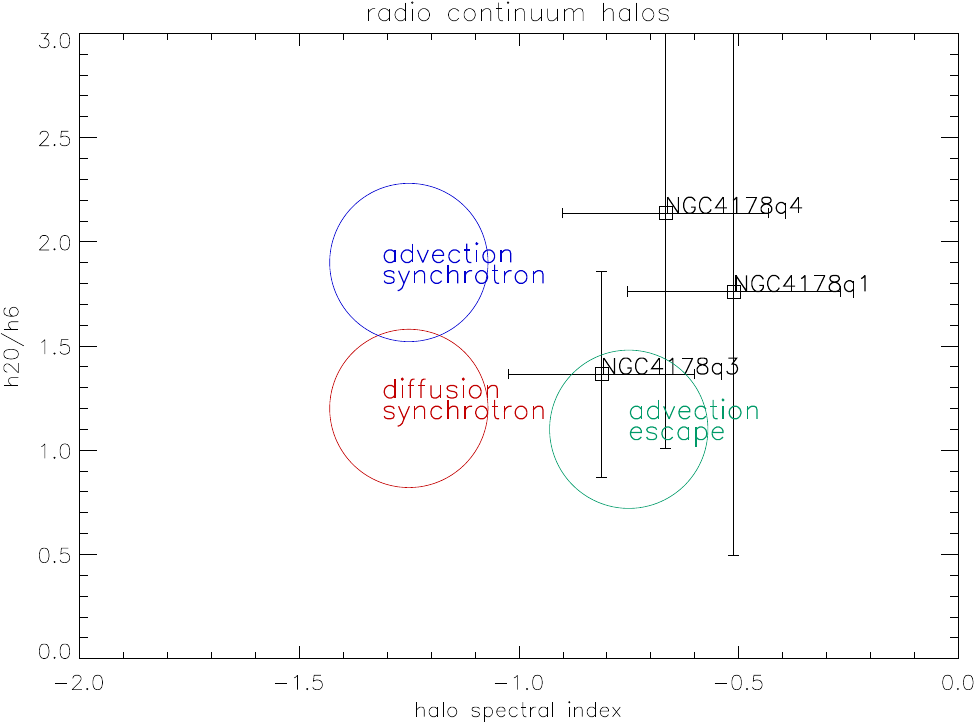}}
  \resizebox{\hsize}{!}{\includegraphics{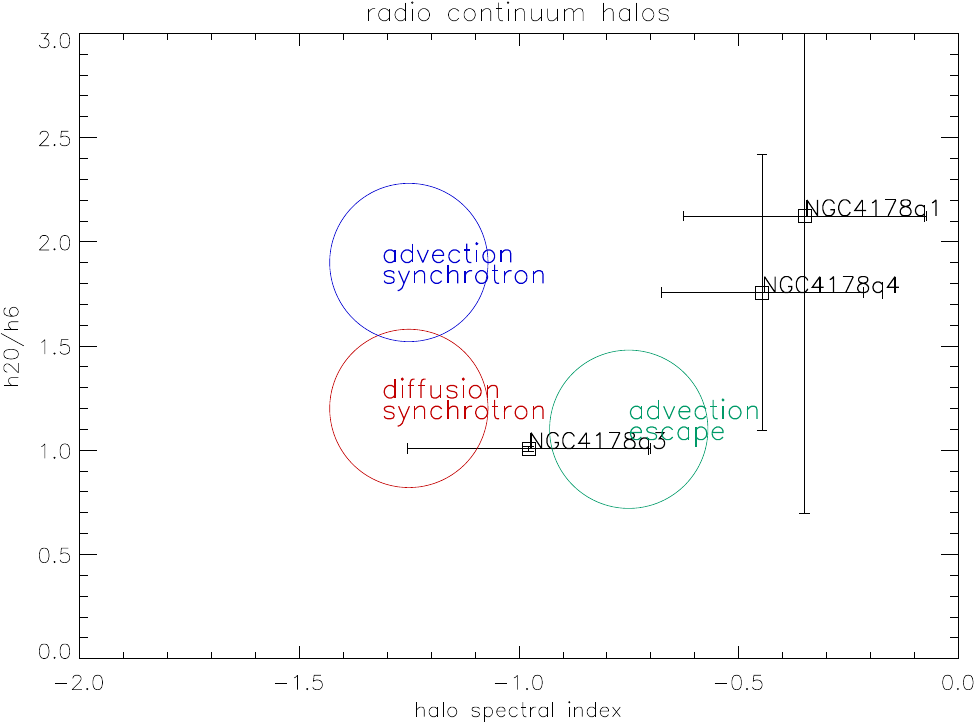}}
  \caption{NGC~4178. Height ratio as a function of the spectral index between 20cm and 6cm; upper panel: mean of all ten models; lower panel:
    mean of the four best-fit models.
  \label{fig:haloclass_n4178}}
\end{figure}

\begin{figure}[ht!]
  \centering
  \resizebox{\hsize}{!}{\includegraphics{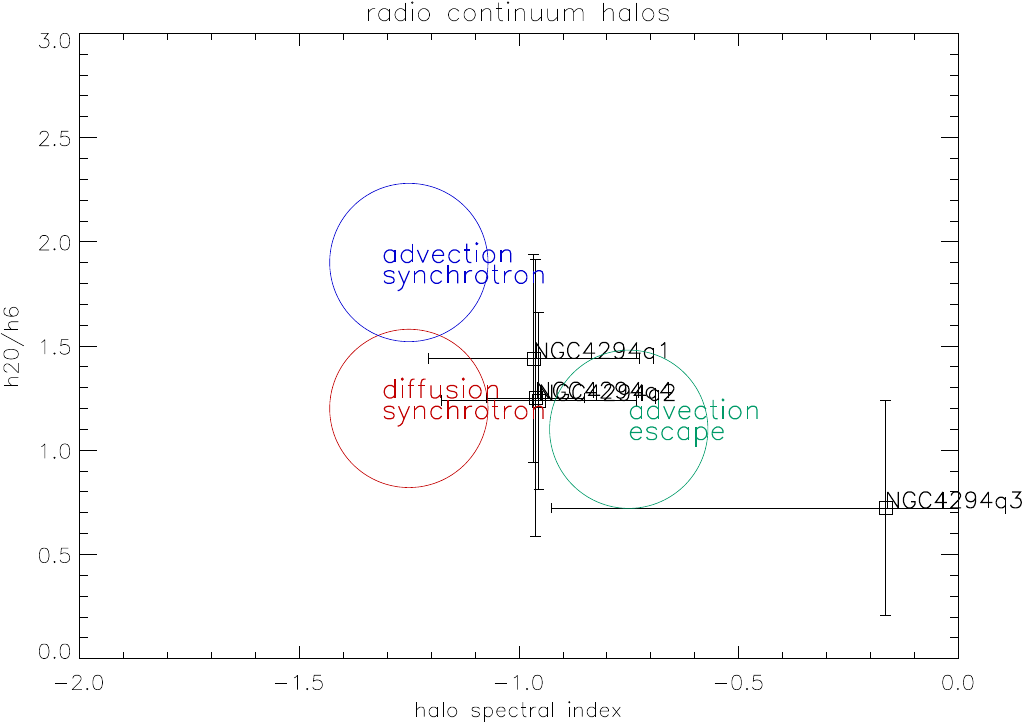}}
  \resizebox{\hsize}{!}{\includegraphics{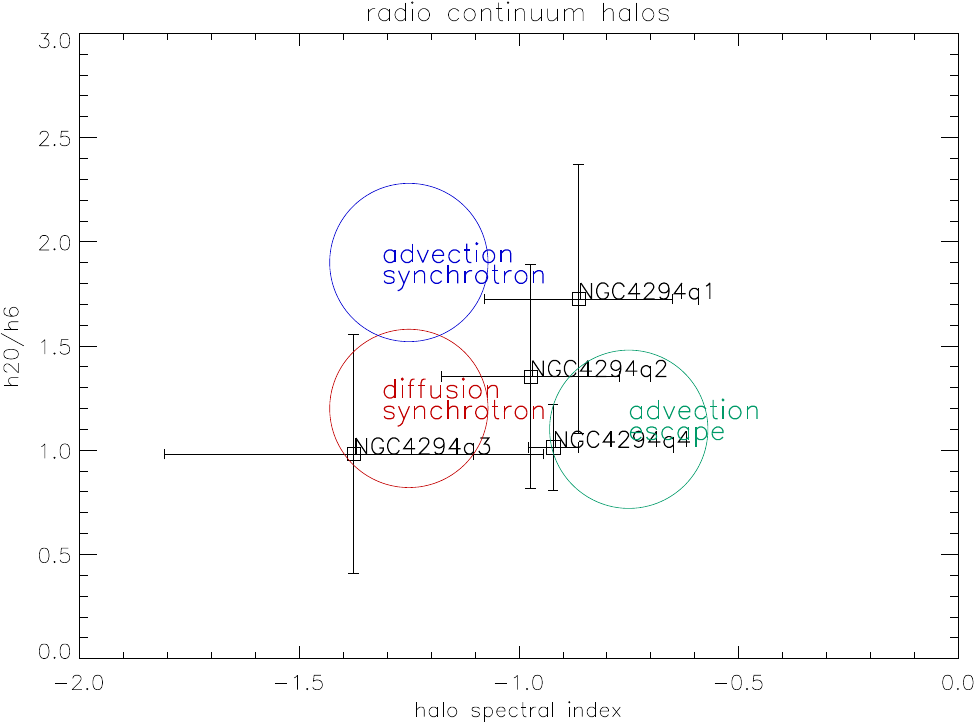}}
  \caption{NGC~4294. Height ratio as a function of the spectral index between 20cm and 6cm; upper panel: mean of all ten models; lower panel:
    mean of the four best-fit models.
  \label{fig:haloclass_n4294}}
\end{figure}

For the other Virgo galaxies the classification is less clear.
When the halo height ratios of a given quadrant is very high, it is not shown in the corresponding $r_H$--SI diagram.
In NGC~4178 the height ratio in the second quadrant is $\sim 8$ because of a large height
at $20$~cm. Whereas the halo of the third quadrant is advection escape-dominated, the conclusions for the first and fourth quadrants are less clear
(Fig.~\ref{fig:haloclass_n4178}).
The halos of the latter quadrants are marginally consistent with being advection escape-dominated within the error bars.

%%%%%%%%%%%%%%%%%%%%%%%%%%%%%%%%%%%%%%%%%%%%%%%%%%%%%%%%%%%%%%%%%%%%%%%%%%%%%%%%%%%%%%%%%%%%%%%%%%%%%%%%%%%%%%%%%%%%%%%%%%%%%%%%%%%%%%%%%%%%%%%%%%%%

%\FloatBarrier

%\FloatBarrier

\begin{figure}[ht!]
  \centering
  \resizebox{\hsize}{!}{\includegraphics{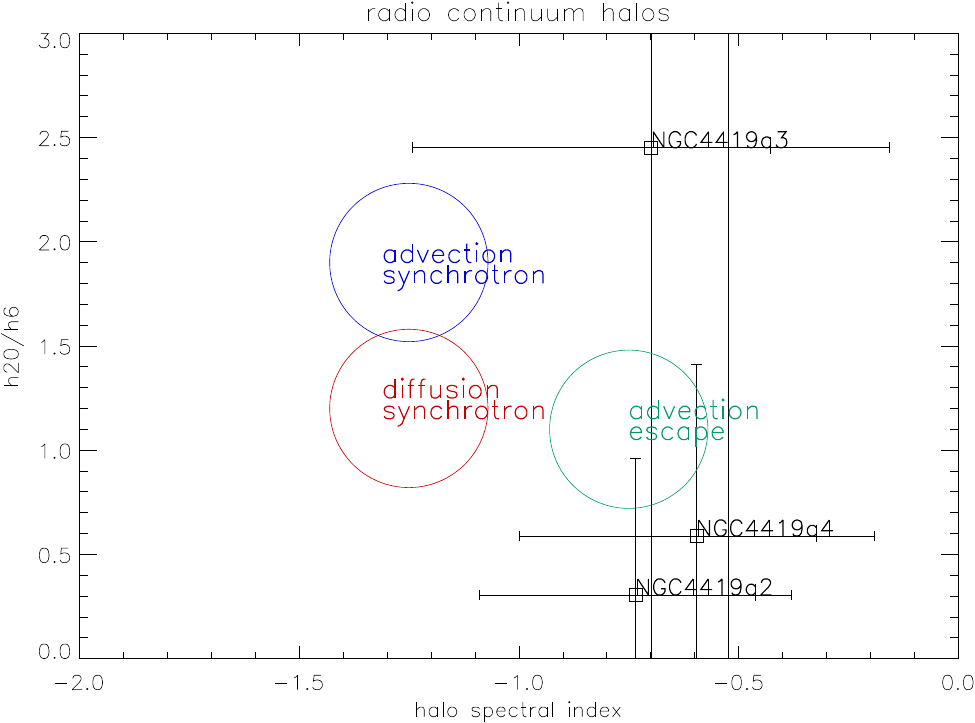}}
  \resizebox{\hsize}{!}{\includegraphics{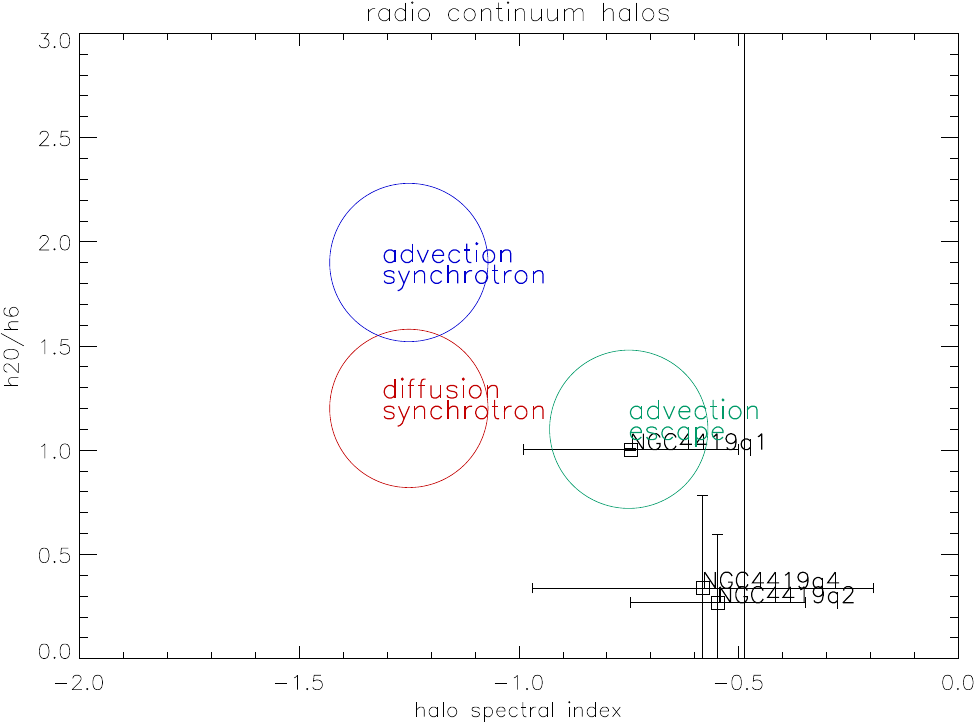}}
  \caption{NGC~4419. Height ratio as a function of the spectral index between 20cm and 6cm; upper panel: mean of all ten models; lower panel:
    mean of the four best-fit models.
  \label{fig:haloclass_n4419}}
\end{figure}
%\FloatBarrier

\begin{figure}[ht!]
  \centering
  \resizebox{\hsize}{!}{\includegraphics{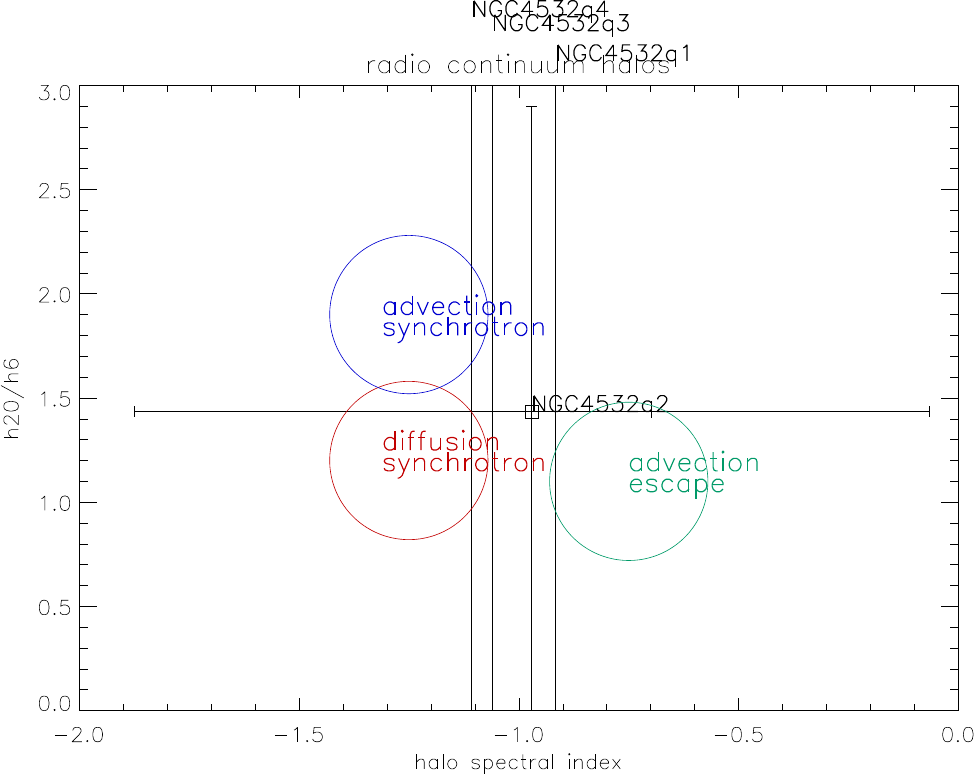}}
  \resizebox{\hsize}{!}{\includegraphics{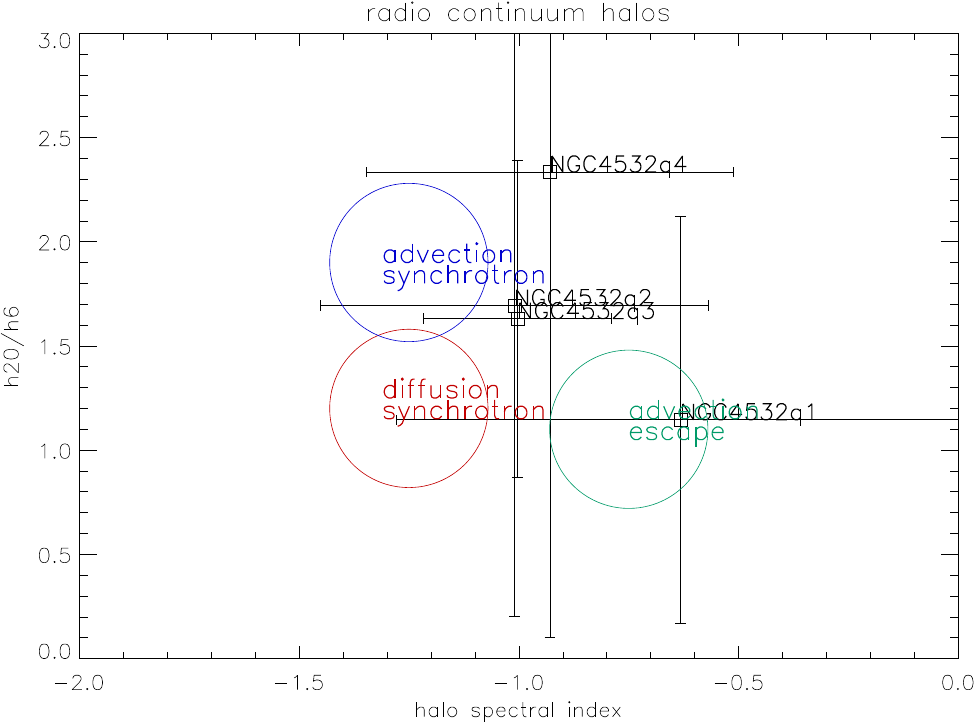}}
  \caption{NGC~4532. Height ratio as a function of the spectral index between 20cm and 6cm; upper panel: mean of all ten models; lower panel:
    mean of the four best-fit models.
  \label{fig:haloclass_n4532}}
\end{figure}
%\FloatBarrier

\begin{figure}[ht!]
  \centering
  \resizebox{\hsize}{!}{\includegraphics{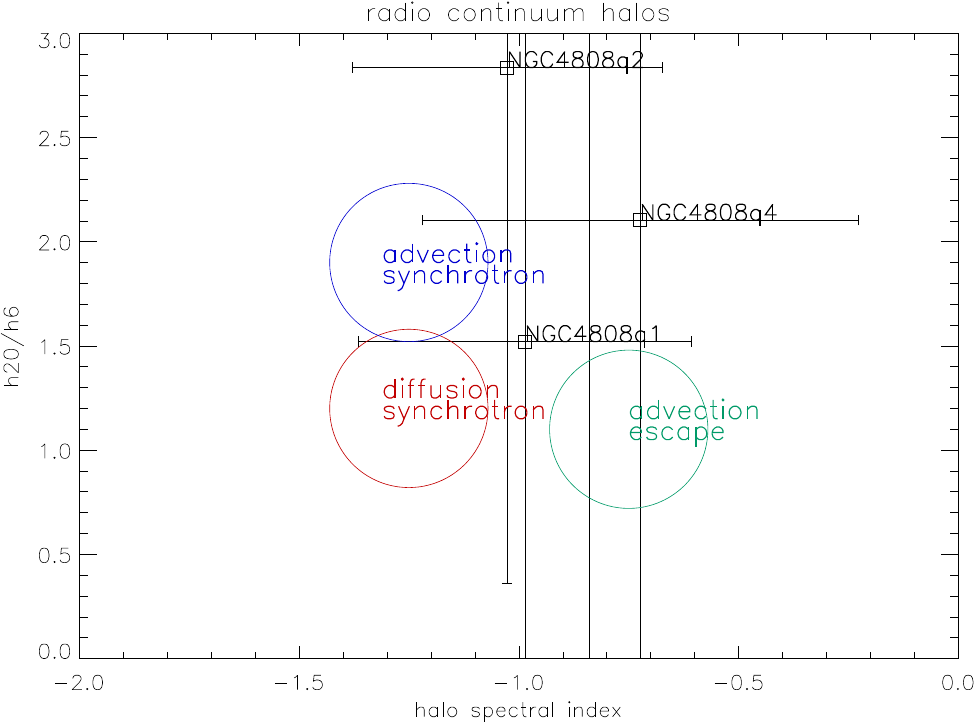}}
  \resizebox{\hsize}{!}{\includegraphics{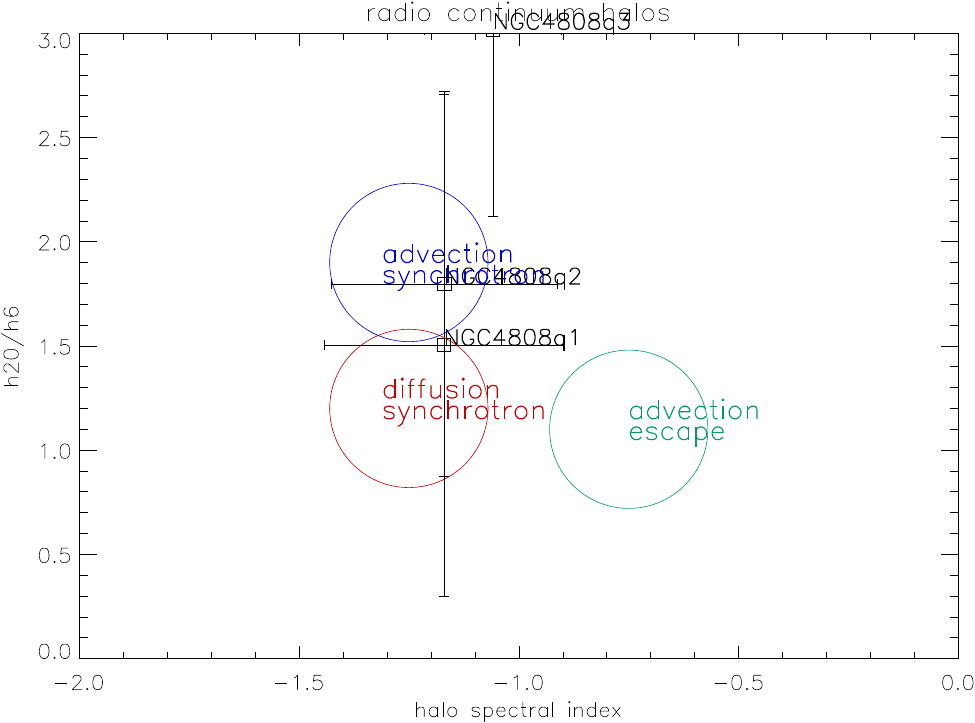}}
  \caption{NGC~4808. Height ratio as a function of the spectral index between 20cm and 6cm; upper panel: mean of all ten models; lower panel:
    mean of the four best-fit models.
  \label{fig:haloclass_n4808}}
\end{figure}
%\FloatBarrier

%%%%%%%%%%%%%%%%%%%%%%%%%%%%%%%%%%%%%%%%%%%%%%%%%%%%%%%%%%%%%%%%%%%%%%%%%%%%%%%%%%%%%%%%%%%%%%%%%%%%%%%%%%%%%%%%%%%%%%%%%%%%%%%%%%%%%%%%%%%%%%%%%%%%

In NGC~4419 the halos of all four quadrants are broadly consistent with being advection escape-dominated (lower panel of Fig.~\ref{fig:haloclass_n4419}).
In NGC~4532 the radio continuum halos of all four quadrants are consistent with being advection escape-dominated or diffusion synchrotron loss-dominated,
with a preference for the former (lower panel of Fig.~\ref{fig:haloclass_n4532}).
The halos of the first and second quadrants in NGC~4808 lie between the regions of advection escape and advection synchrotron dominance
(lower panel of Fig.~\ref{fig:haloclass_n4808}). The halo of the fourth quadrant is probably advection escape-dominated whereas the classification
of the halo in the third quadrant is uncertain. We conclude that, overall, advection escape-dominated radio continuum halos prevail in our sample.

The mean height ratios as a function of the mean halo SI for our Virgo galaxies are shown in Fig.~\ref{fig:finalplot_SIHR}.
Based on the mean values the halos of NGC~4192, NGC~4294, and NGC~4419 are advection escape-dominated.
Despite its large error bar for the height ratio, the halo of NGC~4178 is most probably advection/synchrotron escape-dominated. 
The halos of NGC~4532 and NGC~4808 have rather large error bars in both direction, which prevent their classification based on the
$r_H$--SI diagram.
\begin{figure}[ht!]
  \centering
  \resizebox{\hsize}{!}{\includegraphics{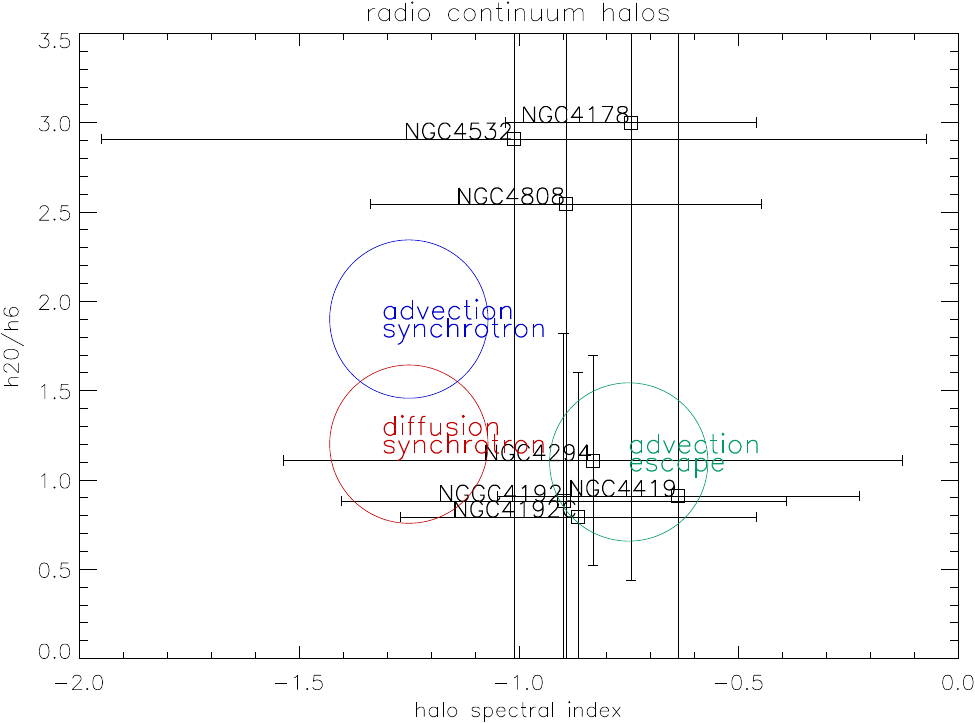}}
   \resizebox{\hsize}{!}{\includegraphics{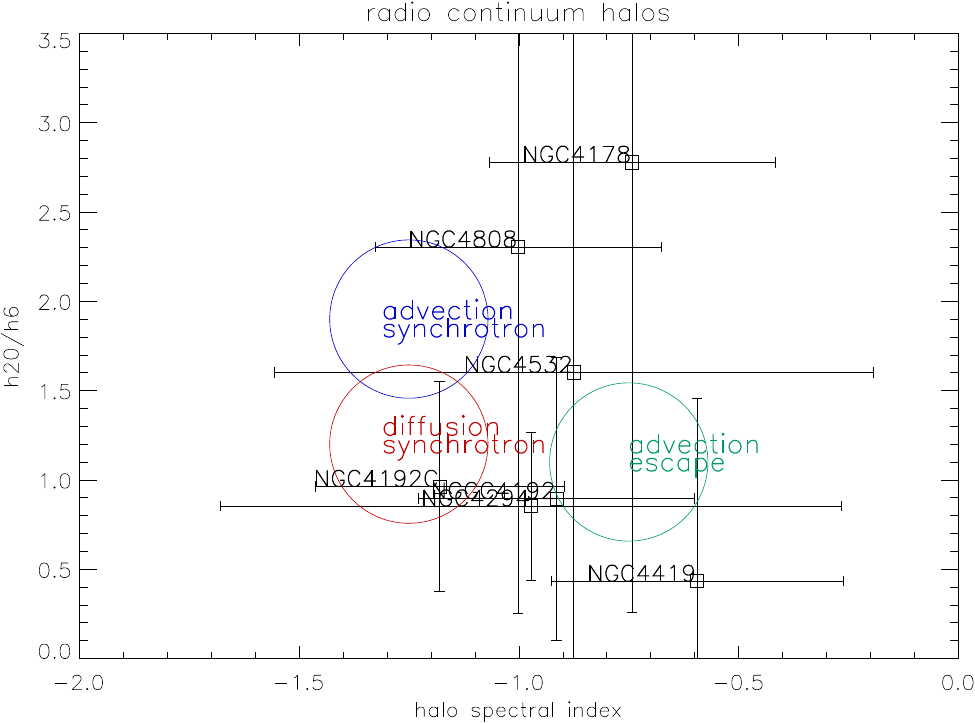}}
  \caption{Quantities averaged over all four quadrants: height ratio at 20cm and 6cm as a function of the spectral index between 20cm and 6cm.
    Upper panel: based on all ten models; lower panel: based on the four best-fit models.
  \label{finalplot_SIHR}}
\end{figure}
%\FloatBarrier

\section{Discussion \label{sec:discussion}}

Radio continuum halo can represent a significant fraction of the total radio continuum emission of a starforming spiral galaxy.
At $20$~cm and $6$~cm between $30$ and $70$\,\% of the total radio continuum emission originate in the halo
(Table~\ref{tab:heights1}; Stein et al. 2019a; Stein et al. 2020). The fraction is higher at low frequencies.
This is in line with the results of radio continuum model of Vollmer et al. (2022, 2025) where the effective height of the
radio continuum emission is that of the gas disk ($H$) and not that of the starforming disk ($l_{\rm driv}$ in their nomenclature).

All galaxies of our sample are located in the Virgo cluster (see Fig.~1 of Vollmer et al. 2013). NGC~4402 and
NGC~4330 show strongly asymmetric radio continuum halos, where one side is compressed by ram pressure from the intercluster medium and the
other side is linked to stripped material (Crowl et al. 2005, Vollmer et al. 2013). The radio continuum halos of NGC~4294, NGC~4532, and
NGC~4808 show a lesser degree of asymmetry. All three galaxies are not H{\sc i}-deficient (Chung et al. 2009). NGC~4294 is part of a galaxy pair
and has an asymmetric outer H{\sc i} distribution. NGC~4532 and NGC~4808 have extended asymmetric H{\sc i} envelopes (Chung et al. 2009).
It thus seems that the cluster environment can lead to strongly asymmetric radio continuum halos via ram pressure, whereas the local environment
(galaxy pairs or external gas accretion) leads to rather weakly asymmtries of the halos.

The ratio between the radio luminosity and the SFR is normal for all galaxies except NGC~4532, for which it is
twice as high as the mean (Fig.~1 of Vollmer et al. 2020 and Table~\ref{tab:sample}).
On the other hand, the mean $6$~cm surface brightness of NGC~4532 and NGC~4808 is about five times higher than that of NGC~4178.
The mean $6$~cm surface brightnesses of NGC~4192, NGC~4294, and NGC~4419 lie between these two extremes (Vollmer et al. 2013).
We conclude that NGC~4532 and NGC~4808 most probably drive galactic winds and their halos are caused by CRe advection.

It turned out that the estimation of the halo parameters is best for large galaxies with a relatively high
radio surface brightness (NGC~4192), which is expected. For the halos of the other galaxies we can use additional information
from large-scale magnetic field structure derived from the polarized $6$~cm radio continuum emission (Fig.~\ref{fig:bvectors}).
The large-scale magnetic field configuration of a disk galaxy can be divided into three main types:
(i) disk type, where the large-scale magnetic field is parallel to the disk plane, (ii) disk-halo type, which shows a characteristic X-structure,
(ii) halo type, where the large-scale magnetic field is perpendicular to the disk plane (see, e.g., Irwin et al. 2024).

The projected magnetic field lines is of disk-halo type in NGC~4192 and NGC~4419, of halo type in the southwestern disk of NGC~4294,
of disk-halo and halo type in NGC~4532, and of halo type in NGC~4808. We interpret the vertical structures of the large-scale magnetic field
within the disk-halo and halo types as a sign of a galactic outflow or wind.
As expected, the two galaxies with the highest radio continuum surface brightnesses and thus SFR surface density show clear signs of galactic winds. 
These outflows lead to an enhanced radio continuum emission with
respect to the SFR only in NGC~4532, which has the highest specific SFR of our galaxy sample (sSFR$=8.2 \times 10^{-10}$~yr$^{-1}$) and might
be qualified as a starburst. Thus, all galaxies except NGC~4178 are expected to harbor an advection-dominated halo.
This in turn means that our estimate of the SI is more robust than that of the height ratio (see Sect.~\ref{sec:physics}).

All halos of our highly-inclined galaxy sample contain information on the underlying star formation distribution within the galactic
disk (see Sect.~\ref{sec:results}). This is expected in an advection-dominated halo, where the CRe are transported into the halo
by a galactic wind or outflow. As for the CHANG-ES sample, our galaxy sample is also dominated by advection escape-dominated
radio continuum halos.

The particularly thin radio disk or halo of NGC~4178 (Fig.~\ref{fig:resultats_finaux_mesquita2}) is remarkable.
Vollmer et al. (2013) did not detect any polarized $6$~cm radio emission in this galaxy. This thinness calls for an explanation.
Another Virgo galaxy in the sample of Vollmer et al. (2013) is NGC~4216. It has the lowest $6$~cm mean surface brightness
of the sample, a factor of $2.5$ lower than that of NGC~4192 and a factor of $1.7$ lower than that of NGC~4178.
Whereas NGC~4216 is strongly H{\sc i} deficient, NGC~4178 is not. In addition, NGC~4178 has a normal specific SFR of sSFR$=2 \times 10^{-10}$~yr$^{-1}$.
The radio-thinness is not due to a diffusion synchrotron loss-dominated halo because the ratio between the $6$~cm vertical height
and diameter are about the same for NGC~4013, which has a diffusion synchrotron loss-dominated halo, and NGC~4217, which has
an advection escape dominated halo. It thus appears that radio-thinness only correlates with the mean $6$~cm surface brightness
and the lack of significant polarized radio continuum emission. 
We can only speculate that NGC~4216 and NGC~4178 lack a strong halo magnetic field. In this case, the halo is probably
diffusion escape-dominated because the SFR surface density is low compared to the average SFR surface density of starforming galaxies in both galaxies.

We can try to connect our flaring radio continuum disks to galactic H{\sc i} an extended diffuse ionized gas (eDIG) disks. 
The H{\sc i} radial profiles appear to have a typical shape, with the ﬂaring increasing linearly with radius
where the stellar disk dominates the local gravitational potential, and steepening to an exponential proﬁle in the outer disk where
the potential is probably dominated by the halo (O'Brien et al. 2010).
Our radio continuum halos reside within the optical disk where stellar disk dominates the local gravitational potential.
One might qualify the inner parts of the thick halos as linearly flaring but most of them flatten at larger radii. The latter
behavior is contrary to that of the H{\sc i} disk. We thus conclude that the exponential vertical H{\sc i} profile at large galactic radii
has a different cause than that of the flattening vertical radio continuum profile. Whereas the latter is caused by star formation
and CRe transport, the latter is most probably caused by external gas accretion.

Concerning the warm ionized gas, Lu et al. (2023) found that the eDIG scale height of $22$ nearby edge-on spiral galaxies is comparable to
that at $20$~cm. In addition, the eDIG is slightly more extended than the neutral gas. However, they did not measure radial profiles of
the H$\alpha$ scale height. It will be interesting to compare the vertical profiles at large galactic radii of the three different gas phases.
Clustered star formation might be a key ingredient to explain the multi-phase nature of the halo gas.
Indeed, the injection spectrum of CRe and CRp of the model of Vollmer et al. (2024), which is needed to explain the integrated
radio continuum emission of local and high-z main sequence and starburst galaxies, is that expected for superbubbles created
by multiple SN remnants (Vieu et al. 2022). Multiple holes, bubbles, worms (Heiles 1984) and chimneys created by the explosions of multiple clustered
SN allow the CRe to travel into the halo region. At the same time, the illuminated walls of these structures constitute
the warm ionized medium or DIG as. This scenario was already proposed by Koo et al. (1992) and is corroborated by more recent work, as for
example, Belfiore et al. (2022).

\section{Conclusions \label{sec:conclusions}}

The vertically extended component of the diffuse radio continuum can represent an important fraction of the total radio continuum emission of
a starforming galaxy. The characteristics of radio continuum halos are traditionally measured in edge-on galaxies (Krause et al. 2018). These 
measurements are limited by the projection of the vertical emission distribution and by the assumption of a constant 
halo scale height. We made an attempt to reconstruct the radial properties of radio continuum halos in nearly edge-on galaxies ($70^{\circ} \le i \le 78^{\circ}$)
where the SFR surface density distribution can still be deprojected and the vertical radio continuum emission is still well distinct from the disk emission. The 
deprojected SFR surface density distribution is convolved with a Gaussian kernel to take CRe diffusion within the galactic disk into
account and a vertical profile of the radio continuum emissivity (Table~\ref{tab:halopresc}) is added to the disk emission. The three
dimensional emission distribution is then projected on the sky and compared to VLA radio continuum observations
at 20 and 6 cm (Vollmer et al 2013). We also used publicly available VLA data from the CHANG-ES
project (Irwin et al. 2012). The comparison is made separately in the four quadrants of the radio continuum images.
Our results are based on the 10 and 4 model with the lowest $\chi^2$. Based on the detailed comparison we conclude that
\begin{enumerate}
\item
  our method is robust because of the consistent overall results for the CHANG-ES and our data of NGC 4192;
\item
  overall the halo emission contains information on the underlying distribution of the SFR surface density;
\item
  the majority of our galaxies show flaring radio continuum halos;
\item
  except for NGC 4178, our Virgo galaxies follow the trend of increasing effective height with increasing radio
 continuum size found by the CHANG-ES collaboration (Krause et al. 2018);
\item
  radio continuum halos can represent a significant fraction of the total radio continuum emission of a starforming
 spiral galaxy. At 20 cm and 6 cm between 30 and 70\,\% of the total radio continuum emission originate in the
 halo (Table 3; Stein et al. 2019a; Stein et al. 2020);
\item
  a halo classification based on the height ratio and SI between $20$ and $6$~cm is proposed;
\item
 if we interpret the vertical structures of the large-scale magnetic field within the disk-halo and halo types as
 a sign of a galactic outflow or wind, all galaxies except NGC 4178 most probably harbor an advection dominated halo.
\end{enumerate}
The exponential vertical H{\sc i} profile at large galactic radii has a different cause than that of the flattening vertical radio
continuum profile. Whereas the latter is caused by star formation and CRe transport, the latter is most probably
caused by external gas accretion. Clustered star formation might be a key ingredient to explain the multi-phase nature of the halo gas.
Multiple holes, bubbles, worms (Heiles 1984) and chimneys created by the explosions of multiple clustered SN allow the CRe to travel into the halo region.
At the same time, the illuminated walls of these structures constitute the warm ionized medium or DIG. It will be interesting to compare
the radial profiles of the energy densities of the warm neutral gas (H{\sc i}), the diffuse ionized gas or warm ionized medium, and the cosmic rays to investigate
if and where the CRe can drive a galactic wind or outflow.

\section{Data availability \label{sec:availability}}

Appendix~C can be found on Zenodo (https://doi.org/10.5281/zenodo.17258433).

\begin{acknowledgements}
We would like to thank the CHANG-ES collaboration for providing the VLA data of NGC~4192.
\end{acknowledgements}

\begin{appendix}

\FloatBarrier
  
\section{Observations}
  
\begin{figure}
  \centering
  \resizebox{\hsize}{!}{\includegraphics{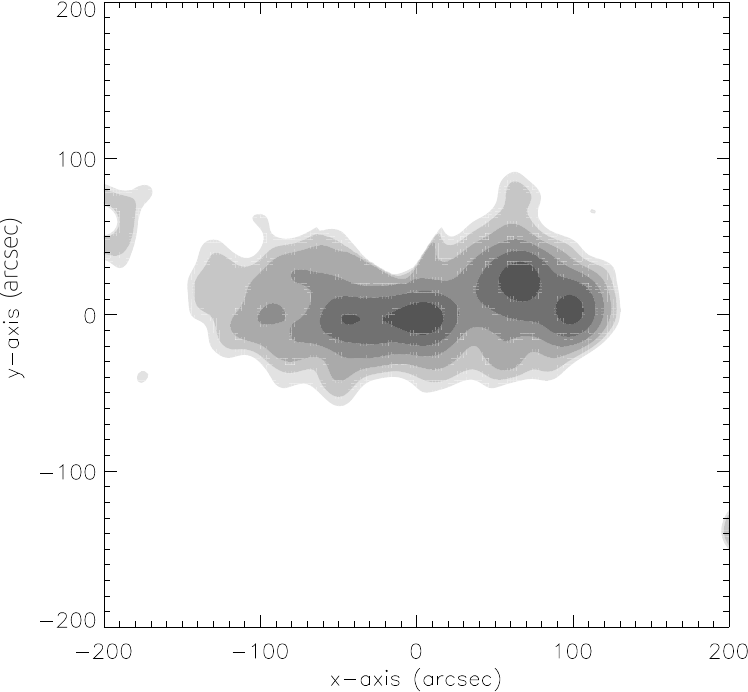}\put(-300,300){\bf \Huge NGC 4178}\includegraphics{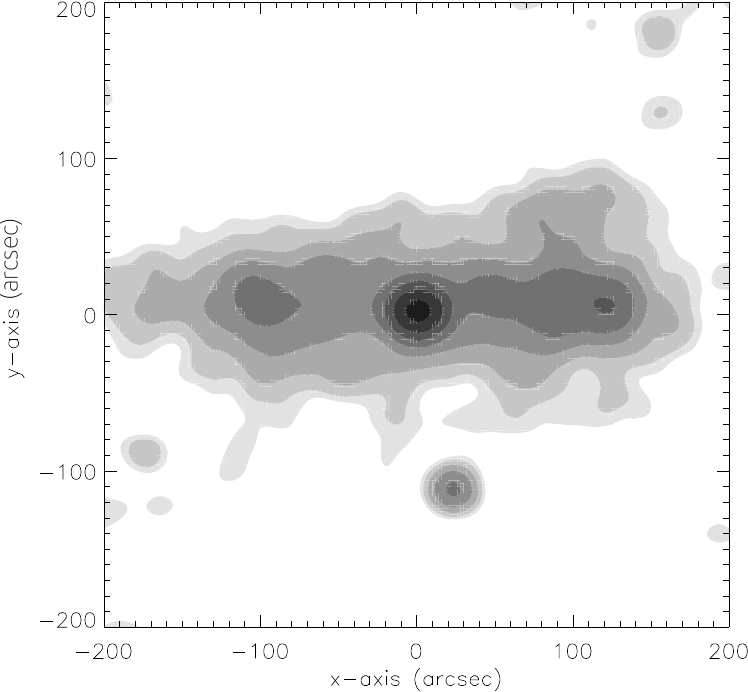}\put(-300,300){\bf \Huge NGC4192}}
  \resizebox{\hsize}{!}{\includegraphics{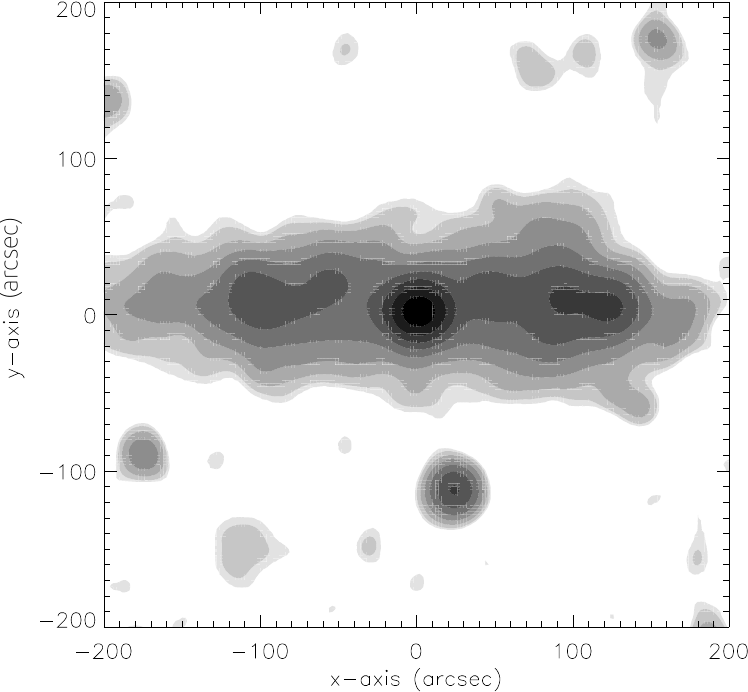}\put(-300,300){\bf \Huge NGC4192 CHANG-ES}\includegraphics{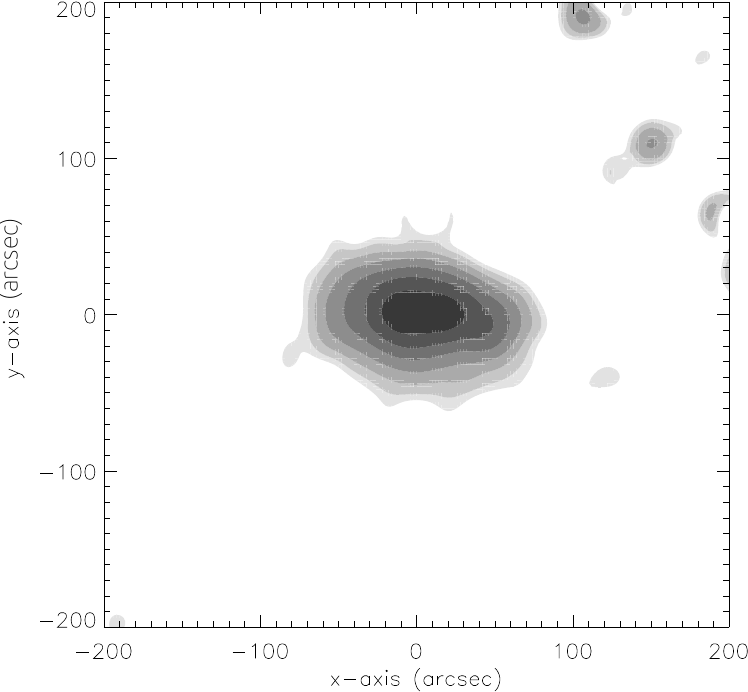}\put(-300,300){\bf \Huge NGC 4294}}
  \resizebox{\hsize}{!}{\includegraphics{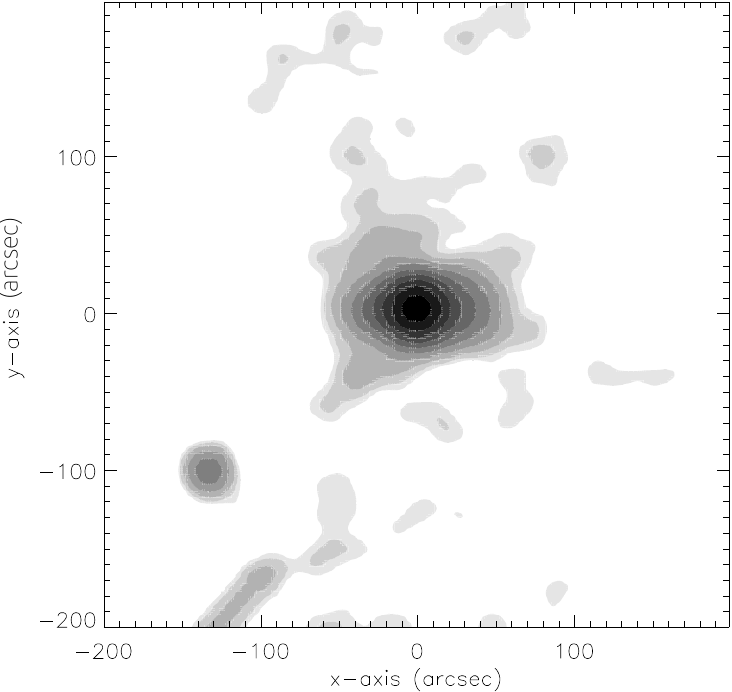}\put(-300,300){\bf \Huge NGC 4419}\includegraphics{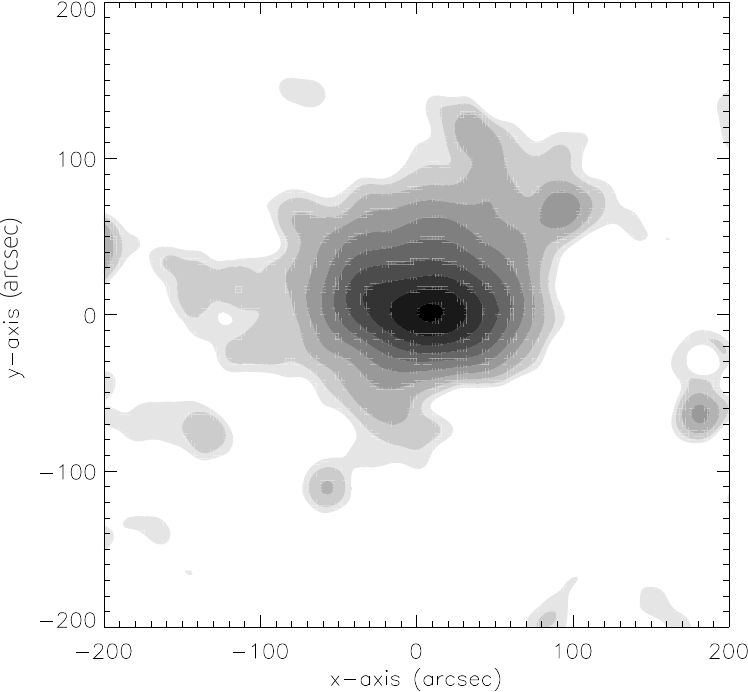}\put(-300,300){\bf \Huge NGC 4532}}
  \resizebox{\hsize}{!}{\includegraphics{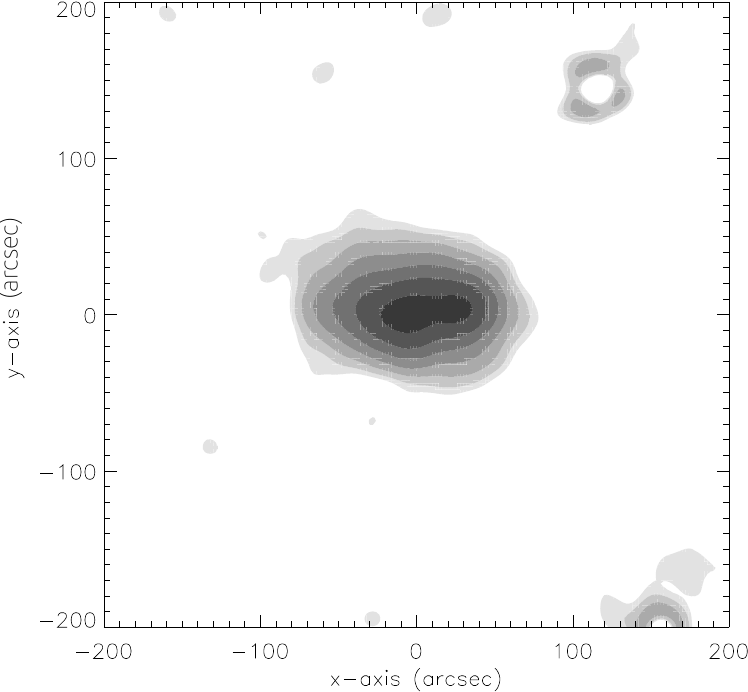}\put(-300,300){\bf \Huge NGC 4808}\textcolor{white}{\rule{\hsize}{\hsize}}}
  \caption{Nonthermal radio continuum emission distributions at $6$~cm. The contours are $(3,5,9,17,33,65,129,257,513,1025) \times \xi$ with $\xi=(11,19,8,11,11,12,20)$~$\mu$Jy/beam for NGC~4178, NGC~4192, NGC~4192CHANGES, NGC~4294, NGC~4419, NGC~4532, and NGC~4808, respectively.
  \label{fig:galaxies_rad}}
\end{figure}
%\FloatBarrier

\begin{figure}
  \centering
  \resizebox{\hsize}{!}{\includegraphics{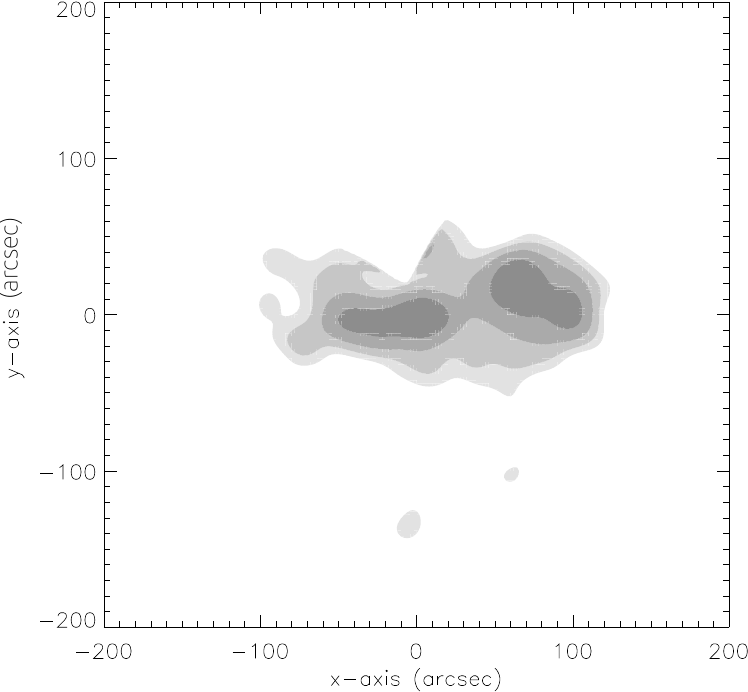}\put(-300,300){\bf \Huge NGC 4178}\includegraphics{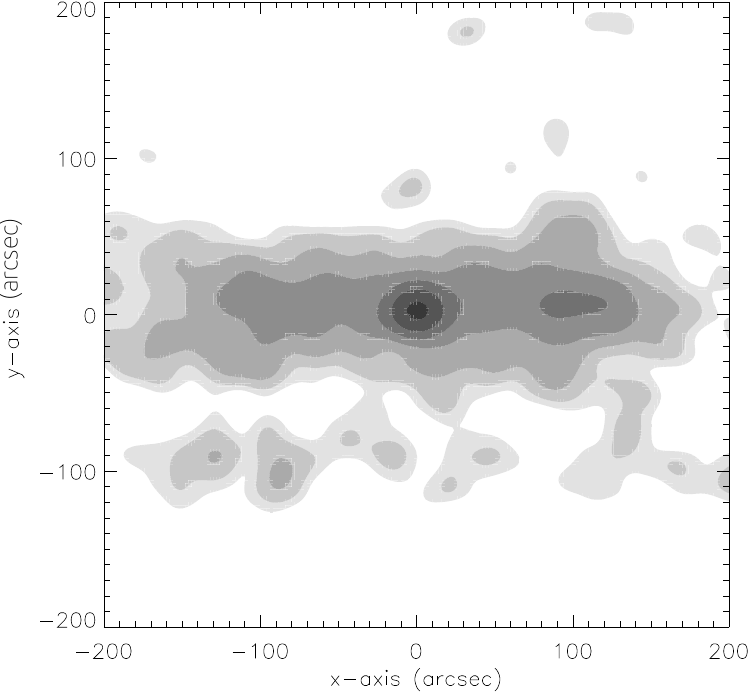}\put(-300,300){\bf \Huge NGC4192}}
  \resizebox{\hsize}{!}{\includegraphics{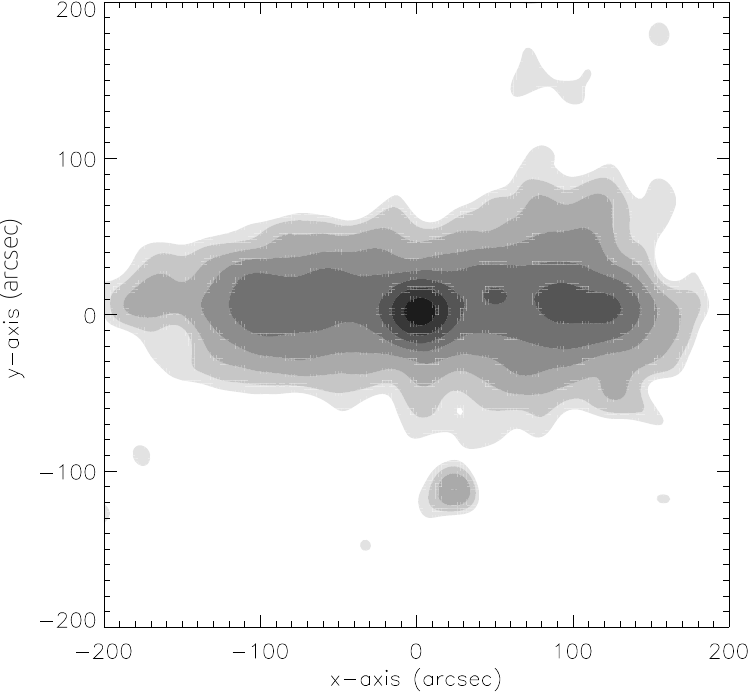}\put(-300,300){\bf \Huge NGC4192 CHANG-ES}\includegraphics{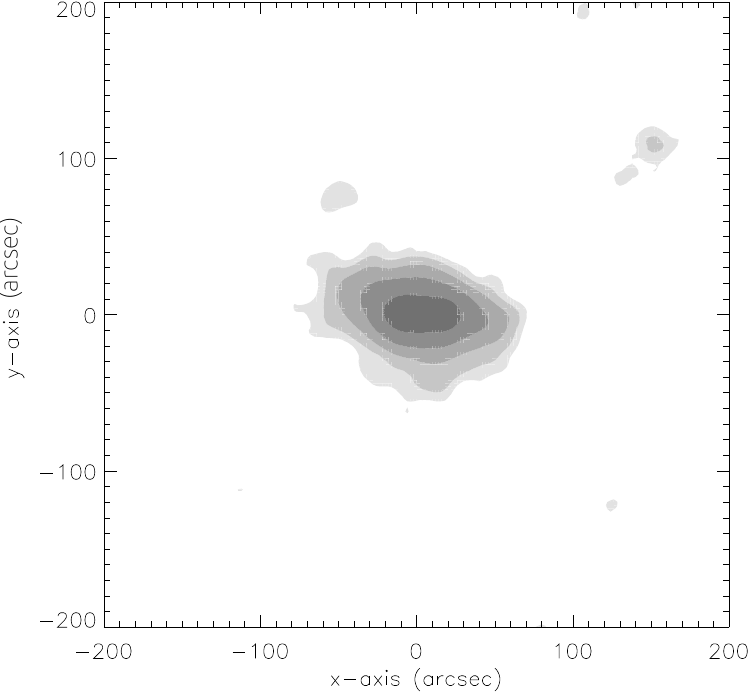}\put(-300,300){\bf \Huge NGC 4294}}
  \resizebox{\hsize}{!}{\includegraphics{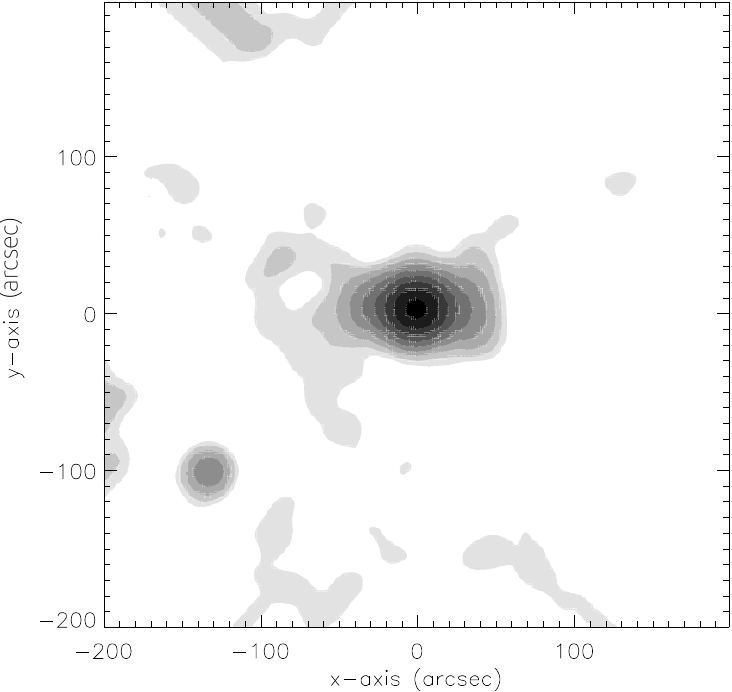}\put(-300,300){\bf \Huge NGC 4419}\includegraphics{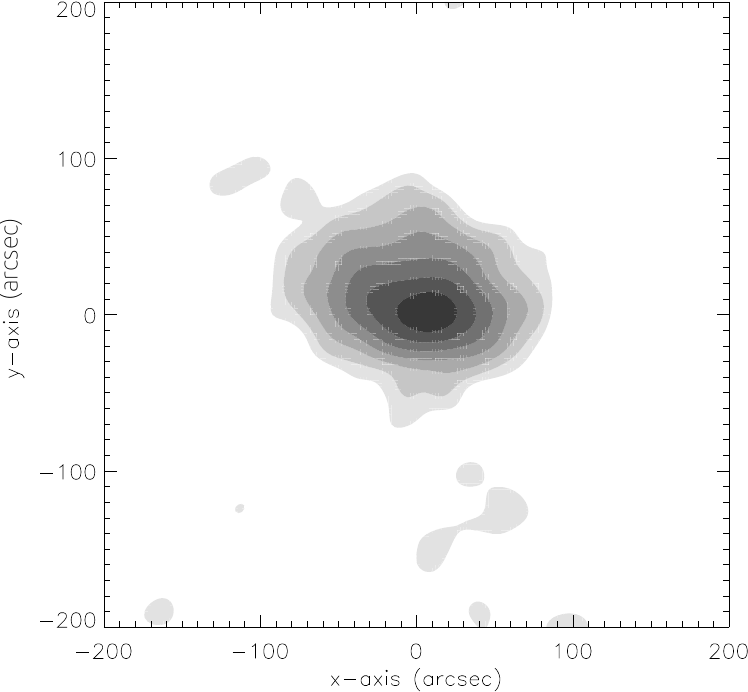}\put(-300,300){\bf \Huge NGC 4532}}
  \resizebox{\hsize}{!}{\includegraphics{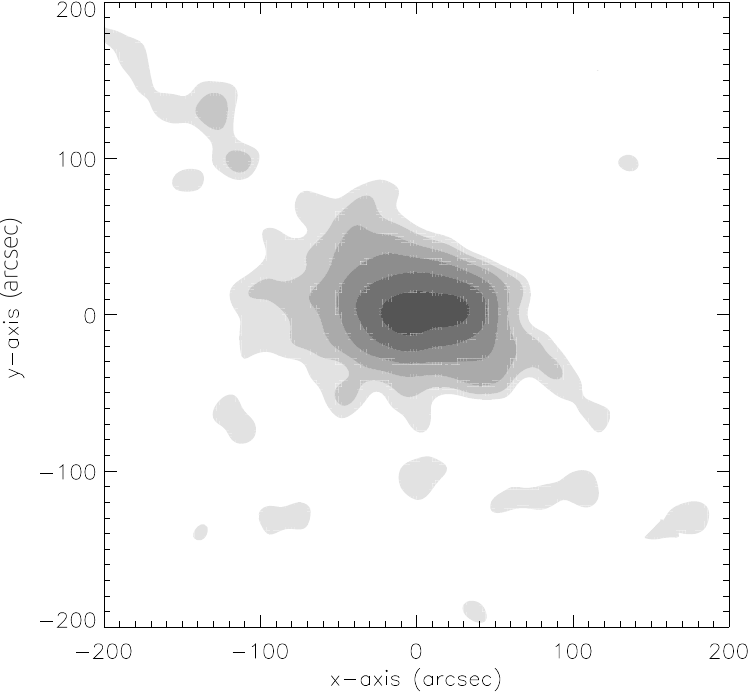}\put(-300,300){\bf \Huge NGC 4808}\textcolor{white}{\rule{\hsize}{\hsize}}}
  \caption{Nonthermal radio continuum emission distributions at $20$~cm. The contours are $(3,5,9,17,33,65,129,257,513,1025) \times \xi$ with $\xi=(69,109,9,100,66,127,90)$~$\mu$Jy/beam for NGC~4178, NGC~4192, NGC~4192CHANGES, NGC~4294, NGC~4419, NGC~4532, and NGC~4808, respectively.
  \label{fig:galaxies_rad20}}
\end{figure}
%\FloatBarrier

\FloatBarrier
\section{Deprojected star formation maps}

\begin{figure*}
  \centering
  \resizebox{\hsize}{!}{\includegraphics{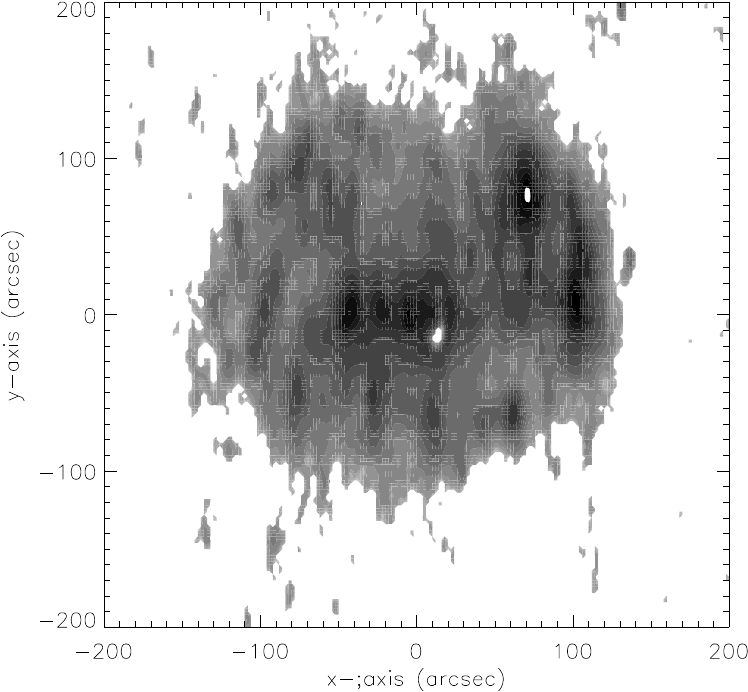}\put(-300,300){\bf \Huge NGC 4178}\includegraphics{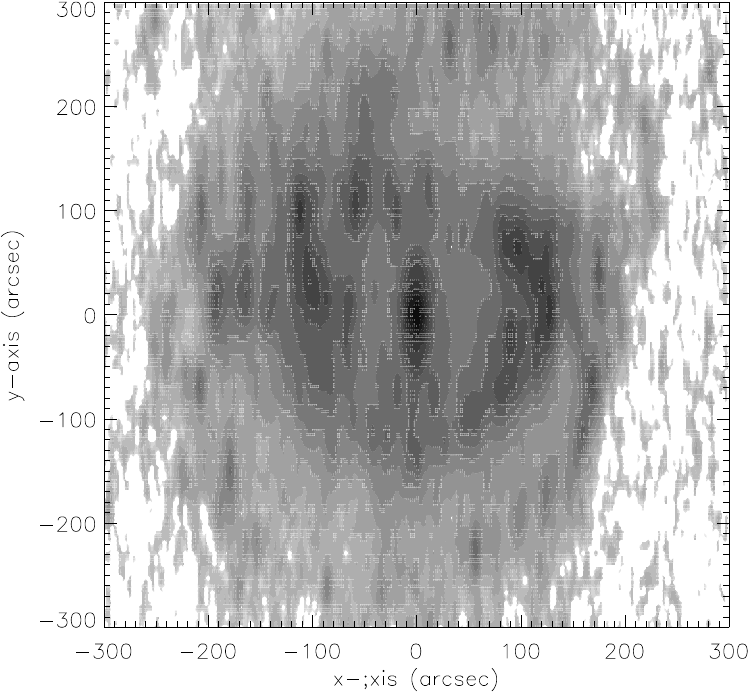}\put(-300,300){\bf \Huge NGC4192}\includegraphics{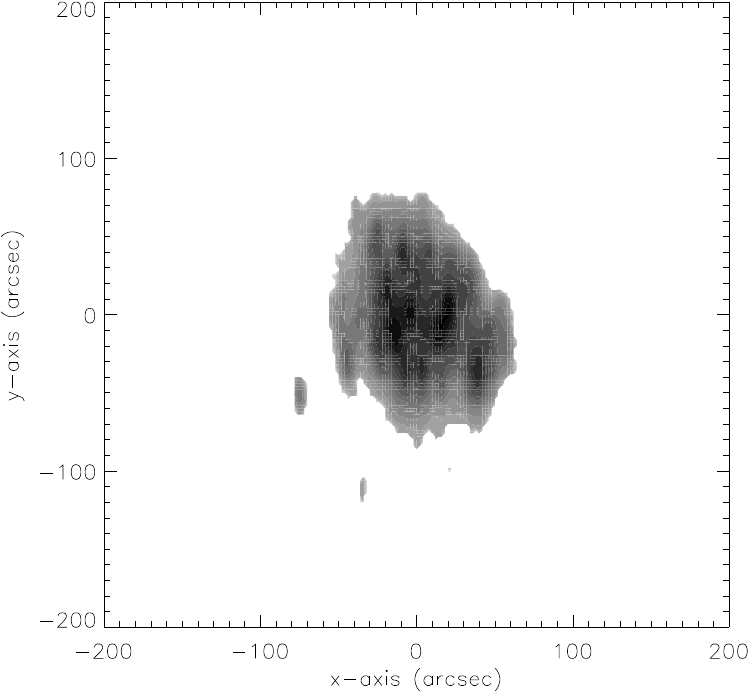}\put(-300,300){\bf \Huge NGC 4294}}
  \resizebox{\hsize}{!}{\includegraphics{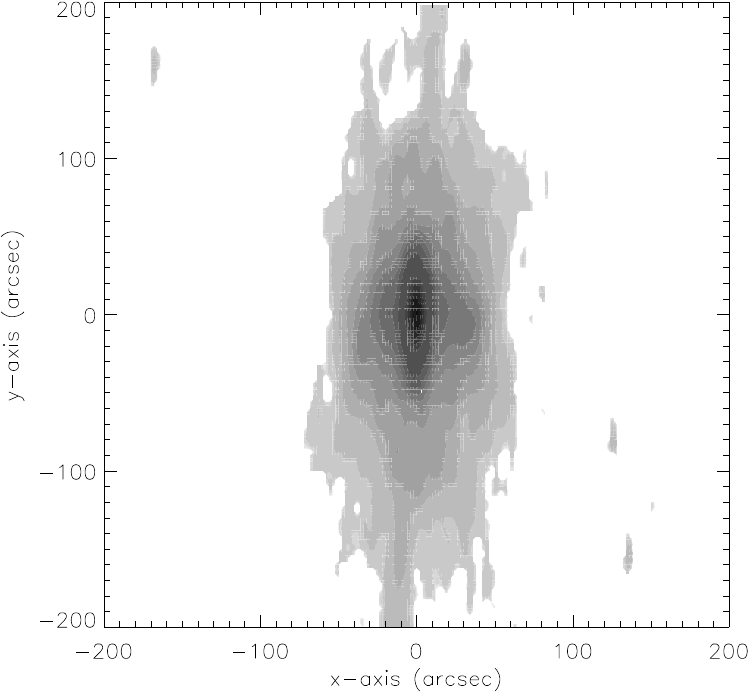}\put(-300,300){\bf \Huge NGC 4419}\includegraphics{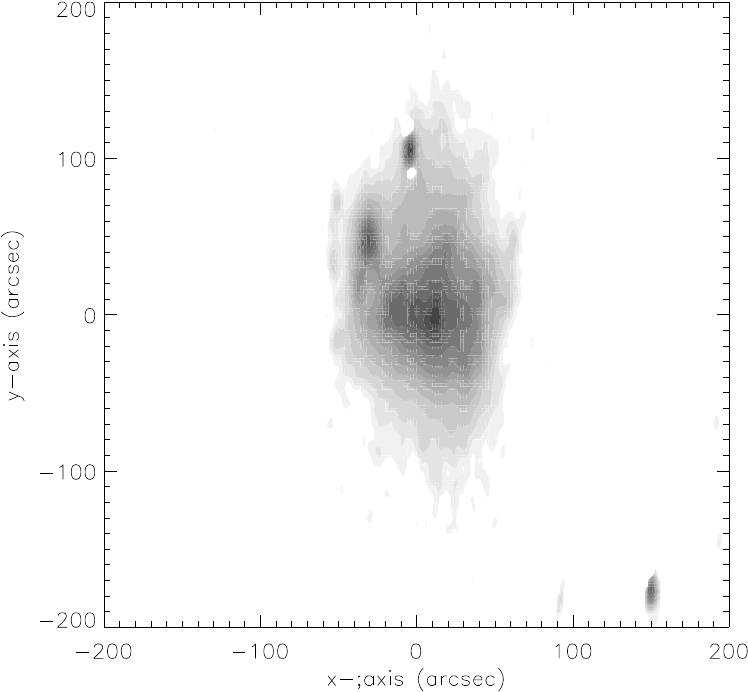}\put(-300,300){\bf \Huge NGC 4532}\includegraphics{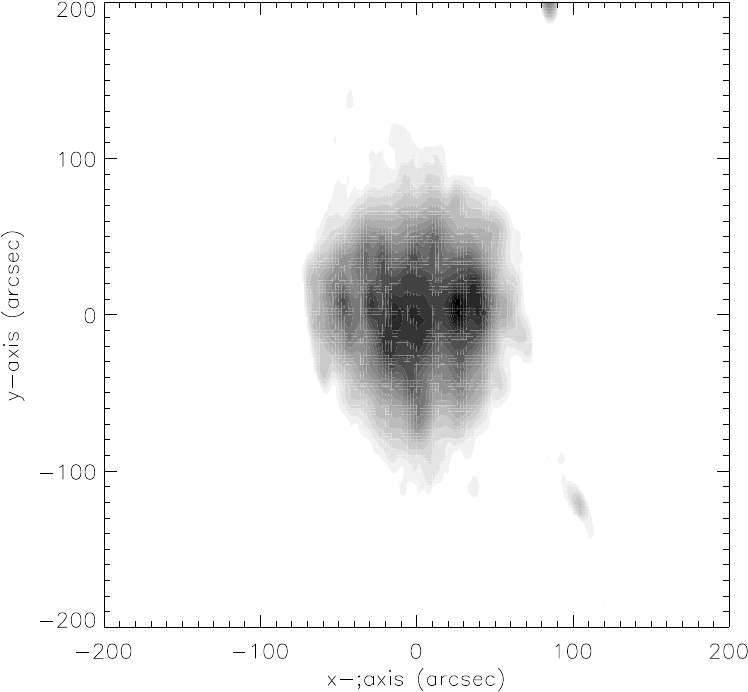}\put(-300,300){\bf \Huge NGC 4808}}
  \caption{Deprojected distributions of the SFR surface density plotted with a square-root transmission function. The units are arbitrary.
  \label{fig:galaxies_sfr}}
\end{figure*}
%\FloatBarrier

\FloatBarrier
\section{Radio continuum halo model \label{sec:halomodels}}

\begin{figure*}
  \centering
  \resizebox{14cm}{!}{\includegraphics{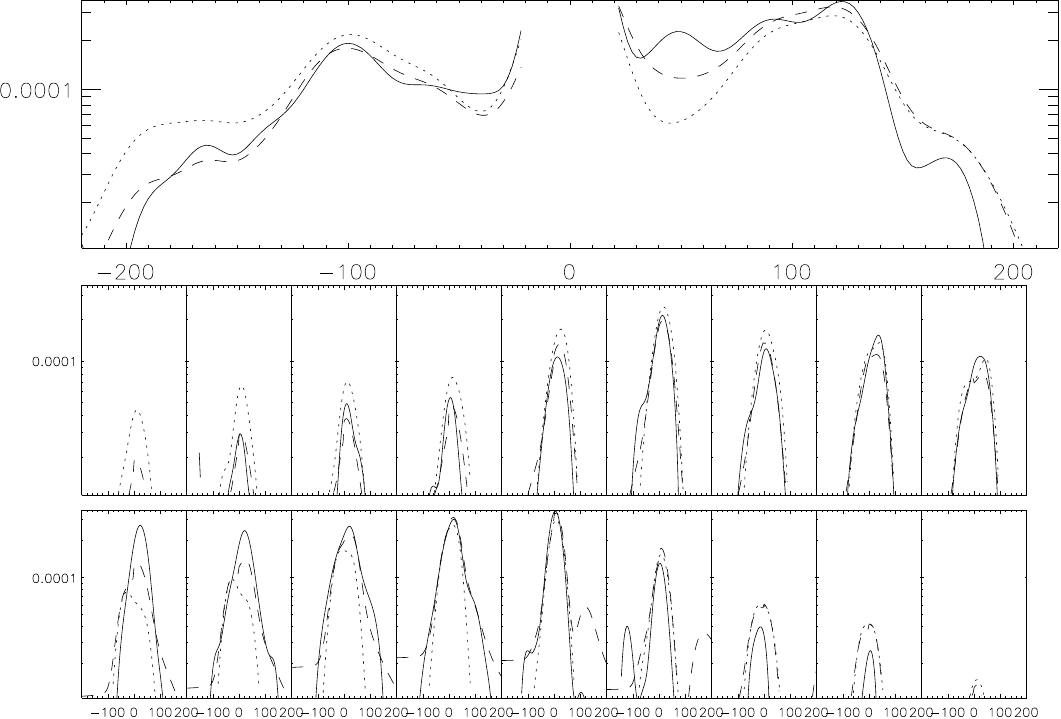}\put(-400,235){\bf \Huge NGC4192 6cm CHANG-ES}\includegraphics{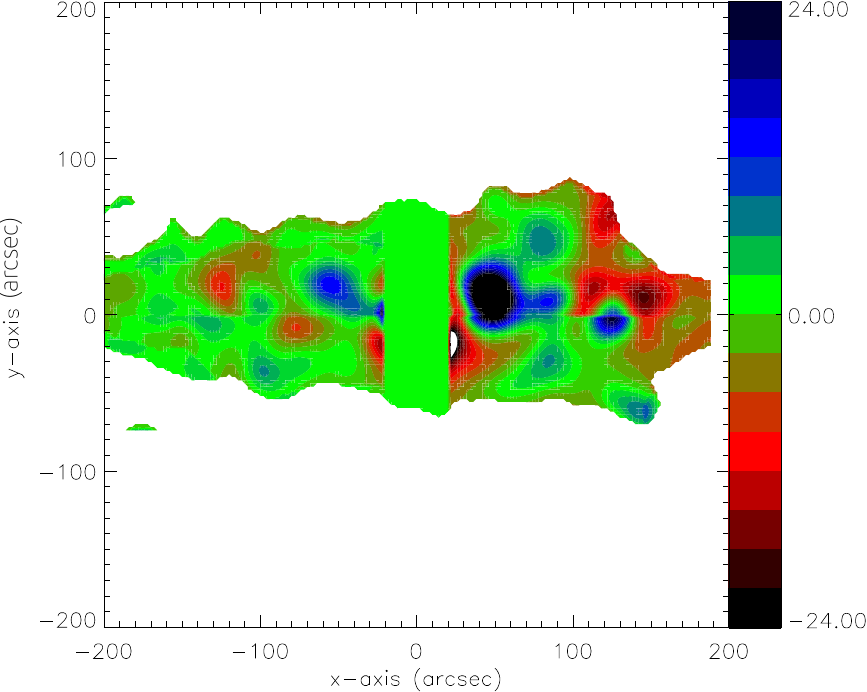}\put(-350,300){\bf \Huge NGC 4192 6cm}\put(-350,270){\bf \Huge CHANG-ES}}
  \resizebox{14cm}{!}{\includegraphics{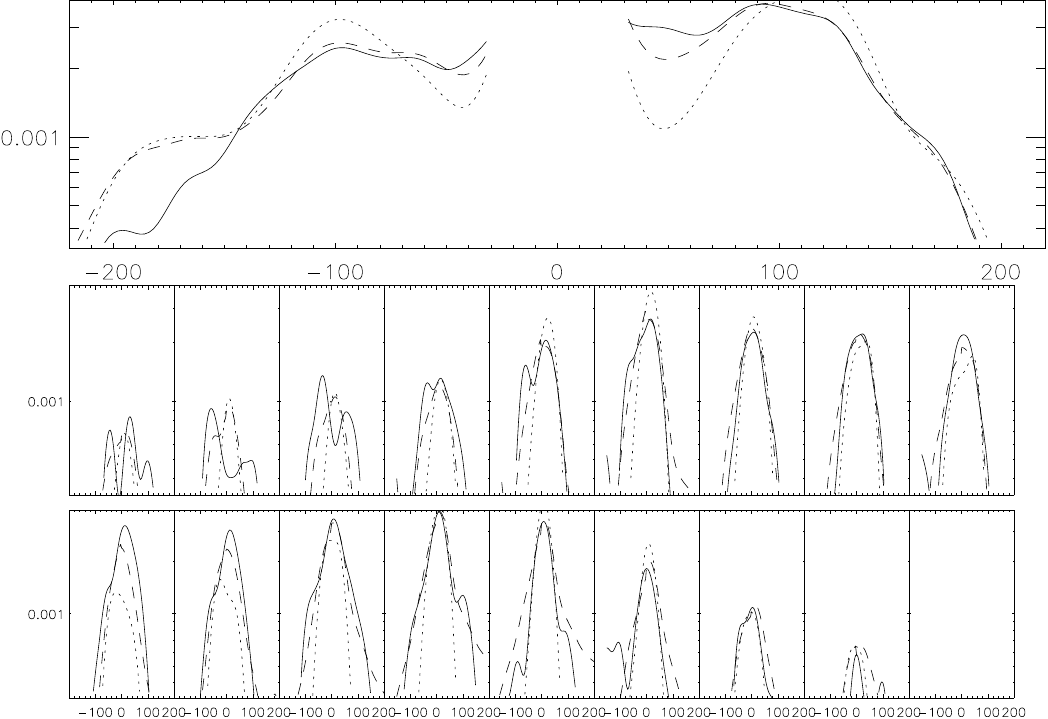}\put(-400,235){\bf \Huge NGC 4192 20cm}\includegraphics{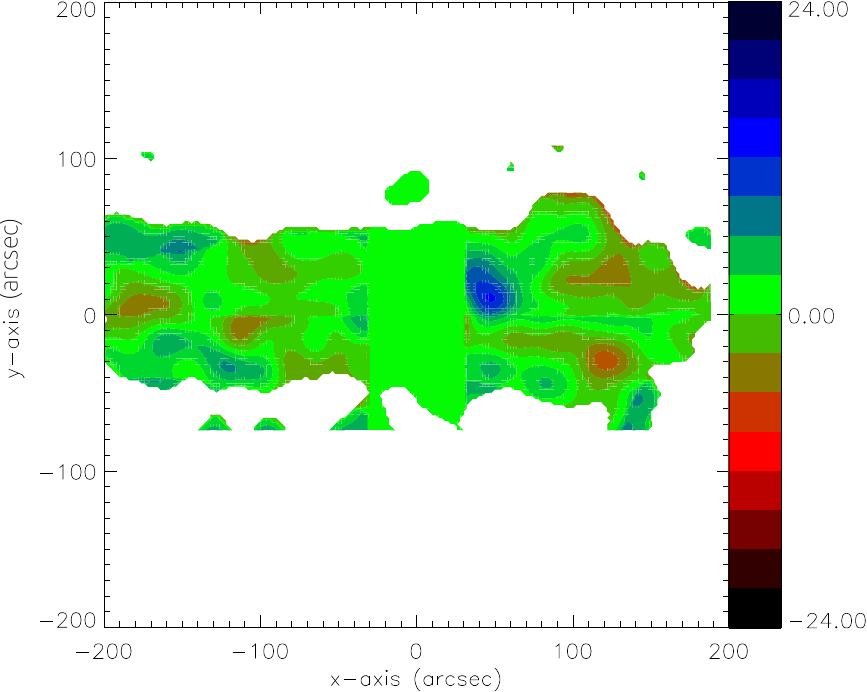}\put(-350,300){\bf \Huge NGC 4192 20cm}}
  \resizebox{14cm}{!}{\includegraphics{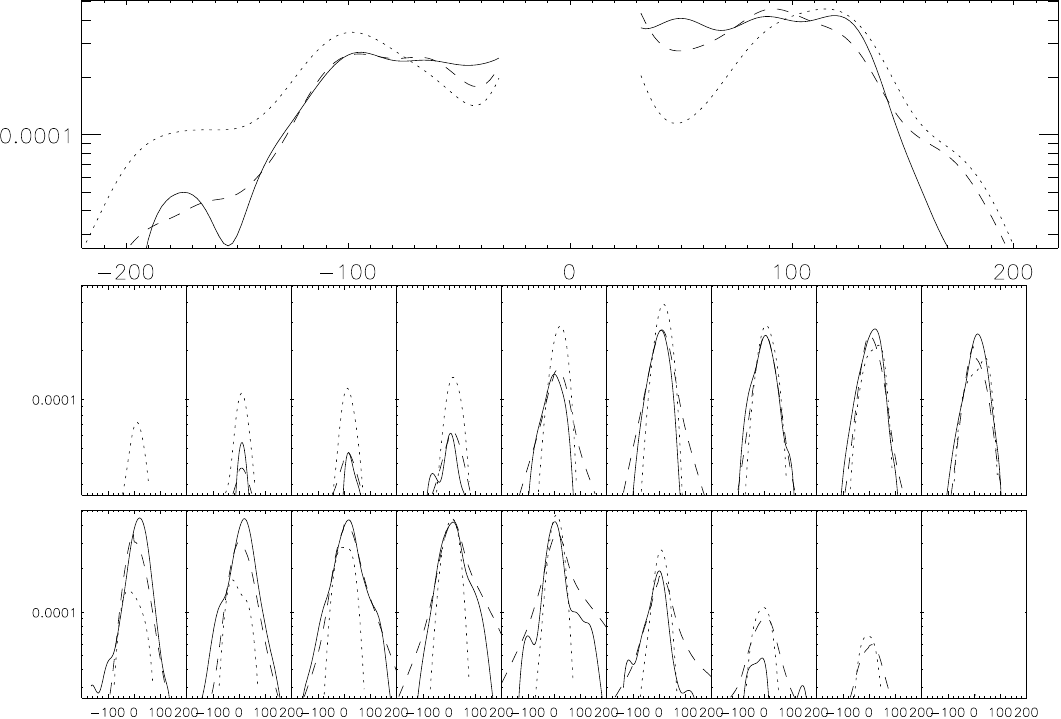}\put(-400,235){\bf \Huge NGC 4192 20cm CHANG-ES}\includegraphics{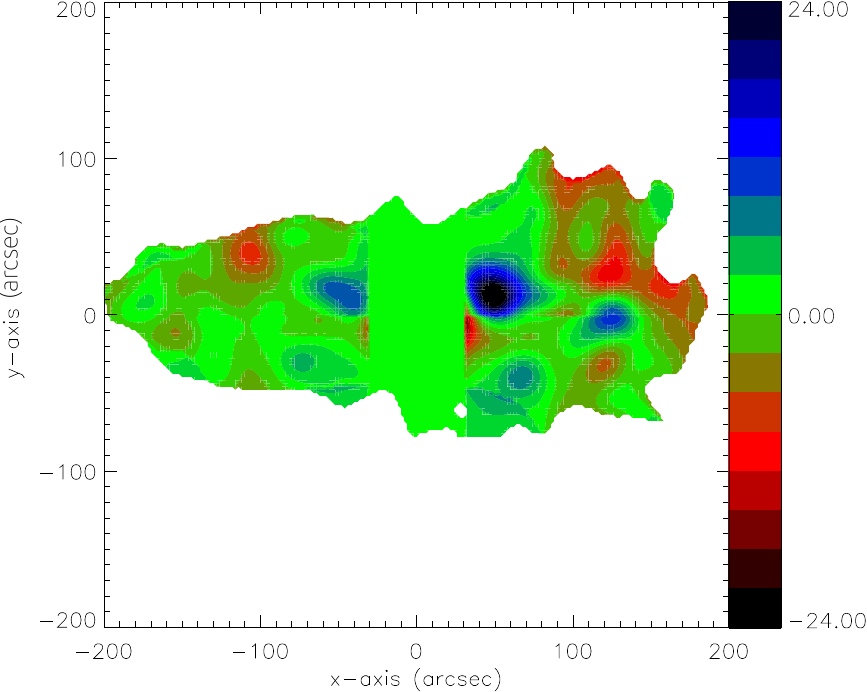}\put(-350,300){\bf \Huge NGC 4192 20cm}\put(-350,270){\bf \Huge CHANG-ES}}
  \caption{Radio continuum halo model of NGC~4192. Left column: profiles along the minor and major axes. Solid lines: observations, dashed lines: model
    including a a thin disk and a halo, dotted lines: model including only a thin disk. All distance are in arcseconds. Right column: maps of the model residuals in units of the rms.  
  \label{fig:ngc4192_profres1}}
\end{figure*} 
%\FloatBarrier

\begin{figure*}
  \centering
  \resizebox{14cm}{!}{\includegraphics{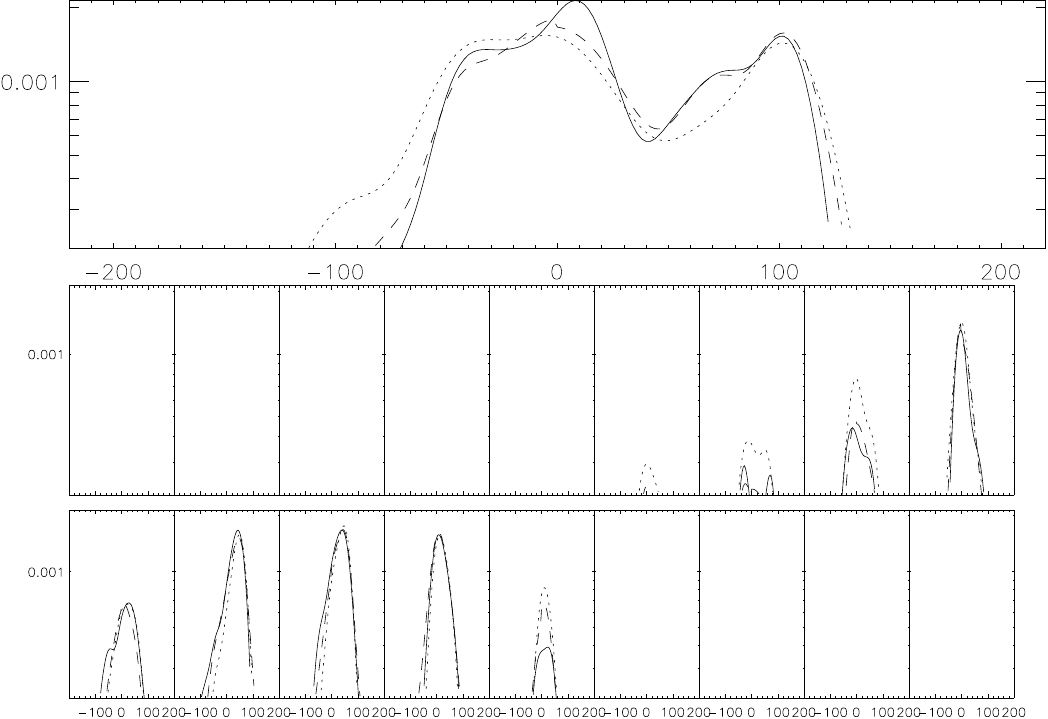}\put(-400,235){\bf \Huge NGC 4178 6cm}\includegraphics{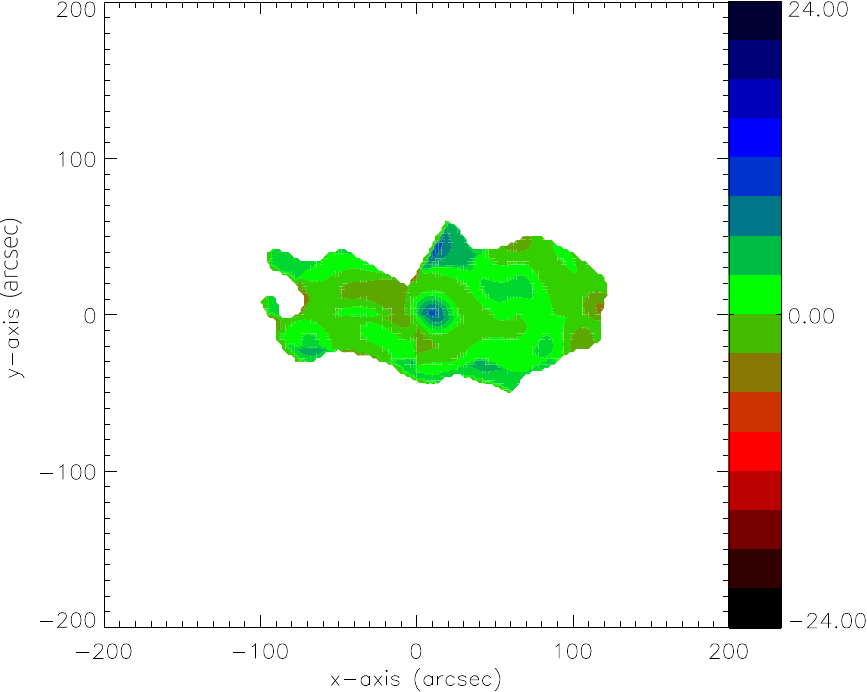}\put(-350,300){\bf \Huge NGC4178 6cm}}
  \resizebox{14cm}{!}{\includegraphics{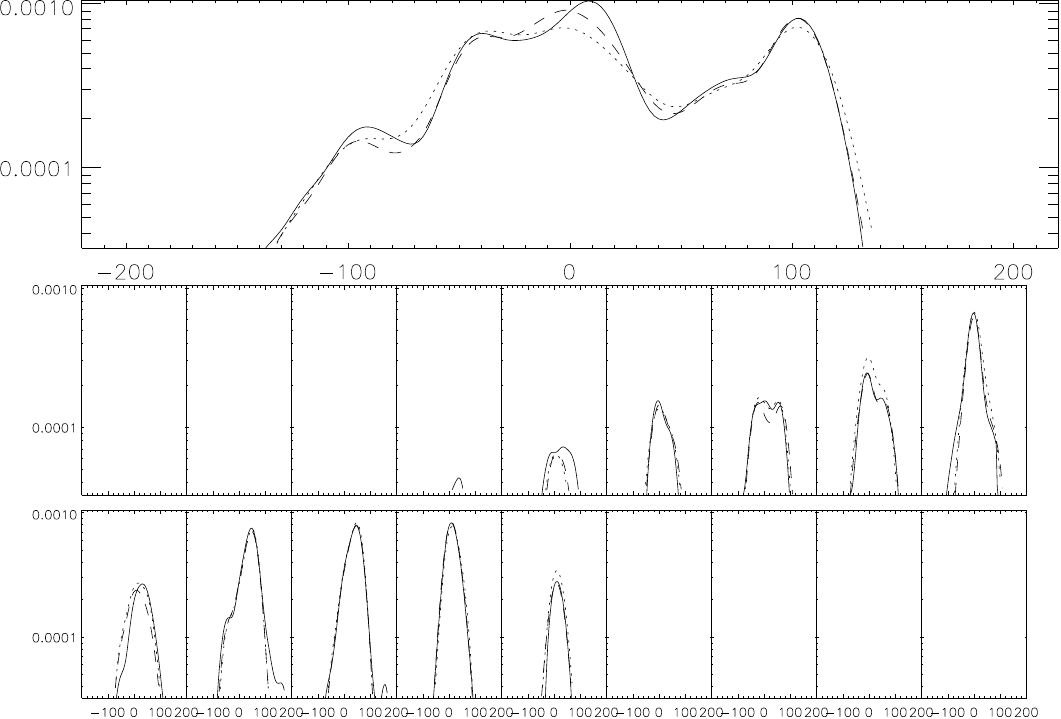}\put(-400,235){\bf \Huge NGC 4178 20cm}\includegraphics{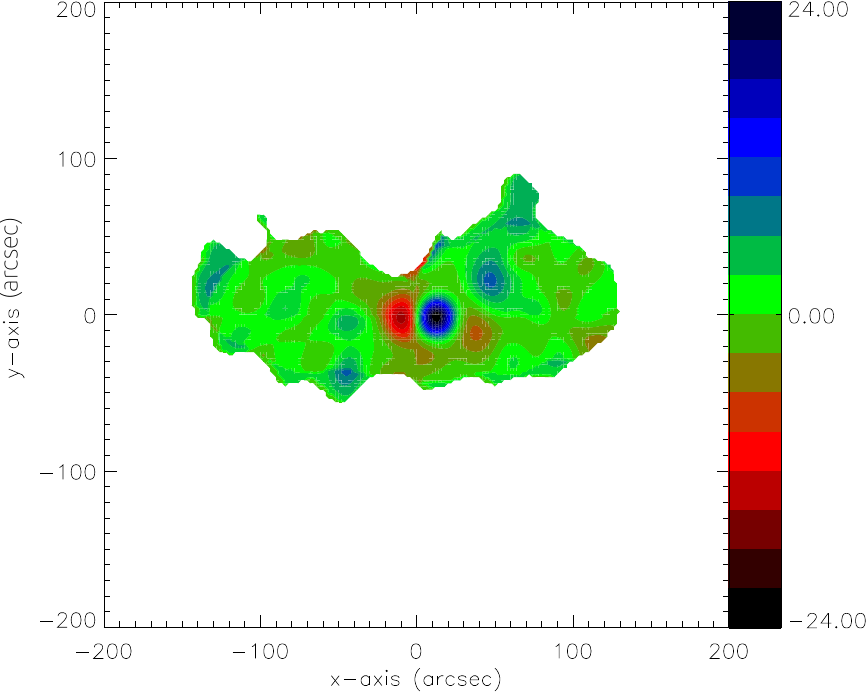}\put(-350,300){\bf \Huge NGC 4178 20cm}}
  \caption{Radio continuum halo model of NGC~4178. Left column: profiles along the minor and major axes. Solid lines: observations, dashed lines: model
    including a a thin disk and a halo, dotted lines: model including only a thin disk. All distance are in arcseconds. Right column: maps of the model residuals in units of the rms.
  \label{fig:ngc4178_profres}}
\end{figure*}
%\FloatBarrier

\begin{figure*}
  \centering
  \resizebox{14cm}{!}{\includegraphics{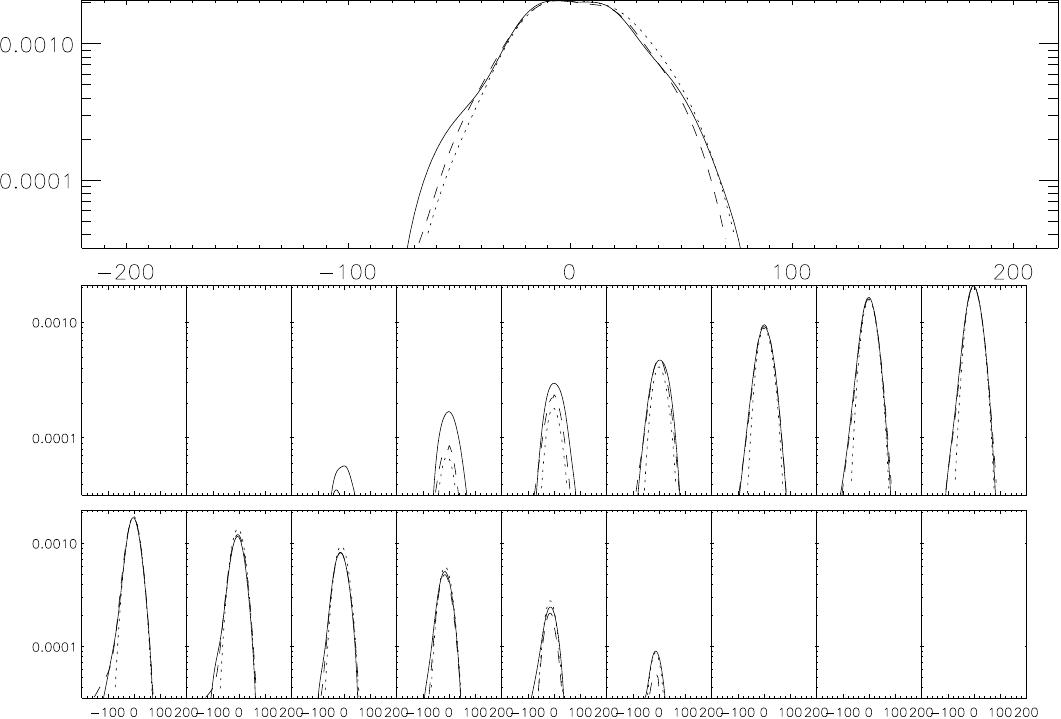}\put(-400,235){\bf \Huge NGC 4294 6cm}\includegraphics{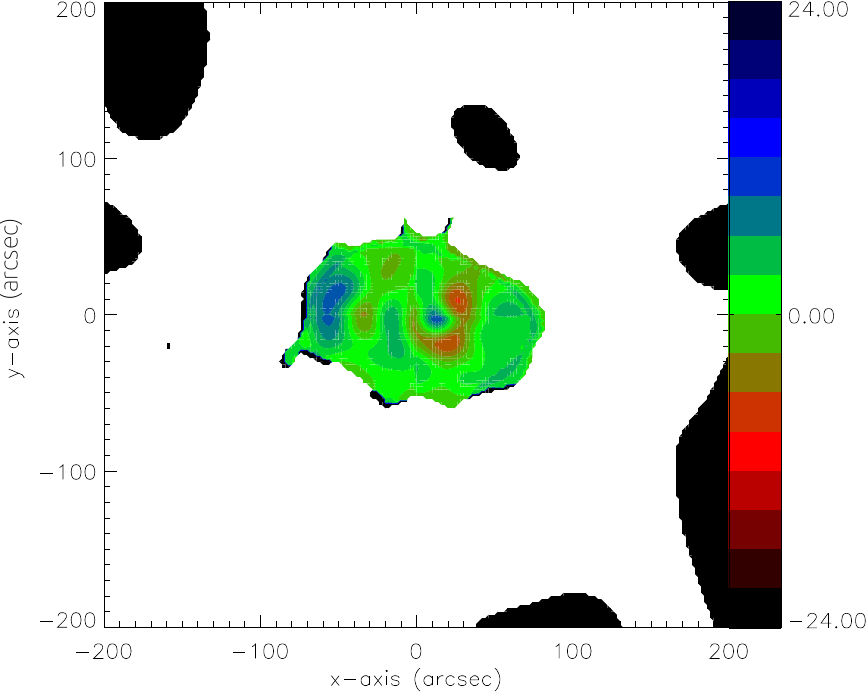}\put(-350,300){\bf \Huge NGC4294 6cm}}
  \resizebox{14cm}{!}{\includegraphics{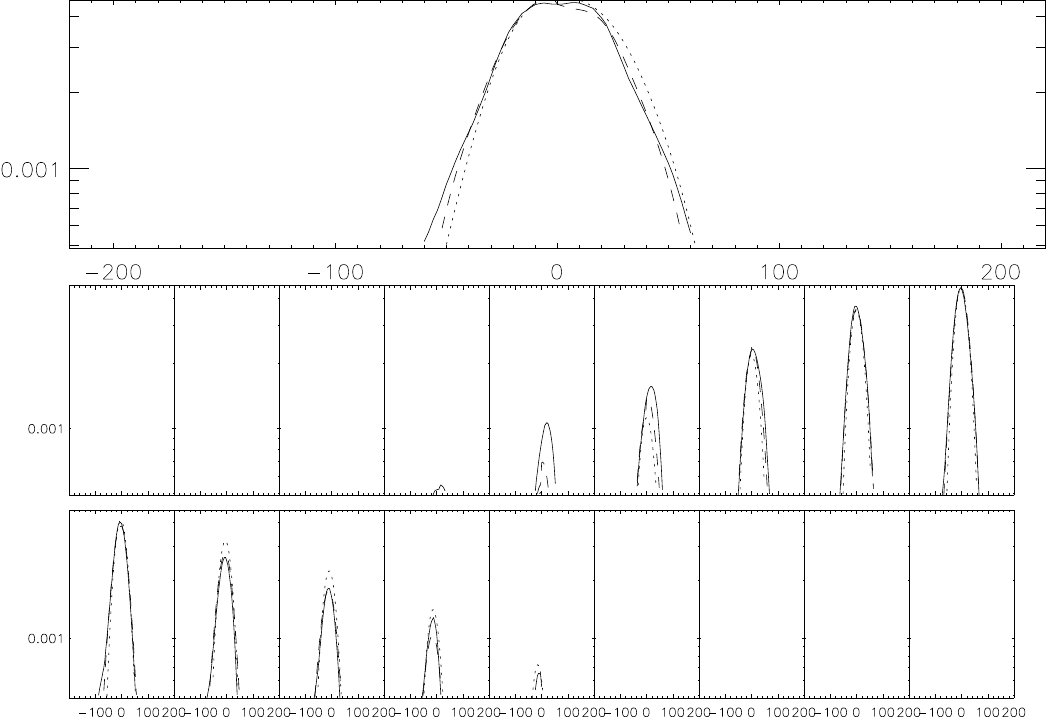}\put(-400,235){\bf \Huge NGC 4294 20cm VIVA}\includegraphics{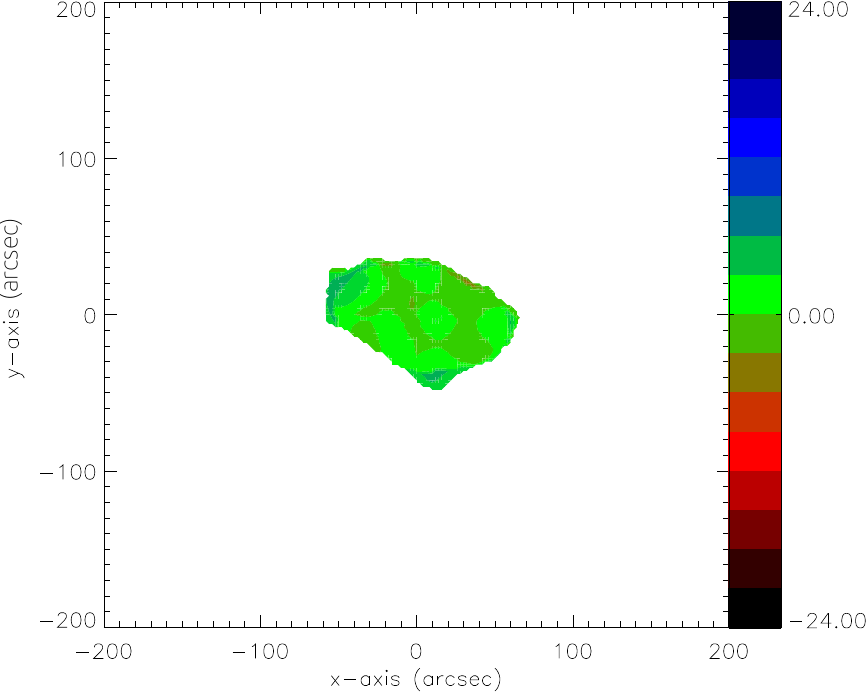}\put(-350,300){\bf \Huge NGC 4294 20cm VIVA}}
  \caption{Radio continuum halo models of NGC~4294. Left column: profiles along the minor and major axes. Solid lines: observations, dashed lines: model
    including a a thin disk and a halo, dotted lines: model including only a thin disk. All distance are in arcseconds. Right column: maps of the model residuals in units of the rms.
  \label{fig:ngc4294_profres}}
\end{figure*}
%\FloatBarrier

\begin{figure*}
  \centering
  \resizebox{14cm}{!}{\includegraphics{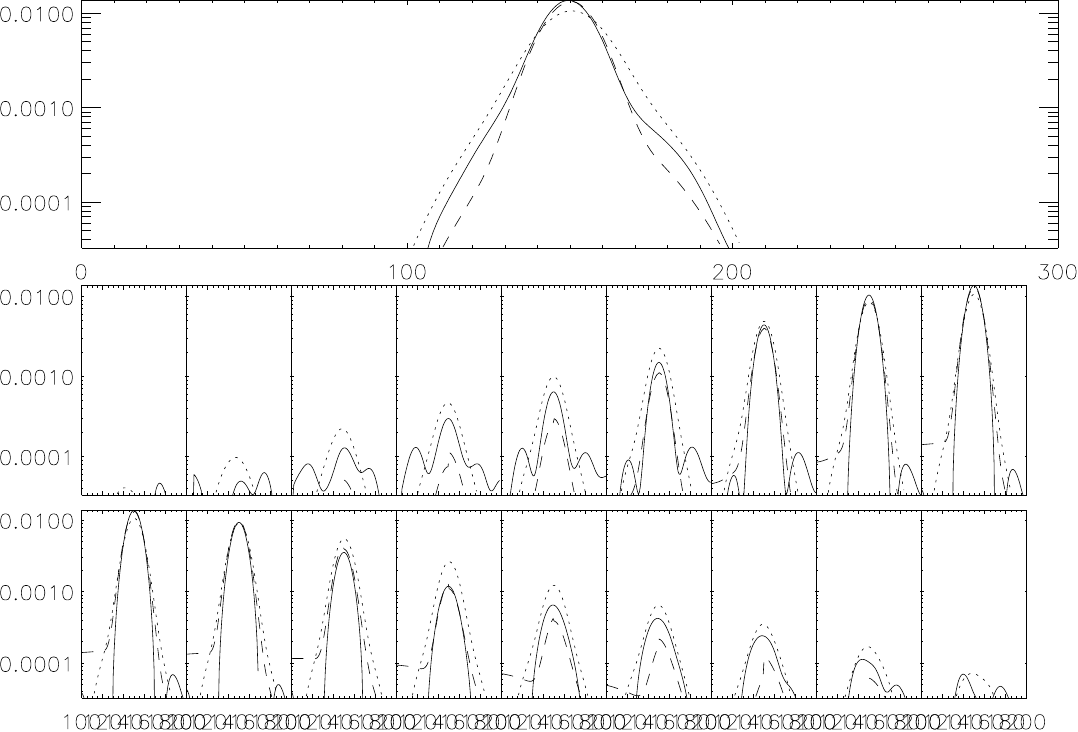}\put(-400,235){\bf \Huge NGC 4419 6cm}\includegraphics{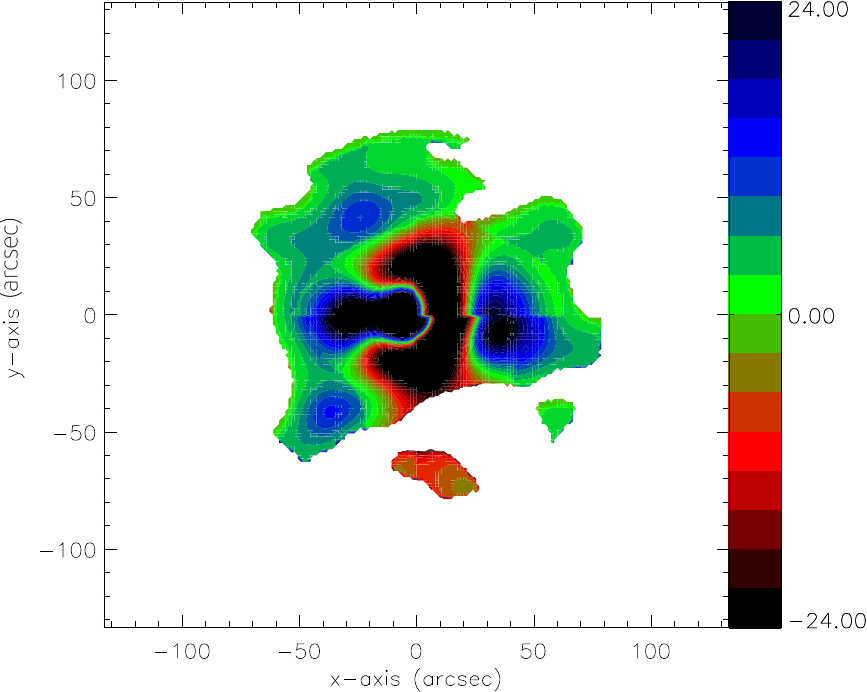}\put(-350,300){\bf \Huge NGC4419 6cm}}
  \resizebox{14cm}{!}{\includegraphics{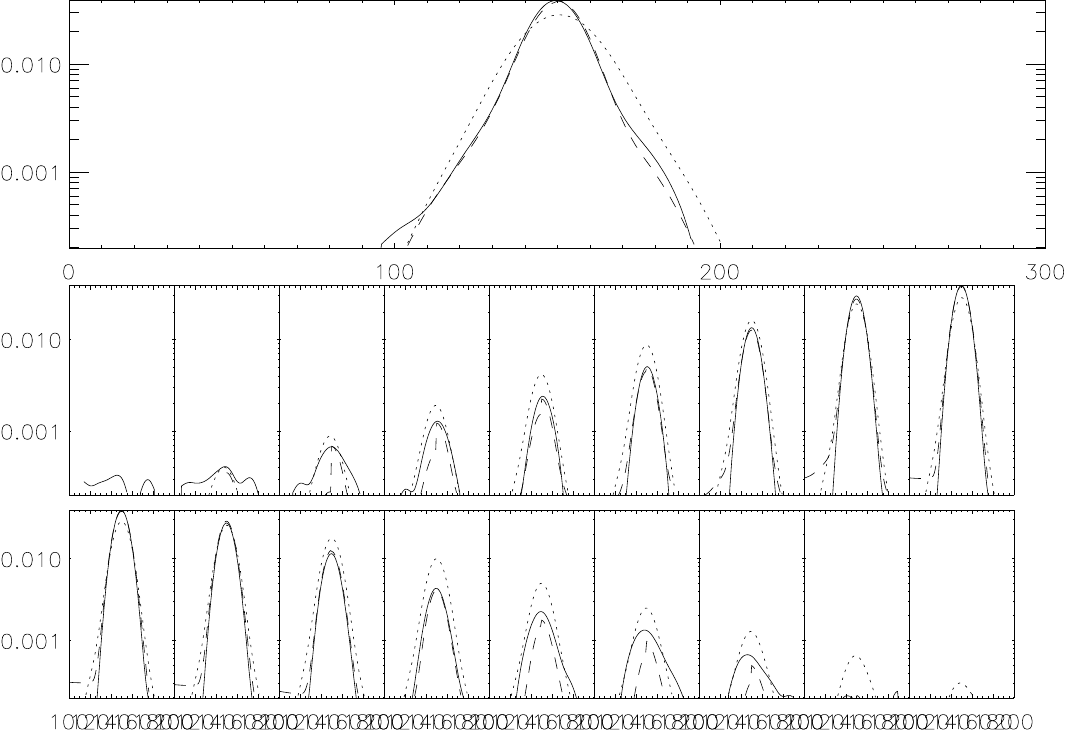}\put(-400,235){\bf \Huge NGC 4419 20cm}\includegraphics{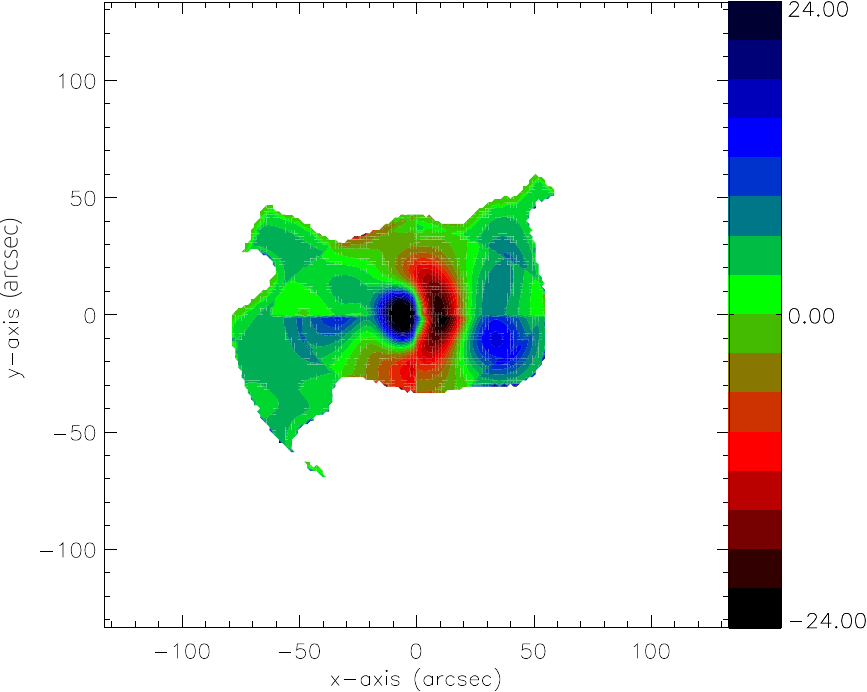}\put(-350,300){\bf \Huge NGC 4419 20cm}}
  \caption{Radio continuum halo models of NGC~4419. Left column: profiles along the minor and major axes. Solid lines: observations, dashed lines: model
    including a a thin disk and a halo, dotted lines: model including only a thin disk. All distance are in arcseconds. Right column: maps of the model residuals in units of the rms.
  \label{fig:ngc4419_profres}}
\end{figure*}
\FloatBarrier

\begin{figure*}
  \centering
  \resizebox{14cm}{!}{\includegraphics{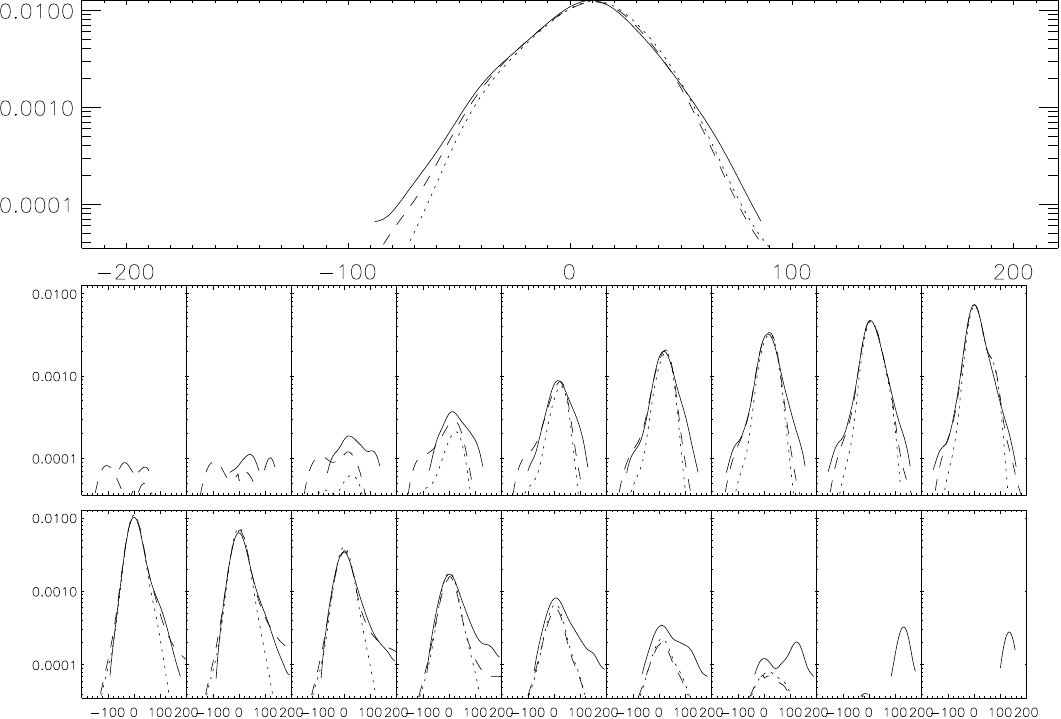}\put(-400,235){\bf \Huge NGC 4532 6cm}\includegraphics{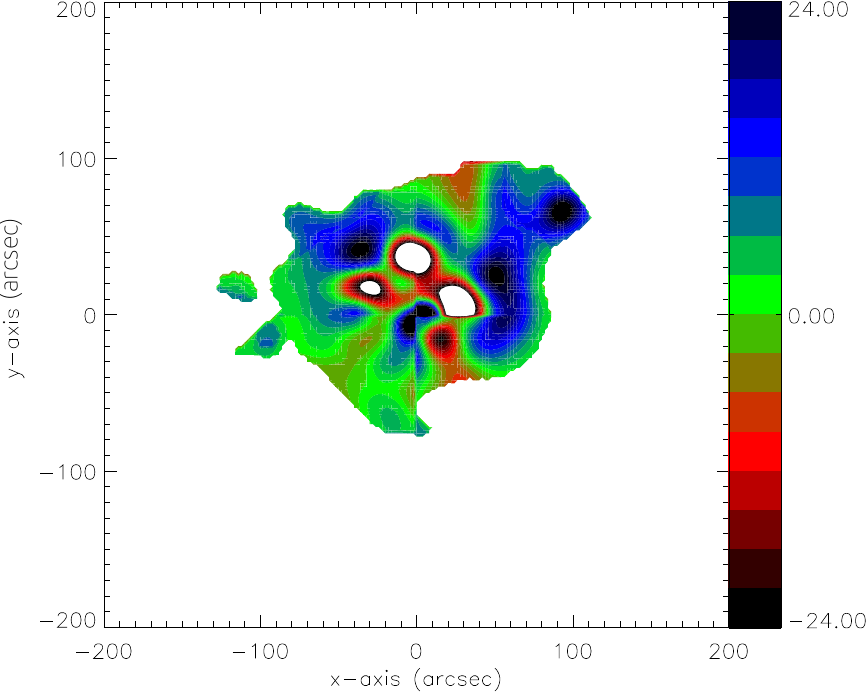}\put(-350,300){\bf \Huge NGC4532 6cm}}
  \resizebox{14cm}{!}{\includegraphics{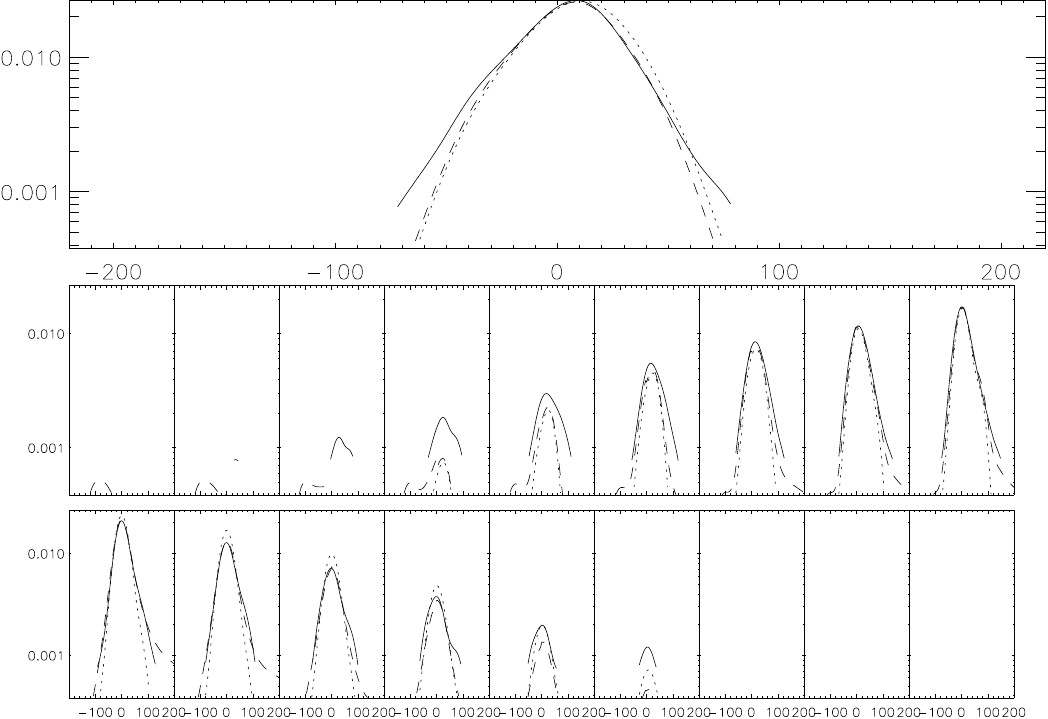}\put(-400,235){\bf \Huge NGC 4532 20cm}\includegraphics{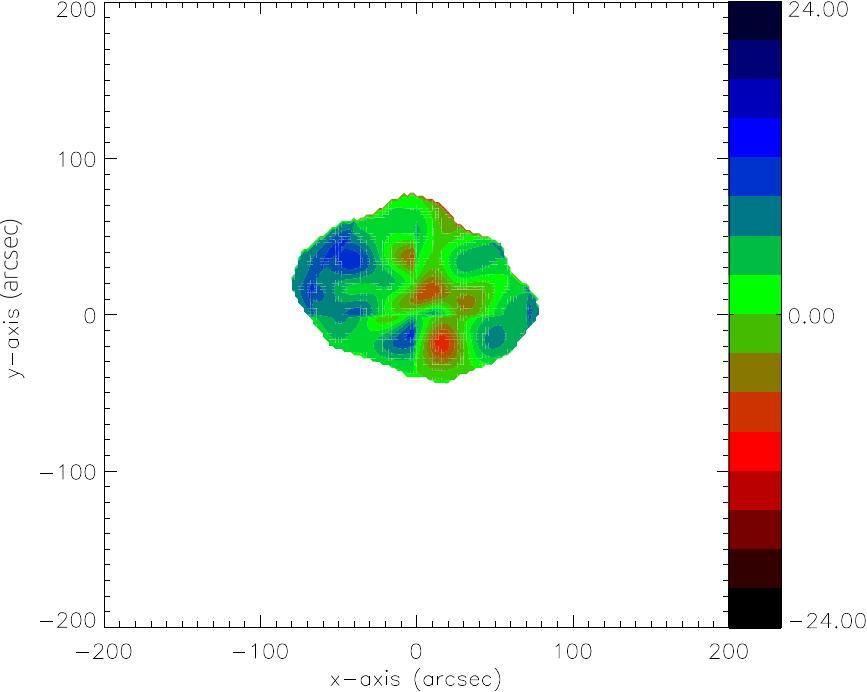}\put(-350,300){\bf \Huge NGC 4532 20cm}}
  \caption{Radio continuum halo models of NGC~4532. Left column: profiles along the minor and major axes. Solid lines: observations, dashed lines: model
    including a a thin disk and a halo, dotted lines: model including only a thin disk. All distance are in arcseconds. Right column: maps of the model residuals in units of the rms.
  \label{fig:ngc4532_profres}}
\end{figure*}
%\FloatBarrier

\begin{figure*}
  \centering
  \resizebox{14cm}{!}{\includegraphics{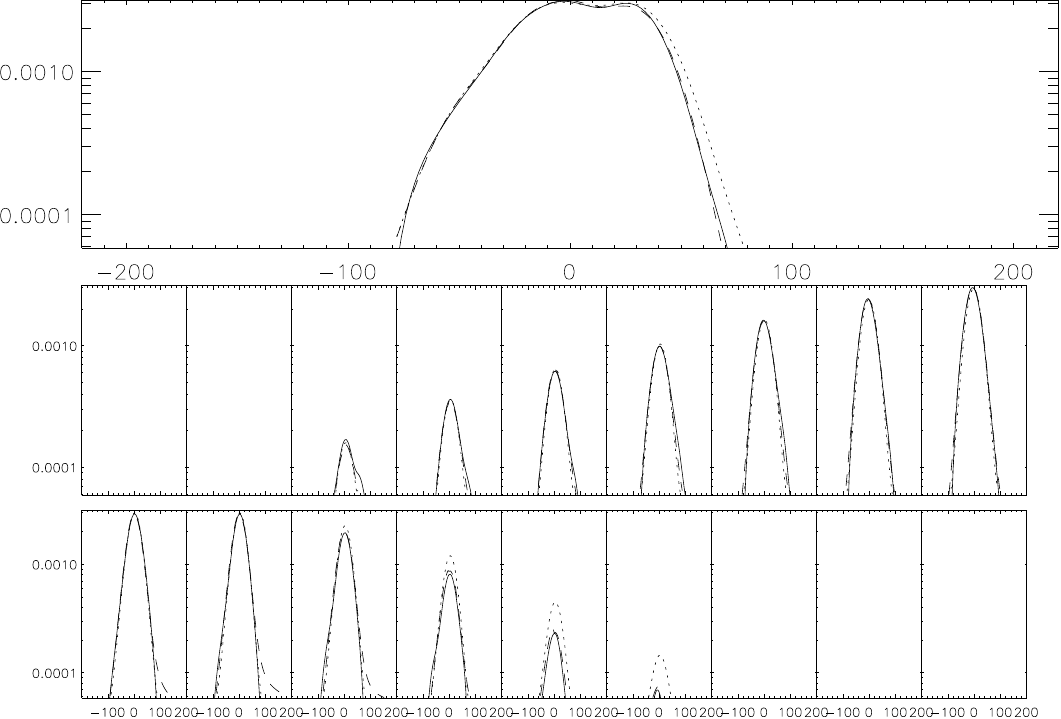}\put(-400,235){\bf \Huge NGC 4808 6cm}\includegraphics{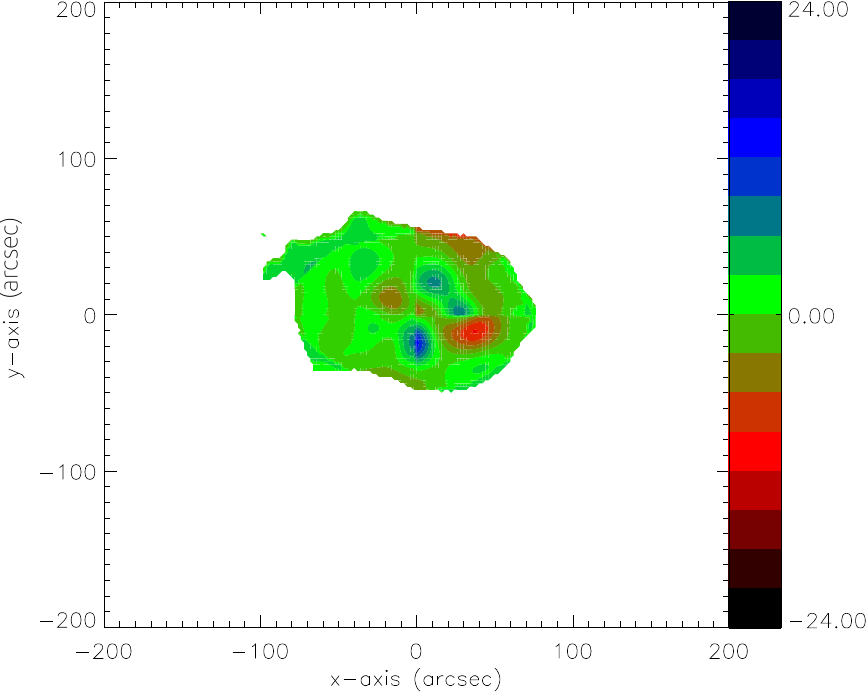}\put(-350,300){\bf \Huge NGC4808 6cm}}
  \resizebox{14cm}{!}{\includegraphics{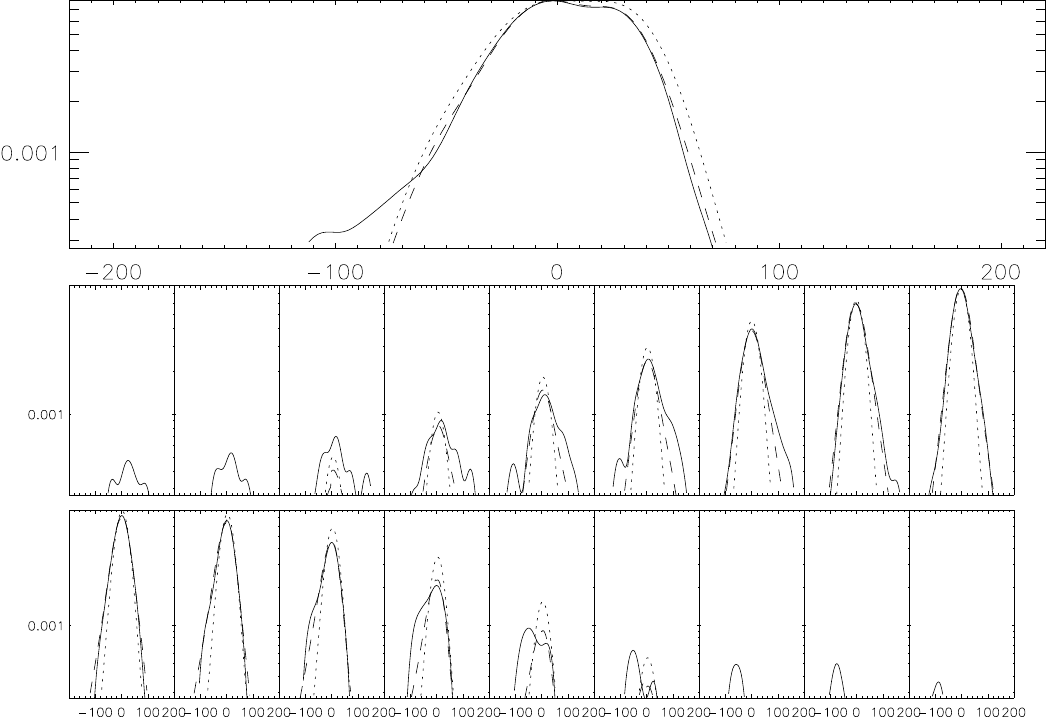}\put(-400,235){\bf \Huge NGC 4808 20cm}\includegraphics{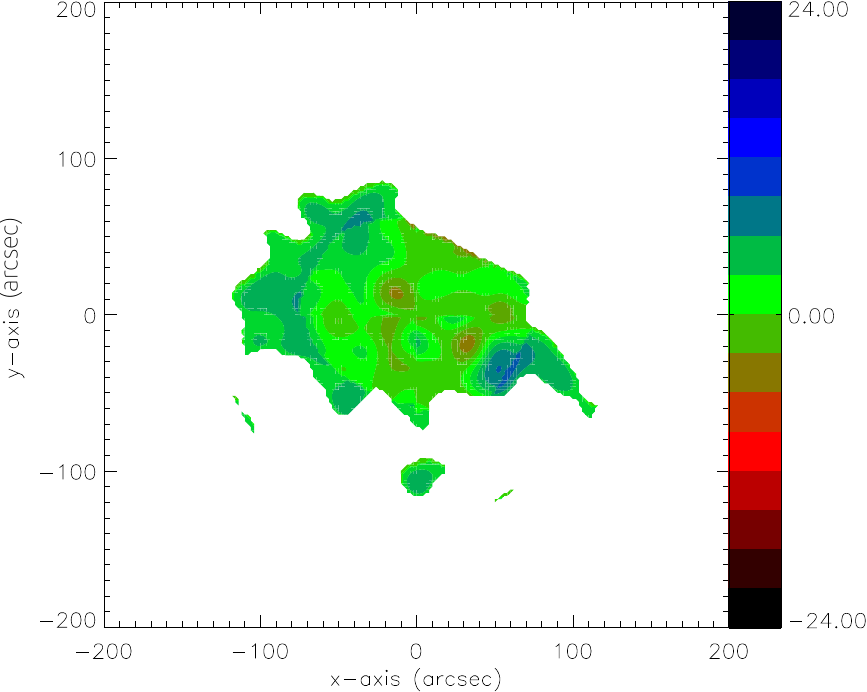}\put(-350,300){\bf \Huge NGC 4808 20cm}}
  \caption{Radio continuum halo models of NGC~4808. Left column: profiles along the minor and major axes. Solid lines: observations, dashed lines: model
    including a a thin disk and a halo, dotted lines: model including only a thin disk. All distance are in arcseconds. Right column: maps of the model residuals in units of the rms.
  \label{fig:ngc4808_profres}}
\end{figure*}
%\FloatBarrier

\FloatBarrier
\onecolumn
\section{Radio halo scale heights}

\begin{figure*}
  \centering
  \resizebox{16cm}{!}{\includegraphics{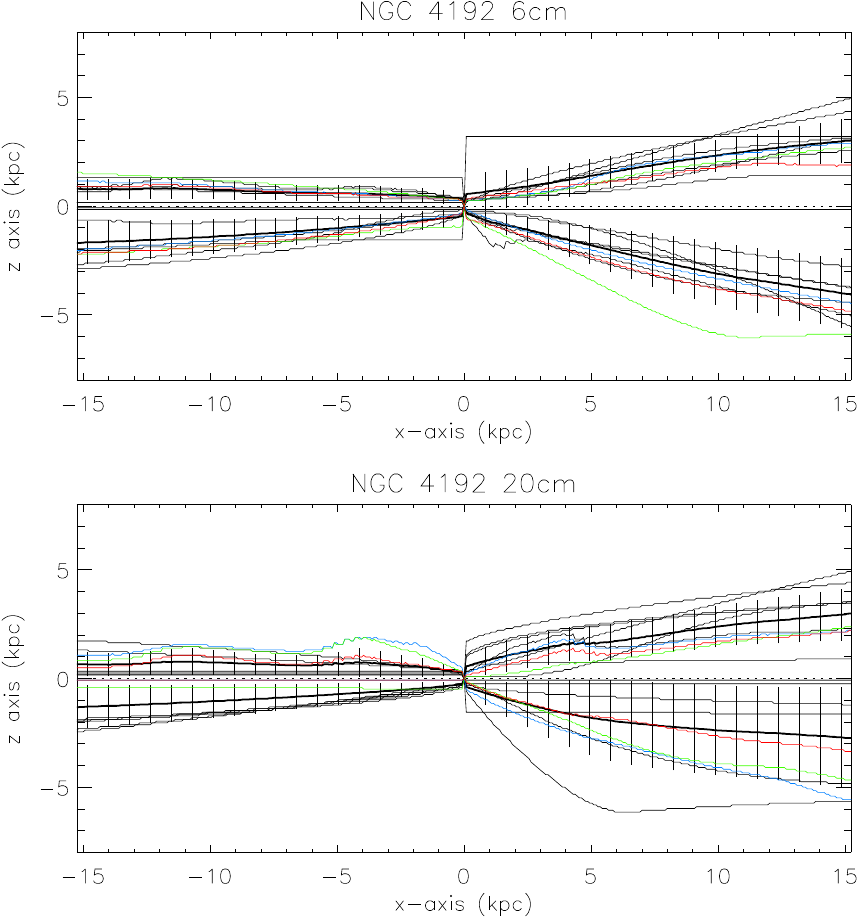}\includegraphics{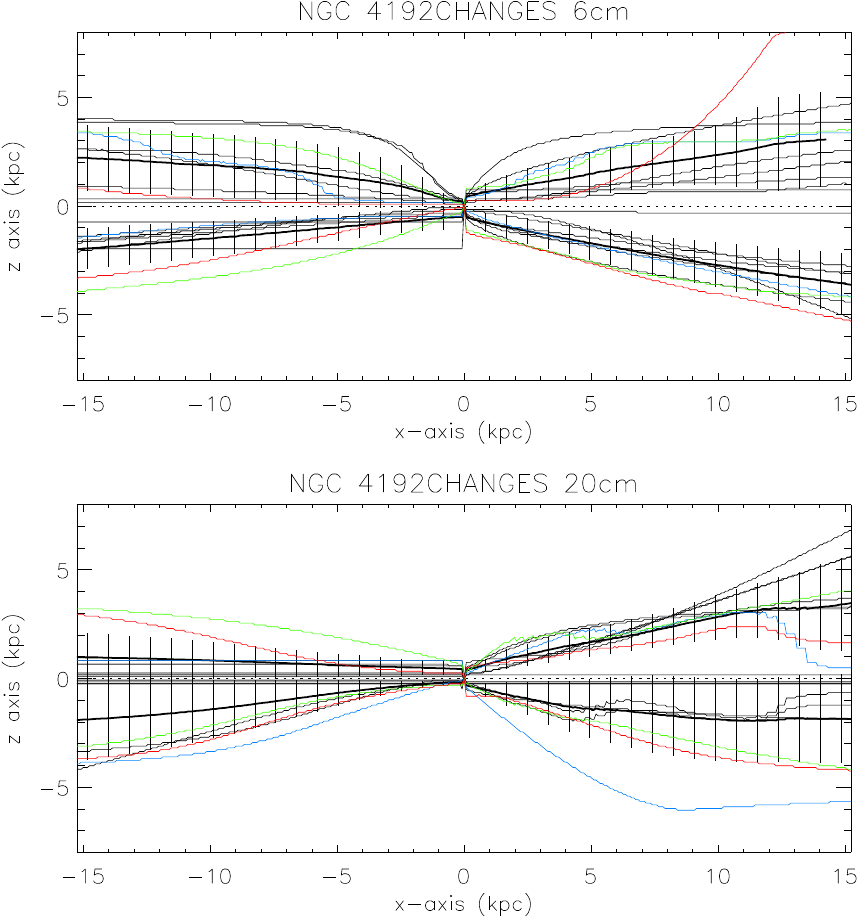}}
  \caption{Radial profiles of the halo scale height of NGC~4192. All ten models. Lower four panels: best four models.
    Models with a low $\chi^2$ are red, those with a high $\chi^2$ are blue.
  \label{fig:ngc4192H20a}}
\end{figure*}
%\FloatBarrier

\begin{figure*}
  \centering
  \resizebox{16cm}{!}{\includegraphics{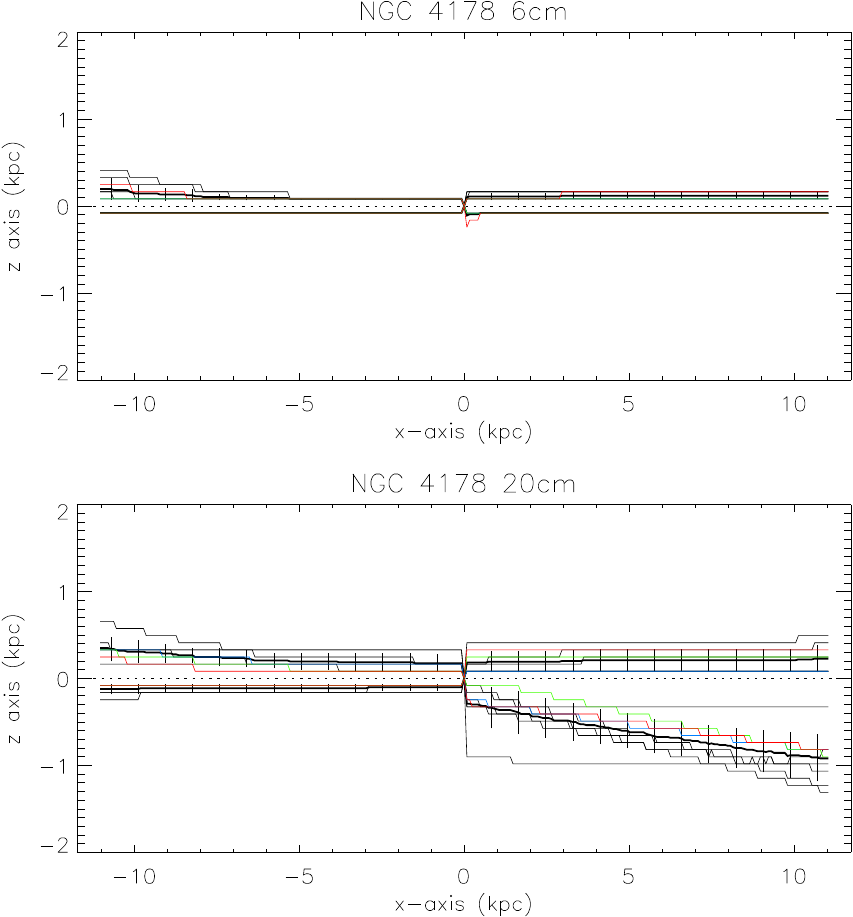}\includegraphics{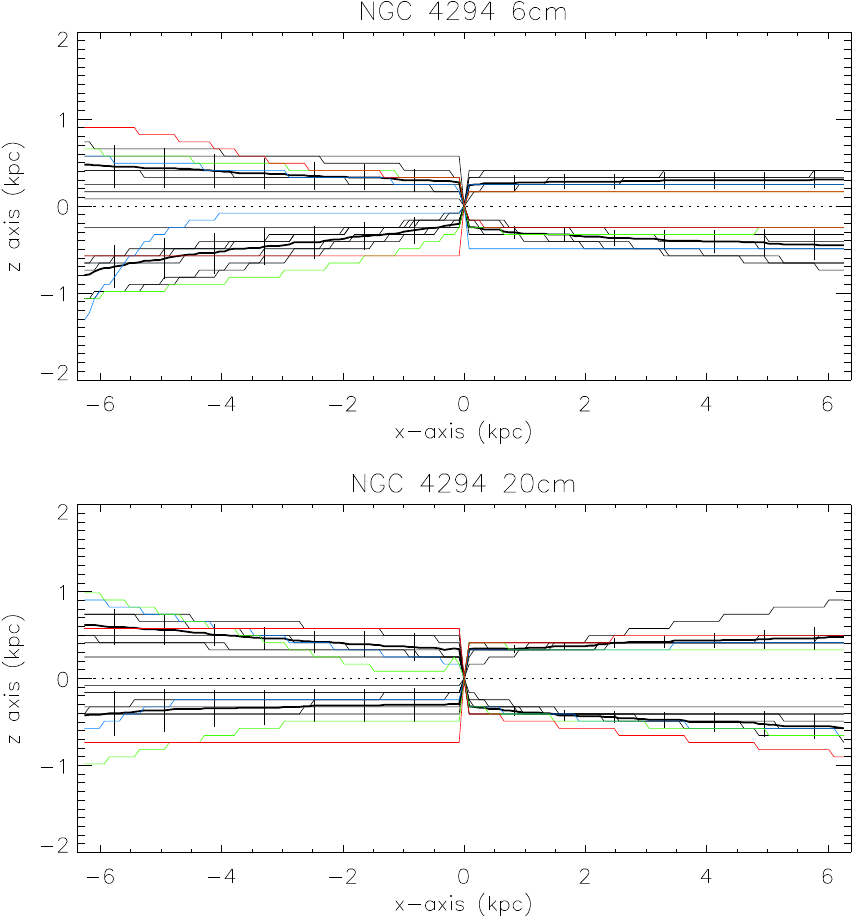}}
  \resizebox{16cm}{!}{\includegraphics{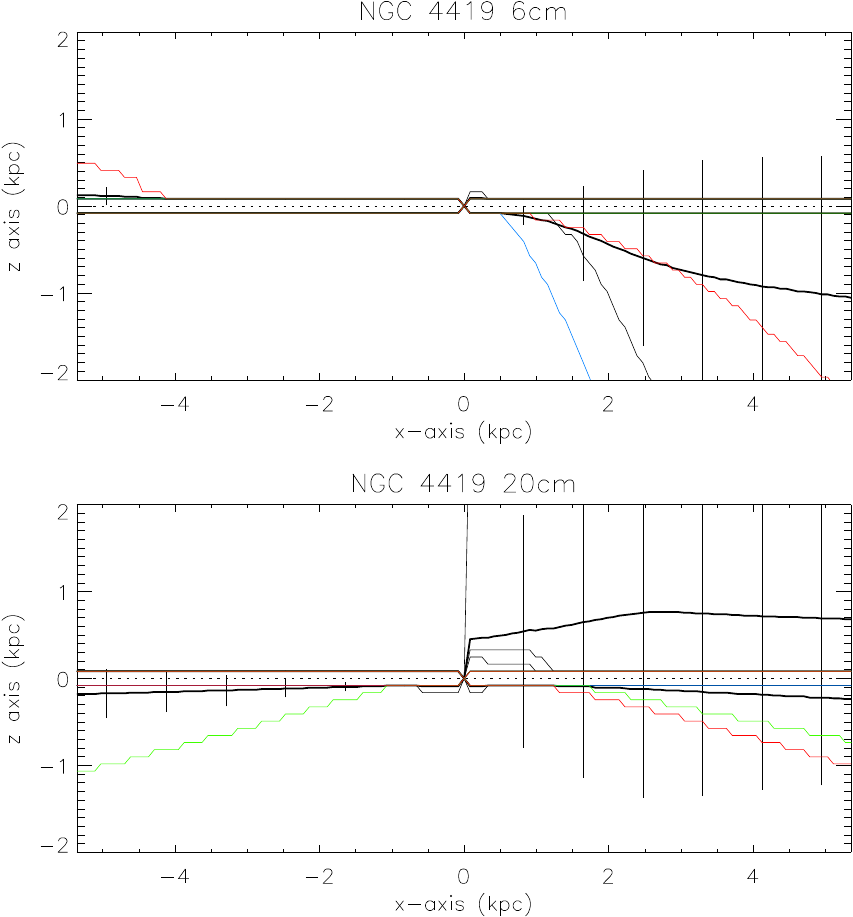}\includegraphics{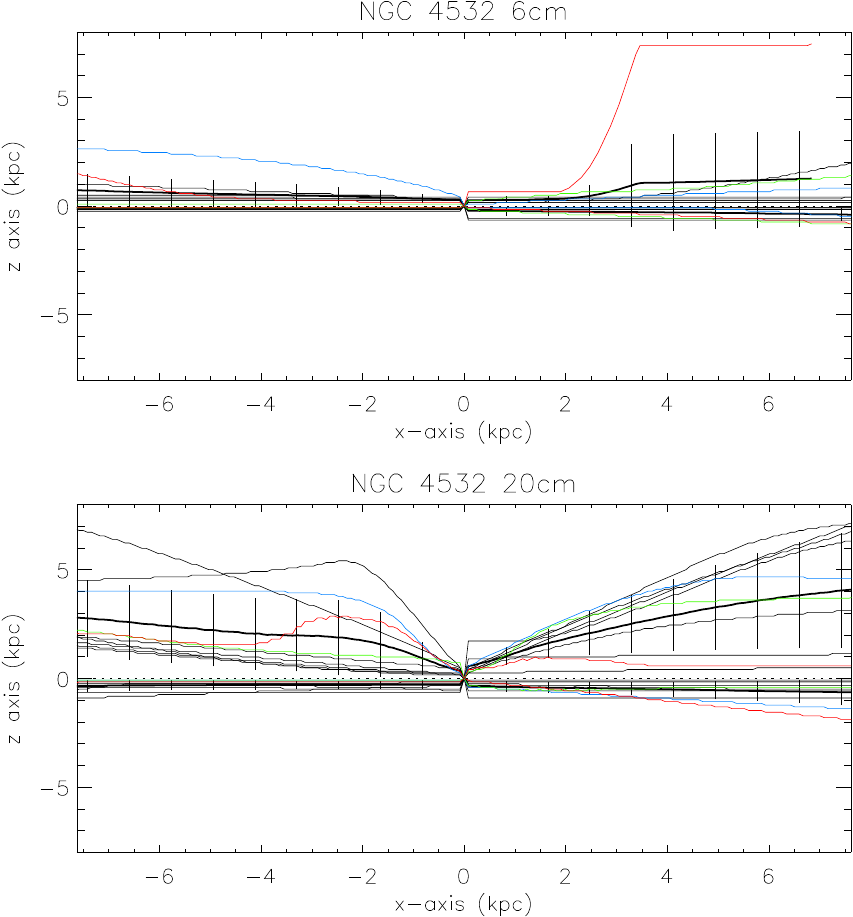}}
  \caption{Radial profiles of the halo scale height of NGC~4178, NGC~4294, NGC~4419, and NGC~4532. All ten models.
    Models with a low $\chi^2$ are red, those with a high $\chi^2$ are blue.
  \label{fig:ngc4178H20b}}
\end{figure*}
%\FloatBarrier

\begin{figure*}
  \centering
  \resizebox{16cm}{!}{\includegraphics{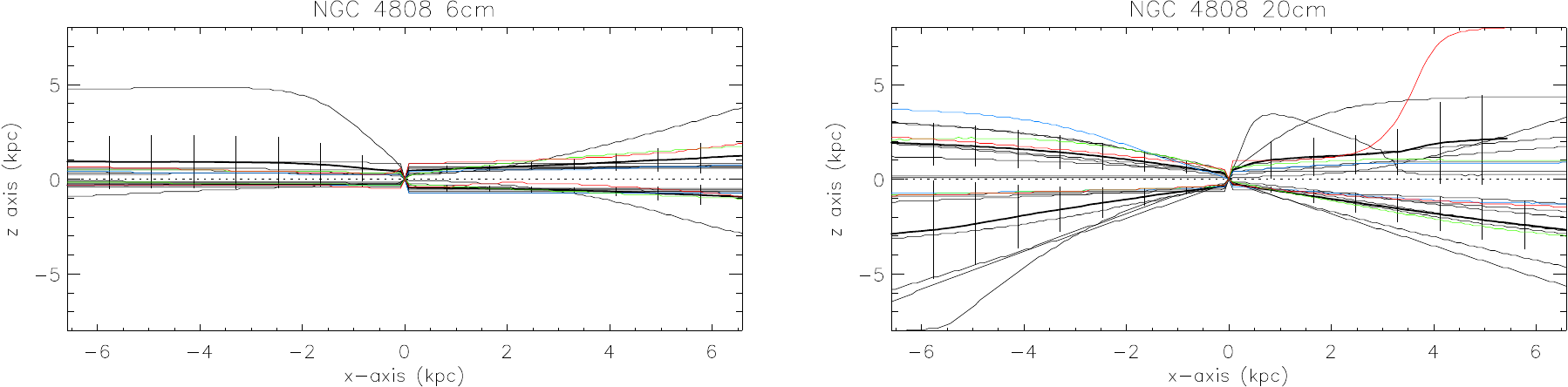}}
  \caption{Radial profiles of the halo scale height of NGC~4808. All ten models. 
    Models with a low $\chi^2$ are red, those with a high $\chi^2$ are blue.
  \label{fig:ngc4808H20}}
\end{figure*}
%\FloatBarrier

\begin{figure*}
  \centering
  \resizebox{14cm}{!}{\includegraphics{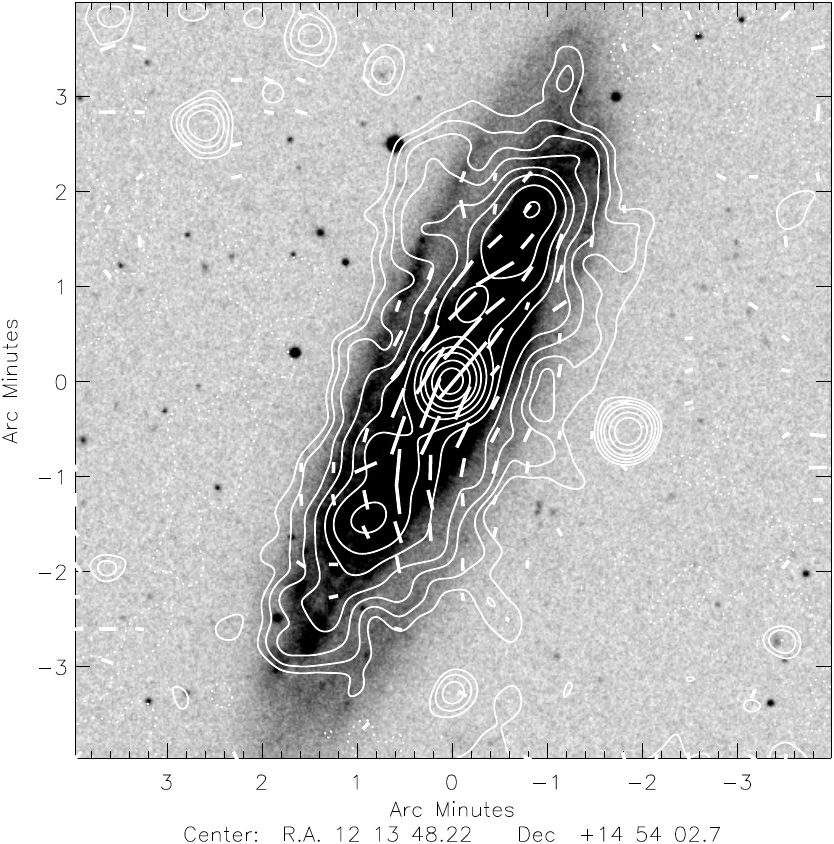}\put(-150,70){\bf \Huge NGC 4192}\includegraphics{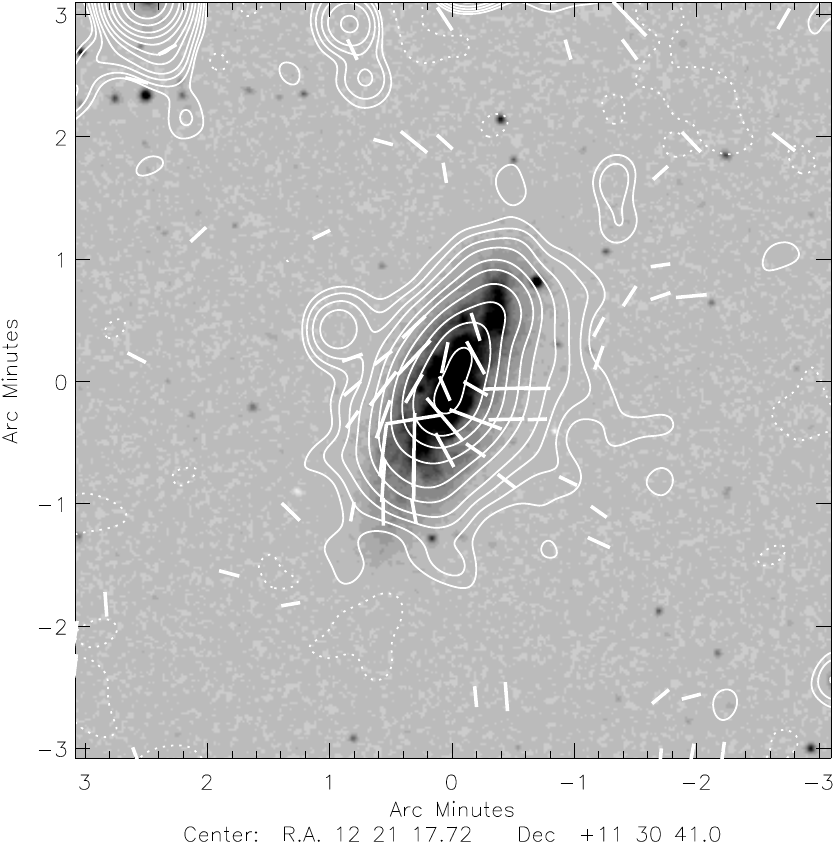}\put(-150,70){\bf \Huge NGC 4294}}
  \resizebox{14cm}{!}{\includegraphics{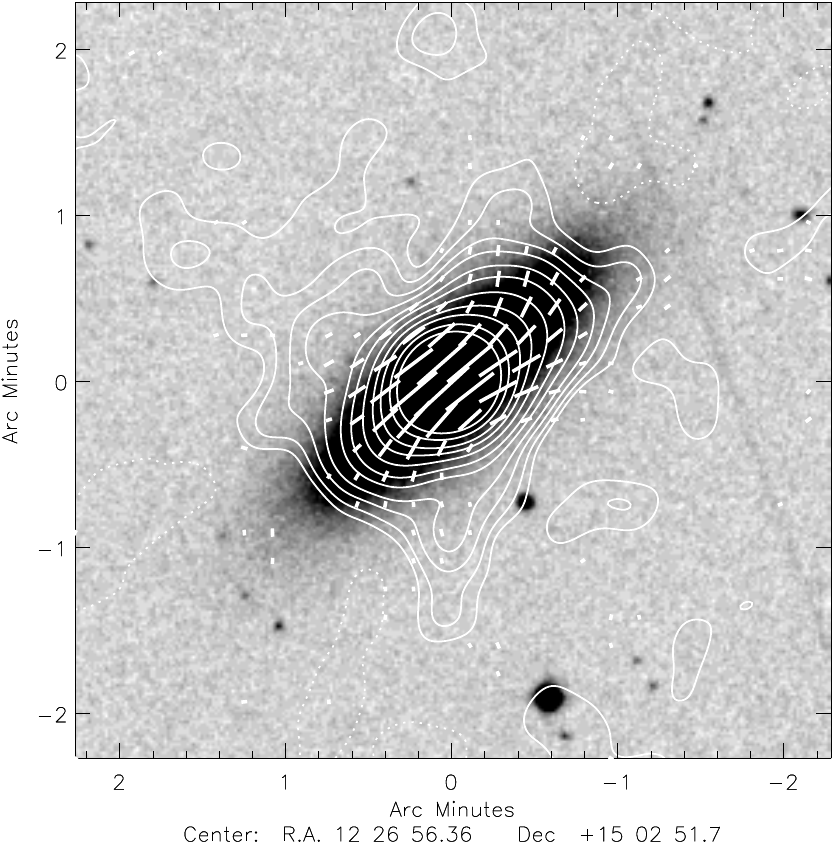}\put(-150,70){\bf \Huge NGC 4419}\includegraphics{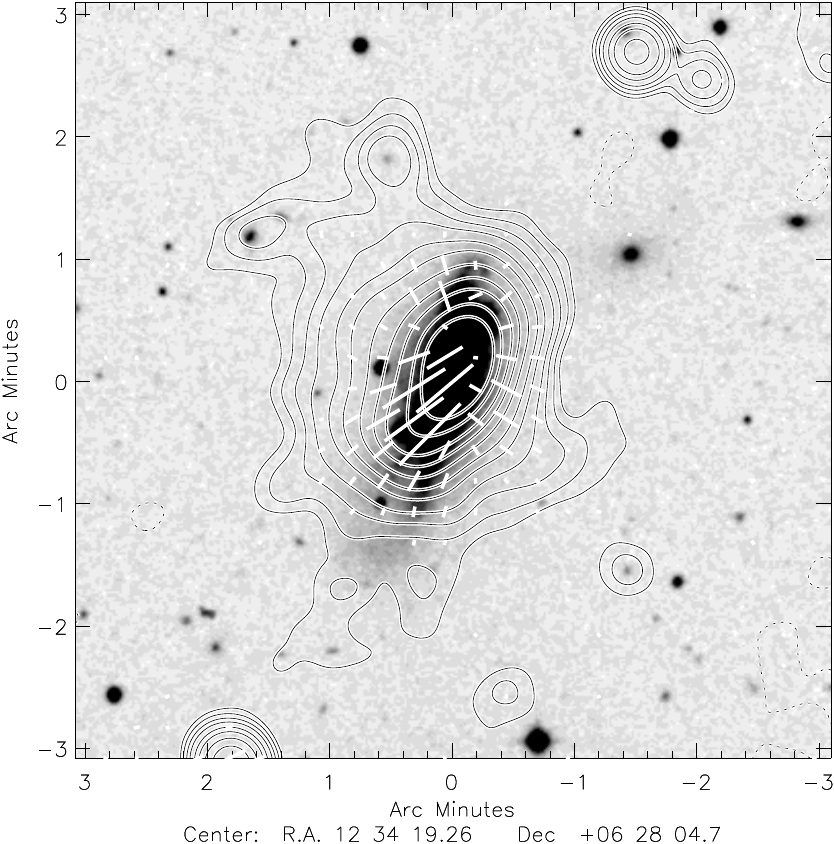}\put(-150,70){\bf \Huge NGC 4532}}
  \resizebox{14cm}{!}{\includegraphics{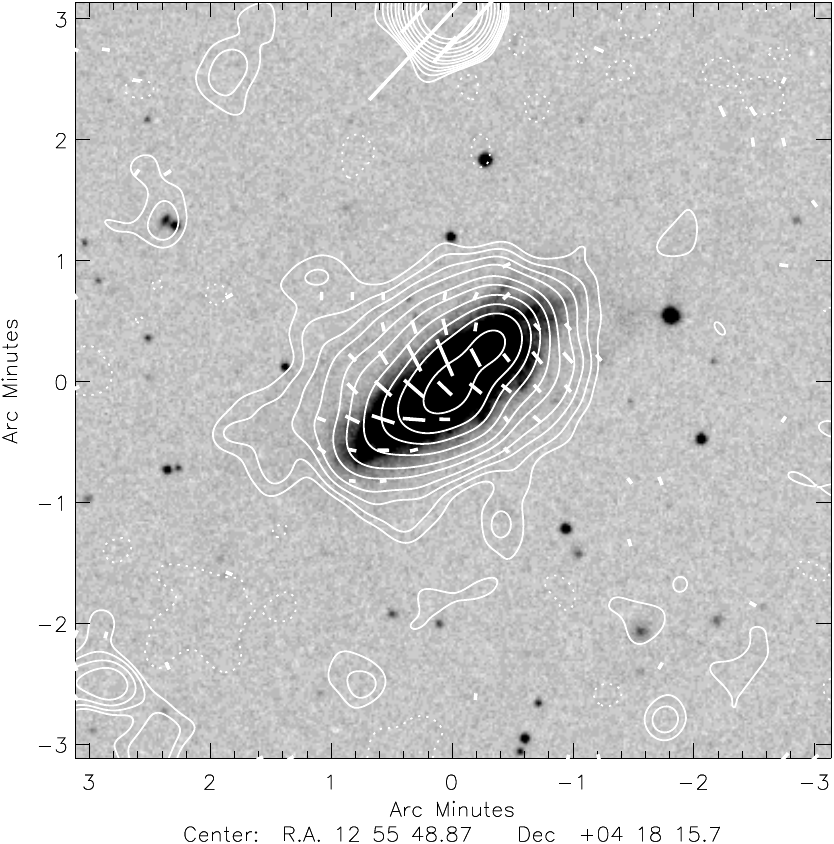}\put(-150,70){\bf \Huge NGC 4808}\textcolor{white}{\rule{14cm}{14cm}}}
  \caption{Total power emission distribution at 6~cm on DSS B band image together with the apparent B vectors (from Vollmer et al. 2013).
    Contour levels are $\xi \times (-3,3,5,8,12,20,30,50,80,120,200,300)$, with $\xi=16$~$\mu$Jy for NGC~4192,
    $\xi=10$~$\mu$Jy for NGC~4294,  $\xi=11$~$\mu$Jy for NGC~4419,  $\xi=18$~$\mu$Jy for NGC~4532, and $\xi=14$~$\mu$Jy
    for NGC~4808.
  \label{fig:bvectors}}
\end{figure*}
\twocolumn

\end{appendix}

\end{document}